\documentclass[a4paper,fleqn]{cas-sc}
\graphicspath{{./figs/}}

\usepackage[numbers]{natbib}

\def\tsc#1{\csdef{#1}{\textsc{\lowercase{#1}}\xspace}}
\tsc{WGM}
\tsc{QE}
\tsc{EP}
\tsc{PMS}
\tsc{BEC}
\tsc{DE}

\usepackage{amsmath}
\usepackage{amssymb}
\usepackage{hyperref}
\usepackage{ulem}
\usepackage[utf8]{inputenc}

\begin{document}
\let\WriteBookmarks\relax
\def\floatpagepagefraction{1}
\def\textpagefraction{.001}
\shorttitle{Dark degeneracy I: Dynamical or interacting dark energy?}
\shortauthors{Rodrigo von Marttens et~al.}

\title [mode = title]{Dark degeneracy I: Dynamical or interacting dark energy?}     

\author[1,2]{Rodrigo~von~Marttens}
\cormark[1]
\ead{rodrigovonmarttens@gmail.com}

\author[1]{Lucas~Lombriser}
\ead{lucas.lombriser@unige.ch}

\author[1]{Martin~Kunz}
\ead{martin.kunz@unige.ch}

\author[3]{Valerio~Marra}
\ead{valerio.marra@me.com}

\author[4,5]{Luciano~Casarini}
\ead{casarini.astro@gmail.com}

\author[2]{Jailson~Alcaniz}
\ead{alcaniz@on.br}

\address[1]{D\'epartement de Physique Th\'eorique, Universit\'e de Gen\`eve, 24 quai Ernest Ansermet, 1211 Gen\`eve 4, Switzerland}
\address[2]{Departamento de Astronomia, Observat\'orio Nacional, 20921-400, Rio de Janeiro, RJ, Brasil}
\address[3]{N\'ucleo Cosmo-ufes \& Departamento de F\'isica, Universidade Federal do Esp\'irito Santo, 29075-910, Vit\'oria, ES, Brasil}
\address[4]{Departamento de F\'isica, Universidade Federal de Sergipe, 49100-000, Aracaju, SE, Brasil}
\address[5]{Institute of Theoretical Astrophysics, University of Oslo, 0315 Oslo, Norway}
\cortext[cor1]{Corresponding author}

\begin{abstract}
We revisit the dark degeneracy that arises from the Einstein equations relating geometry to the total cosmic substratum but not resolving its individual components separately.
We establish the explicit conditions for the dark degeneracy in the fluid description of the dark sector.
At the background level, this degeneracy can be formally understood in terms of a unified dark sector Equation of State (EoS) that depends both on the dynamical nature of the dark energy (DE) as well as on its interaction with the pressureless dark matter.
For linear perturbations, the degeneracy arises for specified DE pressure perturbations (or sound speed, equivalently) and DE anisotropic stress.
Specializing to the degeneracy between non-interacting dynamical DE and interacting vacuum DE models, we perform a parameter estimation analysis for a range of dynamical DE parametrizations, where for illustration we explicitly break the degeneracy at the linear level by adopting a luminal sound speed for both scenarios.
We conduct this analysis using cosmological background data alone and in combination with Planck measurements of the cosmic microwave background radiation.
We find that although the overall phenomenology between the dynamical DE and interacting approaches is similar, there are some intriguing differences.
In particular, there is an ambiguity in the strength of constraints on $\Omega_{m0}$ and $\sigma_8$, which are considerably weakened for interacting vacuum DE, indicating that the dark degeneracy can change the significance of tensions in cosmological parameters inferred from different data sets.
\end{abstract}

\begin{keywords}
cosmology \sep dark energy \sep bayesian inference
\end{keywords}

\maketitle

\section{Introduction}
\label{sec.intro}

The Coma galaxy cluster mass obtained from the virial theorem in the 1930s~\cite{1933AcHPh...6..110Z,1937ApJ....86..217Z}, followed by the study of galaxy rotation curves in the 1970s~\cite{1970ApJ...159..379R} and the distance measurements of type Ia supernovae (SNe) in the 1990s~\cite{Perlmutter:1998np,Riess:1998cb} have brought to light one of the most intriguing problems of modern cosmology: the dark sector. While ordinary matter (baryons) and radiation compose only about 5\% of the current energy content of our Universe, this exotic dark component contributes the other 95\%, dominating the present cosmic substratum.
The dark sector is thought to be composed of two constituents: dark matter and dark energy (DE).
Each is important for understanding phenomena of different nature and scales.
The importance of dark matter lies mainly in structure formation, for example, to allow baryonic structures to become non-linear after decoupling from the photons.
In contrast, DE is the component responsible for the observed late-time accelerated expansion of the Universe.
In the standard cosmological picture, dark matter is described as a cold, i.e., pressureless, matter component (CDM) and DE is described by the cosmological constant $\Lambda$.
The
$\Lambda$CDM model combines the dark sector with baryons and radiation within the context of General Relativity (GR).
It is an impressive fact that this simple description of the Universe reproduces the range of cosmological observables of very different nature using the same set of cosmological parameters, consequently assigning to $\Lambda$CDM the status of a concordance model.

However, even though the $\Lambda$CDM model enjoys a considerable observational success~\cite{Aghanim:2018eyx,Alam:2016hwk,Troxel:2017xyo}, there remain a number of theoretical and observational points that deserve to be thoroughly investigated~\cite{Buchert:2015wwr}.
From the theoretical perspective, the biggest challenge lies in understanding the fundamental nature of these dark components.
This encompasses, for instance, the attempt of associating the DE component with the vacuum energy in the context of quantum field theory (leading to the cosmological constant problem~\cite{Weinberg:1988cp,Martin:2012bt,Lombriser:2019jia}) or the search for particles beyond the Standard Model for describing CDM~\cite{darkmatter}.
From an observational point of view, besides the small-scale problems in galaxy formation~\cite{DelPopolo:2016emo}, recent cosmological observations indicate a considerable discrepancy in $\Lambda$CDM with the parameter constraints from Planck~\cite{Aghanim:2018eyx}, suggesting the possibility that ``new physics'' may need to be accounted for to resolve the tension.
For example, observations of the large-scale structure from weak lensing with the Kilo-Degree Survey (KiDS)~\cite{Hildebrandt:2016iqg,Joudaki:2017zdt} exhibit a $2.6\sigma$ discrepancy with the latest Planck results on the amplitude of matter fluctuations.
In contrast, the joint lensing-galaxy clustering analysis of the Dark Energy Survey (DES)~\cite{Troxel:2017xyo,Abbott:2017wau,Krause:2017ekm} finds a considerably higher value of this amplitude, consistent with the Planck data.
A larger tension currently faced by the $\Lambda$CDM model concerns the measurement of the present value of the Hubble constant $H_0$ describing the expansion rate of the Universe when inferred from late and early physics~\cite{Verde:2019ivm} (cf.~\cite{Lombriser:2019ahl}).
Whereas the last Planck~\cite{Aghanim:2018eyx} result gives $H_{0}=67.36\pm0.54 {\rm \ Km\ s^{-1}\ Mpc^{-1}}$, some local measurements are in more than $4\sigma$ tension with this result, e.g., measurements from type Ia SNe calibrated with cepheids ($H_{0}=74.03\pm1.42 {\rm \ km\ s^{-1}\ Mpc^{-1}}$~\cite{Riess:2019cxk} or $75.66 \pm 1.69 \text{ km s}^{-1} {\rm Mpc}^{-1}$~\cite{Camarena:2019moy}) and measurements from type Ia SNe using tip of the red giant branch stars ($H_{0}=72.4\pm 1.9 {\rm \ km\ s^{-1}\ Mpc^{-1}}$~\cite{Yuan:2019npk,Freedman:2019jwv}).

In trying to solve, or at least alleviate, the theoretical and potentially observational shortcomings of the current concordance model, a range of alternative proposals have been made for the dark sector.
Within the context of GR, two important phenomenological alternatives consist in: ($i$)~the possibility that DE is not characterized by the vacuum equation of state (EoS) ($w_{x}=-1$) but a dynamical (time-dependent) EoS $w_{x}(a)$ instead~\cite{Chevallier:2000qy,Linder:2002et,Barboza:2008rh,Wetterich:2004pv}; ($ii$) the possibility that the dark components are not independent but can exchange energy/momentum with each other~\cite{Marttens:2017njo,Gonzalez:2018rop,Benetti:2019lxu}. 

In this paper, rather than initially focusing on any specific alternative model for describing the dark sector, we will first establish how different models of the dark sector can yield exactly the same cosmological observables such as the Hubble expansion rate at the background level and the gravitational potentials and gravitational waves at the perturbation level.
This is a consequence of the dark degeneracy~\cite{Wasserman:2002gb,Rubano:2002sx,Kunz:2007rk,Aviles:2011ak,Carneiro:2014uua}, the fact that the Einstein equations only constrain the total energy-momentum tensor but not each matter component separately.
More precisely, for $n$ components of the visible sector, the Einstein equations can be expressed as
\begin{equation} \label{einsteindark}
T_{\mu\nu}^{(d)}=\dfrac{G_{\mu\nu}}{\kappa}-\sum_{i=1}^{n} T_{\mu\nu}^{(i)} \,,
\end{equation}
where $\kappa\equiv8\pi G$ with the gravitational constant $G$ and the speed of light in vacuum $c$ set to unity.
The relation implies that for a given theory of gravity (described by the Einstein tensor $G_{\mu\nu}$ or a generalization thereof) and specified components of the visible sector (defining the energy-momentum tensors $T_{\mu\nu}^{(j)}$), there are an infinite number of possibilities of splitting the dark energy-momentum tensor $T_{\mu\nu}^{(d)}$ in Eq.~(\ref{einsteindark}).
The dark degeneracy for the case of scalar-tensor modifications of gravity in the presence of a single CDM component has been studied in Refs.~\cite{Lombriser:2014ira,Lombriser:2015sxa,Lombriser:2016yzn,Sawicki:2016klv}.
As predicted in Ref.~\cite{Lombriser:2015sxa} this degeneracy was broken by the measurement of the speed of gravitational waves with GW170817~\cite{TheLIGOScientific:2017qsa}.
Precisely because of the breaking of the dark degeneracy, this resulted in important implications for modifications of gravity as a candidate for dark energy~\cite{Lombriser:2015sxa,Lombriser:2016yzn}.
In the second part of our series on the dark degeneracy, we will explore how this degeneracy can be restored by allowing for an effective scalar field that modifies gravity and its additional interactions with one or more dark sector components.
Here, we explore the dark degeneracy between models with a CDM component and a generalized DE fluid and models where a CDM component interacts with a DE component with vacuum EoS $w_x=-1$.
We inspect the degeneracy at the background level and at the perturbation level for scalar and tensor fluctuations in order to determine the precise mapping between the non-interacting generalized dynamical DE fluid models and the interacting dark sector models with vacuum DE EoS.
In order to study the impact of the degeneracy at the background level on parameter constraints, we also break the dark degeneracy at the linear level and perform a parameter estimation analysis with current cosmological observations, employing geometrical probes and cosmic microwave background (CMB) data.

The structure of the paper is the following: In Sec.~\ref{sec.gen} we present a unified description for the dark sector that can be used to formally establish the dark degeneracy at the background level in terms of the unified dark EoS.
We also present a general discussion of the most interesting cases for describing CDM and DE components, i.e., CDM with a generalized dynamical dark energy fluid versus an interacting dark sector where the DE EoS is fixed to $w_x=-1$.
Sec.~\ref{sec.equiv} is devoted to obtaining the explicit expressions that map generalized dynamical DE fluids with CDM to interacting dark sector models.
In Sec.~\ref{sec.specific}, for the purpose of illustrating the impact of a degeneracy at the background level, we break the dark degeneracy at the perturbation level.
We choose three specific well known dynamical DE parameterizations ($w$CDM, Chevallier-Polarski-Linder (CPL)~\cite{Chevallier:2000qy,Linder:2002et}, and Barboza-Alcaniz (BA)~\cite{Barboza:2008rh}) and compute their analogous interacting models.
Sec.~\ref{sec.stat} provides a Bayesian statistical analysis of the two different scenarios under the three different parameterizations, employing type Ia SNe, baryonic acoustic oscillations (BAO), Cosmic Chronometers and CMB data.
Finally, Sec.~\ref{sec.conclusions} presents the main conclusions of this work.

\section{General unified/interacting background description for the dark sector}
\label{sec.gen}

With `the cosmological background' we refer to the very large scales of the observable Universe, where the cosmological principle holds, i.e., the scales where matter can statistically be considered homogeneously and isotropically distributed.
In this regime, the space-time can be described by the Friedmann-Lema\^itre-Robertson-Walker (FLRW) line element,
\begin{equation} \label{bgflrw}
ds^{2}=dt^{2}-a^{2}\left(t\right)\left[dr^{2}+r^{2}\left(d\theta^{2}+\sin^{2}\theta d\phi\right)\right]\,.
\end{equation}
In GR, using the flat FLRW metric above, the dynamics of the space-time are described by the usual Friedmann equation,
\begin{equation} \label{friedmann}
3H^{2}=8\pi G\rho\,,
\end{equation}
where $\rho$ is the energy density of the total cosmic substratum, which must take into account all the matter components of the Universe.
We assume four components: radiation, baryons, cold dark matter (CDM) and dark energy (DE). Each one of these
is described as a perfect fluid with equation of state (EoS) $p_{i}=w_{i}\rho_{i}$, where the EoS parameter $w_{i}$ is not necessarily a constant. The radiation component, or relativistic matter, will be denoted by the subindex $r$ and is characterized by $w_{r}=1/3$. The baryonic component will be denoted by the subindex $b$ and is characterized by $w_{b}=0$. CDM will be denoted by the subindex $c$ and, as for the baryons, is also characterized by $w_{c}=0$. But in this case, as will be discussed in more detail in the following sections, these two matter constituents must be considered separately. Lastly, the DE component will be denoted by the subindex $x$, and we consider \textit{a priori} a general time-dependent EoS parameter $w_{x}\left(a\right)$.

In the standard cosmological scenario, all components are considered independent at late times, i.e., there is no physical process providing an energy/momentum exchange between the different components.
Here, we relax this assumption: we allow the possibility of interactions between the dark components, but baryons and radiation shall remain independent and separately conserved.
For radiation and baryons, the background energy conservation equations are given by
\begin{eqnarray}
\dot{\rho}_{r}+4H\rho_{r}=0&\qquad\Rightarrow\qquad&\rho_{r}=\dfrac{3H_{0}^{2}}{8\pi G}\Omega_{r0}\,a^{-4}\,, \label{rhor} \\
\dot{\rho}_{b}+3H\rho_{b}=0&\qquad\Rightarrow\qquad&\rho_{b}=\dfrac{3H_{0}^{2}}{8\pi G}\Omega_{b0}\,a^{-3}\,, \label{rhob}
\end{eqnarray}
where dots denote derivatives with respect to cosmic time and $\Omega_{i}\equiv8\pi G\rho_{i}/3H_{0}^{2}$ is the density parameter associated to the $i$-th matter component.
For a physical quantity, the subscript ${0}$ denotes its present value at $a=a_{0}=1$.

For the dark sector, even if it is physically described by separate components (interacting or not), it is always possible to effectively describe it as a unified fluid with energy density and pressure at the background level respectively given by
\begin{equation} \label{rhop}
\rho_{d}=\rho_{c}+\rho_{x}\qquad\mbox{and}\qquad p_{d}=p_{x}=w_{x}\left(a\right)\rho_{x}\,,
\end{equation}
where the subindex $d$ denotes the unified dark fluid.
At this point, it is convenient to introduce
the ratio between the CDM and DE energy densities $r\equiv\rho_{c}/\rho_{x}$.
One can express the unified dark energy density from the previous relation~\eqref{rhop} in terms of the ratio $r$ together with one of the energy densities of the dark components such that
\begin{equation} \label{rhopd}
\rho_{d}=\rho_{x}\big[1+r\left(a\right)\big]\qquad\mbox{or}\qquad\rho_{d}=\rho_{c}\big[1+r\left(a\right)^{-1}\big]\,.
\end{equation}
Taking advantage of the fact that the unified dark pressure only arises from the DE component and using the first relation in Eq.~\eqref{rhopd}, we can write an effective EoS for the unified dark fluid in terms of the DE EoS parameter and the ratio between CDM and DE energy densities,
\begin{equation} \label{EoS}
p_{d}=w_{d}\left(a\right)\rho_{d}\qquad {\rm with}\qquad w_{d}\left(a\right)=\frac{w_{x}\left(a\right)}{1+r\left(a\right)}\,.
\end{equation}
Since the unified energy density is the sum of the energy densities of the dark components,
even for interacting models, the energy density of the dark sector as a whole must be
conserved, which means that the unified dark fluid must satisfy the background energy conservation equation
\begin{equation} \label{darkconservation}
\dot{\rho}_{d}+3H\big[1+w_{d}\left(a\right)\big]\rho_{d}=0\,.
\end{equation}
For a given $w_{d}\left(a\right)$, one finds the well-known solution
\begin{equation} \label{rhod}
\rho_{d}\left(a\right)=\dfrac{3H_{0}^{2}}{8\pi G}\Omega_{d0}\exp\left[-3\int\frac{1+w_{d}\left(\hat{a}\right)}{\hat{a}}d\hat{a}\right]\,.
\end{equation}

Since the cosmological background expansion
is described by the Hubble  rate via Eq.~\eqref{friedmann}, one can observe that no individual matter component is relevant for the dynamics alone but only in the sum of all energy densities. Thus, at the background level, all the dark sector information is contained in $\rho_{d}$, or equivalently, according to Eq.~\eqref{rhod}, in the unified dark EoS parameter $w_{d}\left(a\right)$.
As can be seen in Eq.~\eqref{EoS}, the unified dark EoS parameter depends directly on the functions $w_{x}\left(a\right)$ and $r\left(a\right)$, where the first one is related to the dynamical nature of the DE component (independently of the other components) and the second one, as will be seen in Sec.~\ref{ssec.intde}, can be associated to an interaction in the dark sector. However, using Eq.~\eqref{EoS}, it is clear that different combinations of $w_{x}\left(a\right)$ and $r\left(a\right)$ can give us exactly the same $w_{d}\left(a\right)$, and consequently the same Hubble  rate. 

It is important to emphasize that this degeneracy does not mean that two different descriptions with the same $w_{d}\left(a\right)$ but different $w_{x}\left(a\right)$ and $r\left(a\right)$ are identical, i.e.\ have the same fluid content, since the dark components evolve differently. For example, whereas CDM evolves with $a^{-3}$ in the dynamical DE parameterization, the interaction affects the CDM evolution in the second case. But they give exactly the same Hubble  rate. Because of that, no observation based on measurements of distances, e.g. type Ia Supernovae, BAO and cosmic chronometers, can distinguish models with the same $w_{d}\left(a\right)$.

In this work we explore this dark sector degeneracy, focusing on two different approaches for the dark sector: the dynamical DE parameterization and interacting models. In the first approach we consider a dynamical nature for DE through a time-dependent EoS parameter $w_{x}\left(a\right)$, but no interaction in the dark sector, whereas in the second approach we consider a decaying vacuum dark energy, i.e., we set the DE EoS parameter to $w_x=-1$, but we allow for a phenomenological interaction between dark components. In order to clearly distinguish both approaches, from now on, all the quantities related to the dynamical DE approach will be denoted with a bar whereas all the quantities related to the interacting dark sector approach will be denoted with a tilde. When a physical quantity is identical in both approaches, neither a bar nor a tilde will be used.

Note that this degeneracy allows to consider more general cases, where both $w_{x}\left(a\right)$ is dynamical and $r\left(a\right)$ is affected by an interaction (models with interaction between CDM and dynamical DE).
We leave an analysis of such scenarios for future work.

\subsection{Dynamical DE parameterization}
\label{ssec.dynde}

The standard approach for modeling alternative DE models  imposes the hypothesis that all components are independently conserved, and introduces a dynamical DE EoS parameter $\bar{w}_{x}\left(a\right)$. A large number of parameterizations for the DE EoS have been proposed and studied in the cosmological literature, here we will focus on the functions proposed in \cite{Chevallier:2000qy,Linder:2002et,Barboza:2008rh}.
In this approach, as each species evolves independently, the background energy conservation equations for the dark sector components are given by\footnote{Note that it is not necessary to denote the Hubble  rate and the unified dark EoS parameter with a bar or tilde because we are interested in the degenerate case where they are identical in both approaches.}
\begin{eqnarray}
\dot{\bar{\rho}}_{c}+3H\bar{\rho}_{c}&=&0 \,, \label{drhoc1} \\
\dot{\bar{\rho}}_{x}+3H\bar{\rho}_{x}\big[1+\bar{w}_{x}\left(a\right)\big]&=&0 \,, \label{drhox1}
\end{eqnarray}
which lead to the following well-known solutions,
\begin{eqnarray}
\bar{\rho}_{c}&=&\dfrac{3H_{0}^{2}}{8\pi G}\bar{\Omega}_{c0}a^{-3} \,, \label{rhoc1} \\
\bar{\rho}_{x}&=&\dfrac{3H_{0}^{2}}{8\pi G}\bar{\Omega}_{x0}\exp\left[-3\int\dfrac{1+\bar{w}_{x}\left(\hat{a}\right)}{\hat{a}}d\hat{a}\right] \,. \label{rhox1}
\end{eqnarray}
Thus, using these results, the ratio between the CDM and DE energy densities and the unified dark EoS parameter are given by
\begin{equation} \label{rw1}
\bar{r}\left(a\right)=\bar{r}_{0}a^{-3}\exp\left[3\int\dfrac{1+\bar{w}_{x}\left(\hat{a}\right)}{\hat{a}}d\hat{a}\right]\qquad\mbox{and}\qquad w_{d}\left(a\right)=\dfrac{\bar{w}_{x}\left(a\right)}{1+\bar{r}_{0}a^{-3}\exp\left[3\int\frac{1+\bar{w}_{x}\left(\hat{a}\right)}{\hat{a}}d\hat{a}\right]}\,,
\end{equation}
where $\bar{r}_{0}=\bar{\rho}_{c0}/\bar{\rho}_{x0}$ is the current value of the ratio $\bar{r}\left(a\right)$.

\subsection{Interacting dark sector models}
\label{ssec.intde}

In a second approach we relax the hypothesis that dark components are independent, and we assume a phenomenological source term responsible for energy transfer between the dark components. Fixing $\tilde{w}_{x}\left(a\right)=-1$, the interaction between CDM and DE appears in the background energy conservation equation as
\begin{eqnarray}
\dot{\tilde{\rho}}_{c}+3H\tilde{\rho}_{c}&=&\tilde{Q}\,, \label{cdmenergyint} \\
\dot{\tilde{\rho}}_{x}&=&-\tilde{Q} \,, \label{deenergyint}
\end{eqnarray}
where $\tilde{Q}$ is a scalar source function that defines a specific interacting model. In Eqs.~\eqref{cdmenergyint} and~\eqref{deenergyint} the direction of the energy transfer depends on the sign of the source term: if $Q$ is positive we have DE decaying into CDM, whereas the opposite occurs if $Q$ is negative. As for the dynamical DE parameterizations, several interacting models have been studied in the literature, most of them being phenomenologically motivated~\cite{Marttens:2016cba}, but there are theoretical arguments that can be invoked to motivate such a coupling \cite{Sola:2013gha,Schutzhold:2002pr}. 

As previously mentioned, the ratio between the CDM and DE energy densities can be associated with an interaction in the dark sector.
To further clarify this point, it is convenient to introduce the derivative of the ratio with respect to cosmic time.
Using Eqs.~\eqref{cdmenergyint} and \eqref{deenergyint}, one finds 
\begin{equation} \label{drq}
\dot{\tilde{r}}=\tilde{r}\left(\dfrac{\dot{\tilde{\rho}}_{c}}{\tilde{\rho}_{c}}-\dfrac{\dot{\tilde{\rho}}_{x}}{\tilde{\rho}_{x}}\right)\quad\Rightarrow\quad\dot{\tilde{r}}=-\tilde{r}\left[Q\left(\dfrac{\tilde{\rho}_{c}+\tilde{\rho}_{x}}{\tilde{\rho}_{c}\, \tilde{\rho}_{x}}\right)+3H\right]\,.
\end{equation}
We make the \textit{ansatz} that the interaction source function depends only on the energy densities of the interacting components, i.e., the interaction term has the form $\tilde{Q}=3H\tilde{R}\left(\tilde{\rho}_{c},\tilde{\rho}_{x}\right)$, where $\tilde{R}\left(\tilde{\rho}_{c},\tilde{\rho}_{x}\right)$ is a general function of the CDM and DE energy densities with the dimensions of an energy density.
Eq.~\eqref{drq} can then be rewritten as
\begin{equation} \label{drtilde}
\dot{\tilde{r}}+3H\tilde{r}\left[\tilde{R}(\tilde{\rho}_{c},\tilde{\rho}_{x})\left(\dfrac{\tilde{\rho}_{c}+\tilde{\rho}_{x}}{\tilde{\rho}_{c}\, \tilde{\rho}_{x}}\right)+1\right]=0\,.
\end{equation}
Assuming that $\tilde{R}(\tilde{\rho}_{c},\tilde{\rho}_{x})$ does not depend on any preferred absolute energy density, e.g.\ $\tilde{R}(\tilde{\rho}_{c},\tilde{\rho}_{x})=\tilde{\rho}_{c0}$, the first term in the square brackets must be a function of $\tilde{r}$ alone.
This includes almost all the interaction terms that have been studied in the literature so far, see Eq.~\eqref{frexemple}.
It is convenient to define the following function~\cite{vonMarttens:2018iav},
\begin{equation} \label{fR}
\tilde{f}\left(\tilde{r}\right)\equiv \tilde{R}(\tilde{\rho}_{c},\tilde{\rho}_{x}) \left(\frac{\tilde{\rho}_{c}+\tilde{\rho}_{x}}{\tilde{\rho}_{c}\, \tilde{\rho}_{x}}\right)\,,
\end{equation}
so that Eq.~\eqref{drtilde} takes the form
\begin{equation} \label{dr}
\dot{\tilde{r}}+3H\tilde{r}\big[\tilde{f}\left(\tilde{r}\right)+1\big]=0 \,.
\end{equation}
Eq.~\eqref{dr} is a differential equation for $\tilde{r}$ that depends on the interaction term. It is important to emphasize that in a phenomenological approach, the choice of an interaction term $\tilde{Q}$ is equivalent to choosing a function $\tilde{f}\left(\tilde{r}\right)$ as these are related via $\tilde{f}\left(\tilde{r}\right)=\tilde{Q}/3H\left[\left(\tilde{\rho}_{c}+\tilde{\rho}_{x}\right)/\tilde{\rho}_{c}\tilde{\rho}_{x}\right]$.
Note that the interacting dark sector approach can also be put into a different perspective, in the sense that one can make an \textit{ansatz} for the function $\tilde{r}\left(a\right)$ and then solve Eq.~\eqref{dr} for $\tilde{f}\left(\tilde{r}\right)$ with
\begin{equation} \label{ftilde}
\tilde{f}\left(\tilde{r}\right)=-\dfrac{\tilde{r}^{\prime}a}{3\tilde{r}}-1\,,
\end{equation}
where the prime denotes a derivative with respect to scale factor.

The unified dark EoS parameter for the interacting dark sector is given by
\begin{equation} \label{wd2}
\tilde{w}_{d}\left(a\right)=-\dfrac{1}{1+\tilde{r}\left(a\right)}\,,
\end{equation}
which follows from Eq.~\eqref{EoS}.
This means that for a given (phenomenologically motivated) interacting model, one can solve Eq.~\eqref{dr} and then compute the Hubble  rate regardless of knowledge of the explicit form of the energy densities of the dark components. A model-independent study of interacting models based on the function $\tilde{r}\left(a\right)$ can be found in~\cite{vonMarttens:2018bvz}.

This approach of using the ratio between the CDM and DE energy densities is particularly convenient for interacting models (that satisfy the condition $\tilde{Q}=3H\tilde{R}$) because instead of having to solve the coupled Eqs.~\eqref{cdmenergyint} and \eqref{deenergyint}, one can solve Eq.~\eqref{dr} in addition to only one of the following relations,
\begin{eqnarray}
\dot{\tilde{\rho}}_{c}+3H\tilde{\rho}_{c}\left(\dfrac{\tilde{f}\left(\tilde{r}\right)}{1+\tilde{r}}+1\right)&=&0\,, \label{rhocf} \\
\dot{\tilde{\rho}}_{x}-3H\tilde{\rho}_{x}\left(\dfrac{\tilde{f}\left(\tilde{r}\right)}{1+\tilde{r}^{-1}}\right)&=&0\,, \label{rhoxf}
\end{eqnarray}
which forms a decoupled system\footnote{The other energy density is obtained trivially from the definition of $r\left(a\right)$.}.
For example, any interacting model described by the interaction function $\tilde{Q}$ below can be related to the function $\tilde{f}$ according to:
\begin{equation} \label{frexemple}
\tilde{Q}=3H\tilde{\rho}_{c}^{\alpha}\tilde{\rho}_{x}^{\beta}\left(\tilde{\rho}_{c}+\tilde{\rho}_{x}\right)^{\sigma}
\quad \Leftrightarrow \quad
\tilde{f}\left(\tilde{r}\right)=\tilde{r}^{\alpha-1}\left(1+\tilde{r}\right)^{\sigma+1}\,,
\end{equation}
where, from dimensional reasoning, the relation $\alpha+\beta+\sigma=1$ must hold.

\section{Explicit equivalence between dynamical DE parameterization and interacting dark sector models}
\label{sec.equiv}

We shall now present the explicit expressions that relate the dynamical DE parameterization and the interacting dark sector approaches such that they become indistinguishable both at the background and linear perturbation level.
From a theoretical point of view, this consists in imposing that the left-hand side of Einstein's field equations must be equivalent and therefore finding the conditions that ensure that the right-hand side is identical in both approaches.
At the background level, one must impose that the Hubble  rate (or equivalently the scale factor as a function of time) is identical, while at the linear level, the potentials $\Psi$ and $\Phi$ must agree between the two approaches.
As will be discussed, at the linear level, we will also inspect the conservation equations to find the full degeneracy.

\subsection{Background level}
\label{ssec.bg}

As aforementioned, given that radiation and baryons evolve according to Eqs.~\eqref{rhob} and \eqref{rhor}, respectively, the Hubble  rate is determined by the unified dark EoS parameter. Then, the idea here is that for a given dynamical DE parameterization, i.e., with spedified DE EoS parameter $\bar{w}_{x}\left(a\right)$, one
can compute the corresponding interaction $\tilde{f}\left(\tilde{r}\right)$. Thus, the condition that must be satisfied is that $w_{d}\left(a\right)$ is the same in both approaches.
Equating the second relation of Eq.~\eqref{rw1} with Eq.~\eqref{wd2}, one obtains the condition
\begin{equation} \label{rtilde}
\tilde{r}\left(a\right)=-\dfrac{1+\bar{r}_{0}a^{-3}\exp\left[3\int\frac{1+\bar{w}_{x}\left(\hat{a}\right)}{\hat{a}}d\hat{a}\right]}{\bar{w}_{x}\left(a\right)}-1\,.
\end{equation}
Eq.~\eqref{rtilde} shows how any dynamical DE parameterization characterized by $\bar{w}_{x}\left(a\right)$ can be mapped into an interacting dark sector description characterized by $\tilde{r}\left(a\right)$ which yields exactly the same background expansion dynamics. Then, the previous result found for $\tilde{r}\left(a\right)$ can be used to find the corresponding interaction term via Eq.~\eqref{ftilde}.

Note that it is not necessary to use Eq.~\eqref{ftilde} combined with the background energy conservation~\eqref{cdmenergyint} and \eqref{deenergyint} to compute the background energy densities for the dark components in the interacting approach.
Given the energy densities of the dark components in the dynamical DE parameterization, which is a consequence of the specific choice of $\bar{w}_{x}\left(a\right)$, it is possible to use the condition that the total unified dark energy density must be equal between the two approaches, ($\rho_{d}=\bar{\rho}_{c}+\bar{\rho}_{x}=\tilde{\rho}_{c}+\tilde{\rho}_{x}$), to obtain the energy densities in the interacting approach.
Using Eqs.~\eqref{rhop} and \eqref{EoS} for both approaches, the equivalence of $w_{d}\left(a\right)$ can be rewritten as
\begin{equation}
\bar{w}_{x}\dfrac{\bar{\rho}_{x}}{\bar{\rho}_{c}+\bar{\rho}_{x}}=-\dfrac{\tilde{\rho}_{x}}{\tilde{\rho}_{c}+\tilde{\rho}_{x}}\,.
\end{equation}
Since the denominators must be identical, one can then conclude that
\begin{eqnarray}
\tilde{\rho}_{c}&=&\bar{\rho}_{c}+\bar{\rho}_{x}\left(1+\bar{w}_{x}\right)\,, \label{rhoc2} \\
\tilde{\rho}_{x}&=&-\bar{w}_{x}\bar{\rho}_{x}\,. \label{rhox2}
\end{eqnarray}
Note that even if the Hubble  rate is the same in both approaches, the current values of the CDM and DE density parameters are different, $\bar{\Omega}_{c0}\neq\tilde{\Omega}_{c0}$ and $\bar{\Omega}_{x0}\neq\tilde{\Omega}_{x0}$. This will become more explicit in Sec.~\ref{sec.stat}, where a statistical analysis will be presented for specific models.

\subsection{Perturbative level}
\label{ssec.pert}

Now that the degeneracy at the background level is established, we repeat an analogous procedure at the perturbative level: we impose that the left-hand side of Einstein's equations must be identical, which implies that $\Phi$ and $\Psi$ are indistinguishable between the two approaches, and we then determine the relations that must be satisfied for the dark sector on the right-hand side of the equations.

\subsubsection{Scalar perturbations}
\label{sssec.scalar}

We first consider scalar perturbations up to linear level, where we adopt the Newtonian gauge for the perturbed FLRW metric with the (gauge invariant) gravitational potentials $\Psi$ and $\Phi$,\footnote{As was the case for the Hubble  rate and the unified dark EoS parameter, in a degenerate scenario, $\Psi$ and $\Phi$ must be identical between the two approaches. Hence, no bar or tilde is required for these quantities as the distinction is not necessary.}
\begin{equation} \label{newmetric}
ds^{2}=a^{2}\left(\tau\right)\left[-\left(1+2\Psi\right)d\tau^{2}+\left(1+2\Phi\right)d\textbf{x}^{2}\right]\,,
\end{equation}
where $\tau$ is the conformal time $dt=a\,d\tau$.

At the linear level, the energy-momentum tensor of the total cosmic fluid is given by
\begin{eqnarray}
\delta T^{0}_{\,\,0}&\equiv&-\delta\rho\,, \label{t00} \\
\delta T^{0}_{\,\,i}&\equiv&\left(\rho+p\right)v_{i}\,, \label{t0i} \\
\delta T^{i}_{\,\,j}&\equiv&\delta p\,\delta^{i}_{\,j}+\Sigma^{i}_{\,j}\,, \label{tij}
\end{eqnarray}
where $\delta\rho$, $v_{i}$, $\delta p$ and $\Sigma^{i}_{\,j}$ are respectively the total energy density perturbation, the total spatial $3$-velocity, total pressure perturbation and the total anisotropic shear perturbation. For scalar perturbations, we can write the total spatial $3$-velocity and the total anisotropic shear perturbation in terms of the divergence of the total spatial $3$-velocity and the anisotropic stress, which in Fourier space are respectively defined as
\begin{eqnarray}
\theta&\equiv&ik^{j}v_{j} \,, \label{thetatot} \\
\left(\rho+p\right)\sigma&\equiv&-\left(\hat{\textbf{k}}^{i}\cdot \hat{\textbf{k	}}_{j}-\dfrac{1}{3}\delta^{i}_{\,j}\right)\Sigma^{j}_{\,\,i} \,. \label{sigmatot}
\end{eqnarray}

The perturbative dynamics of the Universe (space-time + matter content) is then described by the Einstein and conservation equations.
The linear Einstein equations now take the following form,
\begin{eqnarray}
-k^{2}\Phi+3\mathcal{H}\left(-\Phi^{\prime}+\mathcal{H}\Psi\right)&=&4\pi Ga^{2}\delta T^{0}_{\,\,0}\,, \label{eq00} \\
k^{2}\left(-\Phi^{\prime}+\mathcal{H}\Psi\right)&=&4\pi Ga^{2}\left(\rho+p\right)\theta \,, \label{eq0i} \\
-3\Phi^{\prime\prime}+3\mathcal{H}\left(\Psi^{\prime}-2\Phi^{\prime}\right)+3\left(2\dfrac{a^{\prime\prime}}{a}-\mathcal{H}^{2}\right)\Psi-k^{2}\left(\Phi+\Psi\right)&=&4\pi Ga^{2}\delta T^{i}_{\,\,i}\,, \label{eqii} \\
-k^{2}\left(\Phi+\Psi\right)&=&12\pi Ga^{2}\left(\rho+p\right)\sigma\,. \label{eqij}
\end{eqnarray}

For a general interacting fluid, the conservation equations can be written as~\cite{Valiviita:2008iv},
\begin{eqnarray}
\delta\rho^{\prime}_{i}+3\mathcal{H}\left(\delta\rho_{i}+\delta p_{i}\right)+\left(\rho_{i}+p_{i}\right)\left(3\Phi^{\prime}+\theta_{i}\right)&=&aQ_{i}\Psi+a\delta Q_{i}\,, \label{pertenergy} \\
\left[\left(\rho_{i}+p_{i}\right)\theta_{i}\right]^{\prime}+\left(\rho_{i}+p_{i}\right)\left(4\mathcal{H}\theta_{i}-k^{2}\Psi\right)-k^{2}\delta p_{i}+\left(\rho_{i}+p_{i}\right)k^{2}\sigma_{i}&=&aQ_{i}\theta+ak^{2}\mathcal{F}_{i}\,, \label{pertmomentum}
\end{eqnarray}
where the functions $\delta Q_{i}$ and $\mathcal{F}_{i}$ describe the interaction at the linear level, being respectively responsible for the perturbative energy and momentum transfer, and $\theta\equiv\sum\left(\rho_{i}+p_{i}\right)\theta_{i}/\sum\left(\rho_{i}+p_{i}\right)$ is the total velocity of the cosmic substratum.
These equations are general.
For the dynamical DE approach we have $Q_{i}=\delta Q_{i}=\mathcal{F}_{i}=\bar{Q}_{i}=\delta\bar{Q}_{i}=\bar{\mathcal{F}}_{i}=0$, whereas for the interacting dark sector approach $Q_{i}=\tilde{Q}_{i}$, $\delta Q_{i}=\delta\tilde{Q}_{i}$ and $\mathcal{F}_{i}=\tilde{\mathcal{F}}_{i}$. In general, $\delta\tilde{Q}_{i}$ and $\tilde{\mathcal{F}}_{i}$ can be seen as free functions, and also with $\tilde{Q}_{i}$ are phenomenologically motivated.

Considering explicitly all matter species separately, these quantities can be divided as follows,
\begin{equation} \label{fluid}
\delta\rho=\sum_{i}\delta\rho_{i}\quad,\quad\delta p=\sum_{i}\delta p_{i}\quad,\quad\left(\rho+p\right)\theta=\sum_{i}\left(\rho_{i}+p_{i}\right)\theta_{i}\quad\mbox{and}\quad\left(\rho+p\right)\sigma=\sum_{i}\left(\rho_{i}+p_{i}\right)\sigma_{i}\,,
\end{equation}
where the subindex $i$ runs over all matter species ($i=\{r,b,c,x\}$). Since we are particularly interested in the dark components, it is convenient to write here the conservation equations for CDM and DE for both approaches. In the dynamical DE approach, Eqs.~\eqref{pertenergy} and \eqref{pertmomentum} applied to the dark components reduce to
\begin{eqnarray}
\delta\bar{\rho}_{c}^{\prime}+3\mathcal{H}\delta\bar{\rho}_{c}+\bar{\rho}_{c}\left(3\Phi^{\prime}+\bar{\theta}_{c}\right)&=&0 \,, \label{cdmenergybar} \\
\bar{\theta}_{c}^{\prime}+\mathcal{H}\bar{\theta}_{c}-k^{2}\Psi&=&0 \,, \label{cdmmomentumbar} \\
\delta\bar{\rho}_{x}^{\prime}+3\mathcal{H}\left(\delta\bar{\rho}_{x}+\delta\bar{p}_{x}\right)+\bar{\rho}_{x}\left(1+\bar{w}_{x}\right)\left(3\Phi^{\prime}+\bar{\theta}_{x}\right)&=&0 \,, \label{deenergybar} \\
\bar{\theta}_{x}^{\prime}+\dfrac{\bar{w}_{x}^{\prime}}{1+\bar{w}_{x}}\bar{\theta}_{x}+\mathcal{H}\left(1-3\bar{w}_{x}\right)\bar{\theta}_{x}-k^{2}\Psi-\dfrac{k^{2}\delta\bar{p}_{x}}{\bar{\rho}_{x}\left(1+\bar{w}_{x}\right)}+k^{2}\bar{\sigma}_{x}&=&0 \,. \label{demomentumbar}
\end{eqnarray}
In comparison, in the interacting approach, the analogous relations are
\begin{eqnarray}
\delta\tilde{\rho}_{c}^{\prime}+3\mathcal{H}\delta\tilde{\rho}_{c}+\tilde{\rho}_{c}\left(3\Phi^{\prime}+\tilde{\theta}_{c}\right)&=&-a\tilde{Q}\Psi-a\delta\tilde{Q}\,, \label{cdmenergytilde} \\
\tilde{\theta}_{c}^{\prime}+\mathcal{H}\tilde{\theta}_{c}-k^{2}\Psi&=&-\dfrac{a}{\tilde{\rho}_{c}}\left(\tilde{Q}\tilde{\theta}+k^{2}\mathcal{\tilde{F}}\right)\,, \label{cdmmomentumtilde} \\
\delta\tilde{\rho}_{x}^{\prime}+3\mathcal{H}\left(\delta\tilde{\rho}_{x}+\delta\tilde{p}_{x}\right)&=&a\tilde{Q}\Psi+a\delta\tilde{Q}\,, \label{deenergytilde}
\end{eqnarray}
where, in order to ensure that the dark sector as a whole is conserved, we choose $\tilde{Q}=\tilde{Q}_{x}=-\tilde{Q}_{c}$, $\delta\tilde{Q}=\delta\tilde{Q}_{x}=-\delta\tilde{Q}_{c}$ and $\mathcal{\tilde{F}}=\mathcal{\tilde{F}}_{x}=-\mathcal{\tilde{F}}_{c}$. In this case, since we require $\tilde{w}_{x}=-1$, the DE momentum conservation does not provide the dynamics for the DE velocity, but it can be seen as an algebraic constraint for the perturbative momentum exchange due to the interaction,
\begin{equation} \label{Ftilde}
\mathcal{\tilde{F}}=-\dfrac{1}{a}\left(\delta\tilde{p}_{x}+a\tilde{Q}\dfrac{\tilde{\theta}}{k^{2}}\right) \,.
\end{equation}
Note that by definition in the interacting approach there is no DE contribution in $\tilde{\theta}$.

We again assume that quantities related to baryons and radiation are equivalent between the two approaches,
but quantities related to CDM and DE differ.
In general, as can be seen from the relations \eqref{fluid}, we have eight degrees of freedom for describing the dark sector at linear level, \{$\delta\rho_{c}$, $\delta\rho_{x}$, $\delta p_{c}$, $\delta p_{x}$, $\theta_{c}$, $\theta_{x}$, $\sigma_{c}$, $\sigma_{x}$\}, but we assume as an \textit{ansatz} that CDM has a vanishing pressure perturbation and anisotropic stress in both approaches, $\delta\bar{p}_{c}=\delta\tilde{p}_{c}=\bar{\sigma}_{c}=\tilde{\sigma}_{c}=0$, which reduces the number of ``dark'' degrees of freedom to six.

From a formal point of view, the degeneracy between dynamical DE parameterizations and interacting dark sector models at the linear level means a set of explicit relations that
map the set \{$\delta\bar{\rho}_{c}$, $\delta\bar{\rho}_{x}$, $\delta\bar{p}_{x}$, $\bar{\theta}_{c}$, $\bar{\theta}_{x}$, $\bar{\sigma}_{x}$\} to the corresponding set \{$\delta\tilde{\rho}_{c}$, $\delta\tilde{\rho}_{x}$, $\delta\tilde{p}_{x}$, $\tilde{\theta}_{c}$, $\tilde{\theta}_{x}$, $\tilde{\sigma}_{x}$\} such that both descriptions yield exactly the same results for $\Psi$ and $\Phi$.
Given a degeneracy at the background level,
the Einstein equations~\eqref{eq00}, \eqref{eq0i}, \eqref{eqij} and \eqref{eqii} provide
the following constraints,
\begin{eqnarray}
\delta\bar{\rho}_{c}+\delta\bar{\rho}_{x}&=&\delta\tilde{\rho}_{c}+\delta\tilde{\rho}_{x} \,, \label{deltarhodark} \\
\bar{\rho}_{c}\bar{\theta}_{c}+\bar{\rho}_{x}\left(1+\bar{w}_{x}\right)\bar{\theta}_{x}&=&\tilde{\rho}_{c}\tilde{\theta}_{c} \,, \label{thetadark} \\
\delta\bar{p}_{x}&=&\delta\tilde{p}_{x}\,, \label{deltapdark} \\
\bar{\rho}_{x}\left(1+\bar{w}_{x}\right)\bar{\sigma}_{x}&=&0 \,. \label{sigmadark}
\end{eqnarray}
In the particular case where $\tilde{w}_{x}=-1$,
these constraints simplify\footnote{The general case that relates the dynamical and interacting descriptions of the dark sector is presented in Appendix~\ref{ap.general}.}.
In this scenario, it can be seen that the Einstein equations do not impose constraints on the DE velocity and the DE anisotropic stress of the interacting approach.
More specifically, combining Eqs.~\eqref{thetadark} and \eqref{sigmadark} with the fact that the conservation equations~\eqref{deenergytilde} and \eqref{Ftilde} do not prescribe any dynamics for the interacting DE velocity and anisotropic stress implies that $\tilde{\theta}_{x}$ and $\tilde{\sigma}_{x}$ are free functions but without relevant physical meaning.
For example, they can be set to zero without loss of generality.
Furthermore, according to Eq.~\eqref{sigmadark}, the degeneracy is only possible if $\bar{\sigma}_{x}$ is identically zero.

Like the condition on $\bar{\sigma}_{x}$, Eq.~\eqref{deltapdark} sets a requirement on the physical properties of the dark sector constituents.
Just as with $w$ at the background level, $\delta p$ and $\sigma$ describe the nature of the fluid at the level of the linear perturbations.
Eq.~\eqref{thetadark} also has a simple structure and provides an explicit relation between a single linear physical quantity in the interacting approach with two linear quantities in the dynamical DE approach.
Combining condition~\eqref{thetadark} with Eq.~\eqref{cdmmomentumtilde} and using Eqs.~\eqref{rhoc2}, \eqref{rhox2}, \eqref{cdmmomentumbar}, \eqref{demomentumbar}, \eqref{Ftilde} and \eqref{deltapdark},
one finds that Eq.~\eqref{thetadark}
implies the CDM momemntum conservation of the interacting scenario.

Finally, only Eq.~\eqref{deltarhodark} remains to be discussed.
We still have two degrees of freedom, $\delta\tilde{\rho}_{c}$ and $\delta\tilde{\rho}_{x}$, which means that it is not possible to relate $\delta\tilde{\rho}_{c}$ and $\delta\tilde{\rho}_{x}$ individually to physical quantities of the dynamical DE approach.
However, when solving Eq.~\eqref{deltarhodark} for $\delta\tilde{\rho}_{c}$ and replacing the result in the interacting CDM energy conservation equation~\eqref{cdmenergytilde}, then using the other constraints from the Einstein equations, the background degeneracy relations and the conservation equations in the dynamical DE approach, one recovers the interacting DE energy conservation. The interacting CDM energy conservation can be obtained with an analogous procedure, starting by solving Eq.~\eqref{deltarhodark} for $\delta\tilde{\rho}_{x}$. Physically, this means that if conditions~\eqref{thetadark}, \eqref{deltapdark} and the background degeneracy relations~\eqref{rhoc2} and \eqref{rhox2} are satisfied, the solutions of the energy conservation equations for $\delta\tilde{\rho}_{c}$ and $\delta\tilde{\rho}_{x}$ are already compatible with the dark degeneracy.
Here it is important to emphasize that the dark degeneracy is completely independent of the choice of $\delta\tilde{Q}$ and $\mathcal{\tilde{F}}$.

\subsubsection{Tensor perturbations}
\label{sssec.tensor}

The recent first direct measurements of gravitational waves (GWs), inaugurated with GW150914~\cite{Abbott:2016blz} emitted by a black hole merger, have
opened a new frontier for observational astrophysics and cosmology~\cite{Berti:2015itd}.
In particular, the gravitational wave event GW170817~\cite{TheLIGOScientific:2017qsa} from a neutron star merger with its electromagnetic counterpart GRB170817A~\cite{Monitor:2017mdv} has provided strong constraints on the possible physics of the dark sector~\cite{Lombriser:2015sxa,Lombriser:2016yzn,Creminelli:2017sry,Sakstein:2017xjx,Baker:2017hug,Boran:2017rdn,Amendola:2017orw,Visinelli:2017bny,Crisostomi:2017lbg}.
GWs can also be used to measure a variation of the Planck mass~\cite{Lombriser:2015sxa,Belgacem:2017ihm,Amendola:2017ovw,Belgacem:2018lbp,Belgacem:2019pkk,Dalang:2019fma} and test the late-time cosmology~\cite{Petiteau:2011we,Du:2018tia,Tamanini:2016zlh} through Standard Sirens~\cite{Holz:2005df,Palmese:2019ehe}. Constraints that can be obtained from Standard Sirens have also been discussed for dynamical DE~\cite{DelPozzo:2011yh,DiValentino:2017clw} and interacting models~\cite{Dalang:2019fma,Yang:2019bpr,Yang:2019vni,Cai:2017yww,Caprini:2016qxs}. Finally, there is hope that GW data can shed light on the $H_{0}$ tension~\cite{MacLeod:2007jd,Seto:2017swx,Chen:2017rfc,Feeney:2018mkj,Wang:2018lun,DiValentino:2018jbh} (cf.~\cite{Lombriser:2019ahl}).

We shall hence briefly inspect the tensor perturbations for both approaches to
specify the conditions that maintain the dark degeneracy.
Considering only tensor perturbations, space-time is described by the gauge-independent metric given by the line element
\begin{equation} \label{tensormetric}
ds^{2}=a^{2}\left(\tau\right)\left[-d\tau^{2}+\left(\delta_{ij}+h_{ij}\right)dx^{i}dx^{j}\right]\,,
\end{equation}
where $h_{ij}$ is divergenceless (transverse) and traceless. With this metric, the Einstein equations for tensor perturbations take the form~\cite{Piattella:2018hvi}
\begin{equation} \label{einsteintensor}
h_{ij}^{\prime\prime}+2\mathcal{H}h_{ij}^{\prime}+k^{2}h_{ij}=16\pi Ga^{2}\Pi_{ij}\,,
\end{equation}
where $\Pi_{ij}$ is the tensor part of the anisotropic stress.

Imposing that the geometrical part of Eq.~\eqref{einsteintensor}
is equivalent
between the dynamical DE and interacting scenarios, it is straightforward to conclude that the only necessary condition for ensuring the dark degeneracy is that the tensor anisotropic stress is the same in both approaches ($\bar{\Pi}_{ij}=\tilde{\Pi}_{ij}$). Since the tensor anisotropic stress is absent in both cases, the dark degeneracy for tensor perturbations is a trivial consequence of the background degeneracy.

\section{Specific cases}
\label{sec.specific}

We shall now apply the background degeneracy relations identified in Sec.~\ref{ssec.bg} to formulate interacting models that yield an equivalent background dynamics to some popular dynamical DE parametrizations. We will be interested in the differences that these scenarios can exhibit in the linear perturbations. The linear degeneracy would require a specific choice for the DE sound speed that satisfies the degeneracy conditions presented in Sec.~\ref{sssec.scalar}. This degeneracy is broken here by the assumption that the DE comoving sound speed is constant and equal to unity for both approaches ($\bar{c}_{s}^{2}=\tilde{c}_{s}^{2}=1$). This is, for instance, also the case in quintessence DE. The luminal sound speed is moreover a standard assumption for dynamical DE and interacting dark sector models~\cite{vonMarttens:2018iav,Valiviita:2008iv,Majerotto:2009np,CalderaCabral:2009ja,Asghari:2019qld}. Note, however, that in general this does not need to be the case and the sound speed can be seen as a free parameter. Furthermore, for linear interacting contributions we set $\delta\tilde{Q}=0$ and we impose that there is no momentum transfer in the CDM frame, which leads to $k^{2}\mathcal{\tilde{F}}=\theta-\theta_{c}$.

It is convenient to rewrite the conservation equations in terms of the density contrast and the comoving sound speed, which are respectively defined as\footnote{The definition for the comoving sound speed is only valid in Newtonian gauge.}
\begin{equation}
\delta_{i}\equiv\dfrac{\delta\rho_{i}}{\rho_{i}}\qquad\mbox{and}\qquad c_{s(i)}^{2}\equiv\dfrac{\delta p^{(c)}_{i}}{\delta\rho^{(c)}_{i}}=\dfrac{\delta p_{i}+p^{\prime}_{i}v_{i}}{\delta\rho_{i}+\rho^{\prime}_{i}v_{i}} \,,
\end{equation}
where $\delta p^{(c)}_{i}$ and $\delta\rho^{(c)}_{i}$ are respectively the DE comoving pressure perturbation and the DE comoving energy density perturbation. Using these definitions, Eq.~\eqref{pertenergy} takes the form
\begin{eqnarray} 
\delta^{\prime}_{i}+3\mathcal{H}\left(c_{s(i)}^{2}-w_{i}\right)\delta_{i}+\left(1+w_{i}\right)\left(\theta_{i}+3\Phi^{\prime}\right) && \nonumber \\
+3\mathcal{H}\left[3\mathcal{H}\left(1+w_{i}\right)\left(c_{s(i)}^{2}-w_{i}\right)+w^{\prime}_{i}\right]\dfrac{\theta_{i}}{k^{2}}&=&\dfrac{aQ_{i}}{\rho_{i}}\left[\Psi-\delta_{i}+3\mathcal{H}\left(c_{s(i)}^{2}-w_{i}\right)\dfrac{\theta_{i}}{k^{2}}+\dfrac{\delta Q_{i}}{Q_{i}}\right] \,, \label{deltaenergy}
\end{eqnarray}
whereas, for $w_{i}\neq-1$, Eq.~\eqref{pertmomentum} can be rewritten as
\begin{equation} \label{thetamomentum}
\theta^{\prime}_{i}+\mathcal{H}\left(1-3c_{s(i)}^{2}\right)\theta_{i}-\dfrac{c_{s(i)}^{2}}{1+w_{i}}k^{2}\delta_{i}+k^{2}\sigma_{i}-k^{2}\Psi=\dfrac{aQ_{i}}{\rho_{i}\left(1+w_{i}\right)}\left[\theta-\left(1+c_{s(i)}^{2}\right)\theta_{i}-\dfrac{k^{2}\mathcal{F}_{i}}{Q_{i}}\right] \,.
\end{equation}
As already discussed, in the case where $w_{i}=-1$, the velocity $\theta_{i}$ is not physically relevant and can be set to zero even if the fluid interacts.

In Secs.~\ref{ssec.wcdm},~\ref{ssec.cpl} and~\ref{ssec.ba}, we will present interacting models that are degenerate with three popular dynamical DE parameterizations: $w$CDM, CPL and BA . These parameterizations are characterized by the EoS parameters $w_{0}$ and $w_{a}$. Note that, no bar/tilde notation is needed here for $w_{0}$ and $w_{a}$ because, using Eqs.~\eqref{rhoc2} and \eqref{rhox2}, it is possible to write the solutions of the interacting scenario using those of the dynamical DE, which are consequently specified by the same parameters. Although, it is important to emphasize that in the interacting case $w_{0}$ and $w_{a}$ are not DE EoS parameters but parameters that characterize a specific interaction. Recall that by construction $\tilde{w}_{x}=-1$, independently of the value of $w_{0}$ and $w_{a}$. In all interacting models, the case where $w_{0}=-1$ and $w_{a}=0$ will always lead to a vanishing interaction and consequently to the $\Lambda$CDM model.

For all the cases, we present the explicit solutions of the background dynamics for both approaches.
To illustrate the differences between them, we use (a suitably modified version of) {\sc class}~\cite{Blas:2011rf} for computing background and linear physical quantities.
For ease of reading, in all figures, we shall present results associated with the dynamical DE approach in red whereas results illustrating the interacting approach are presented in blue.
Ratios between quantities from the different approaches are shown in black. When results for both approaches are presented in the same figure, we allow ourselves to avoid the bar/tilde notation (for example, in the axes), but it is implicit that each result corresponds to the scenario it has been computed for. We remind the reader that the dark degeneracy is only maintained at the background level. It is broken at the perturbative level by the choice of DE sound speed equal to unity for both cases. For this reason, there will be differences in observables that depend on the perturbations. Of course, even at the background level the degeneracy implies that the expansion rate is the same in both cases, not that the composition in terms of dark matter and dark energy is the same.

Since the main purpose of this section is to show qualitatively the difference between both approaches, in all plots of Secs.~\ref{ssec.wcdm},~\ref{ssec.cpl} and~\ref{ssec.ba} we fix the cosmological parameter as follows:
\begin{equation} \label{parameters}
H_{0}=70\quad,\quad w_{0}=-0.9\quad,\quad w_{a}=-0.1\quad,\quad \bar{\Omega}_{c0}=0.25\quad\mbox{and}\quad \tilde{\Omega}_{c0}=0.32\,.
\end{equation}
The choice of different values for $\bar{\Omega}_{c0}$ and $\tilde{\Omega}_{c0}$ was made using Eq.~\eqref{rhoc2} in order to show degenerate cases. Apart from the construction of the degenerate models, we would also like to understand how important the breaking of the dark degeneracy at the perturbation level is. To this end, we provide a full Bayesian statistical analysis in Sec.~\ref{sec.stat}.

\subsection{$w$CDM parameterization}
\label{ssec.wcdm}

The first case we consider is the dynamical DE parameterization $\bar{w}_{x}\left(a\right)=w_{0}$, where $w_{0}$ is a constant but not necessarily -1.
This is the so-called $w$CDM model. Using Eqs.~\eqref{rhoc1} and \eqref{rhox1}, the energy densities of the dark components of the dynamical DE parameterization are given by
\begin{equation} \label{wcdmrhobar}
\bar{\rho}_{c}=\dfrac{3H_{0}^{2}}{8\pi G}\bar{\Omega}_{c0}a^{-3}\qquad\mbox{and}\qquad\bar{\rho}_{x}=\dfrac{3H_{0}^{2}}{8\pi G}\bar{\Omega}_{x0}a^{-3\left(1+w_{0}\right)}\,.
\end{equation}
Then, using Eqs.~\eqref{rhoc2} and \eqref{rhox2}, the equivalent energy densities in the interacting approach are
\begin{equation} \label{wcdmrhotilde}
\tilde{\rho}_{c}=\dfrac{3H_{0}^{2}}{8\pi G}\left[\bar{\Omega}_{c0}+\bar{\Omega}_{x0}\left(1+w_{0}\right)a^{-3w_{0}}\right]a^{-3}\qquad\mbox{and}\qquad\tilde{\rho}_{x}=-\dfrac{3H_{0}^{2}}{8\pi G}\Omega_{x0}w_{0}a^{-3\left(1+w_{0}\right)} \,.
\end{equation}

Using Eqs.~\eqref{wcdmrhobar} and~\eqref{wcdmrhotilde}, it is straightforward to obtain the ratio between the CDM and DE energy densities for each case,
\begin{equation} \label{rwcdm}
\bar{r}\left(a\right)=\bar{r}_{0}a^{3w_{0}}\qquad\mbox{and}\qquad\tilde{r}\left(a\right)=-1-\dfrac{1+\bar{r}_{0}a^{3w_{0}}}{w_{0}}\,,
\end{equation}
where $\bar{r}_{0}=\bar{\rho}_{c0}/\bar{\rho}_{x0}=\bar{\Omega}_{c0}/\bar{\Omega}_{x0}$.

Finally, combining Eq.~\eqref{ftilde} with the solution for $\tilde{r}\left(a\right)$ given by the second term of Eq.~\eqref{rwcdm}, one can compute the interaction term associated with the $w$CDM model,
\begin{equation} \label{frwcdm}
f\left(\tilde{r}\right)=-1-\dfrac{\bar{r}_{0}w_{0}a^{3w_{0}}}{1+w_{0}+\bar{r}_{0}a^{3w_{0}}}\,.
\end{equation}
\begin{figure}
\centering
\includegraphics[width=0.49\columnwidth, trim={0.7cm 0 2.1cm 0}, clip]{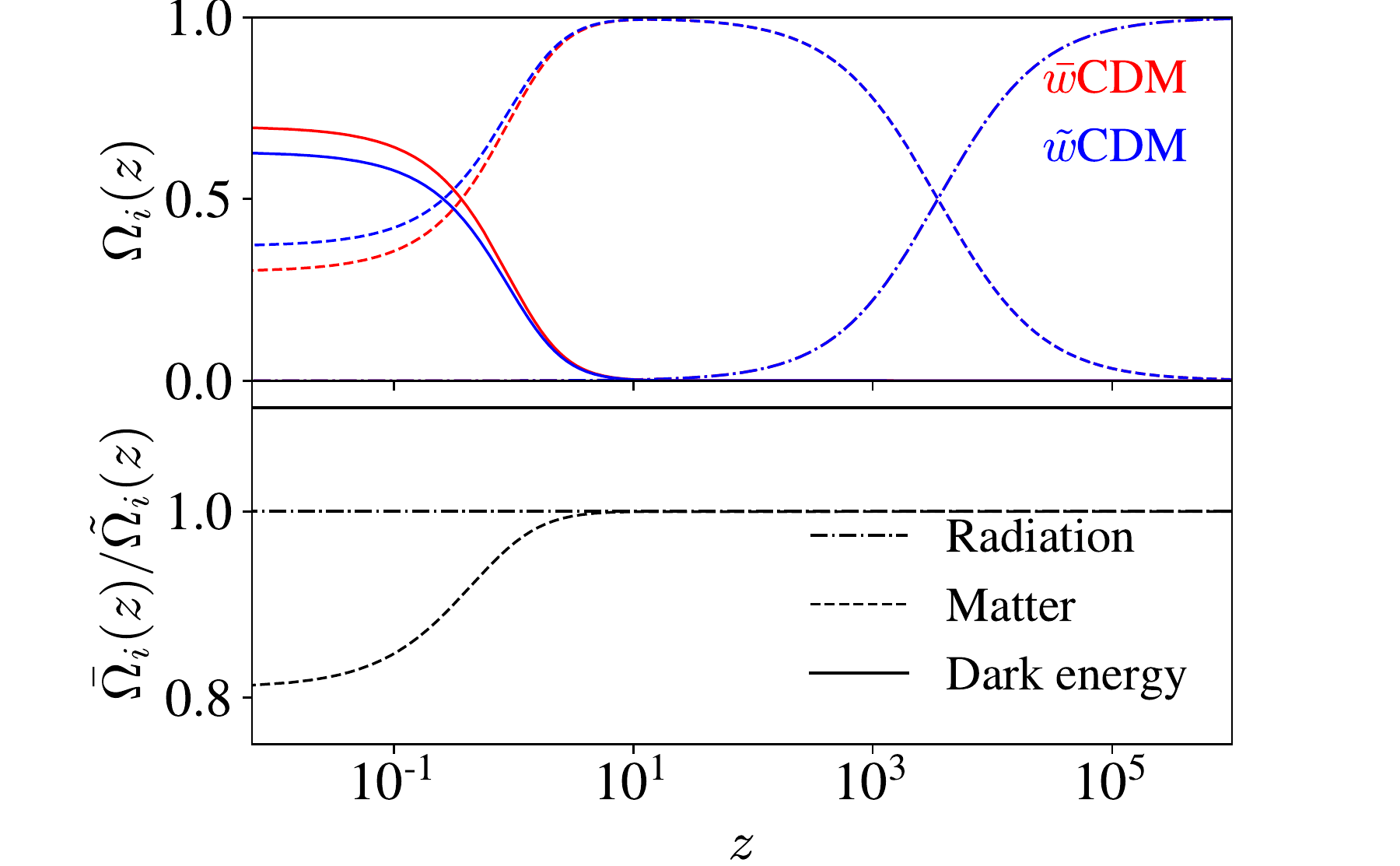} 
\includegraphics[width=0.49\columnwidth, trim={0.7cm 0 2.1cm 0}, clip]{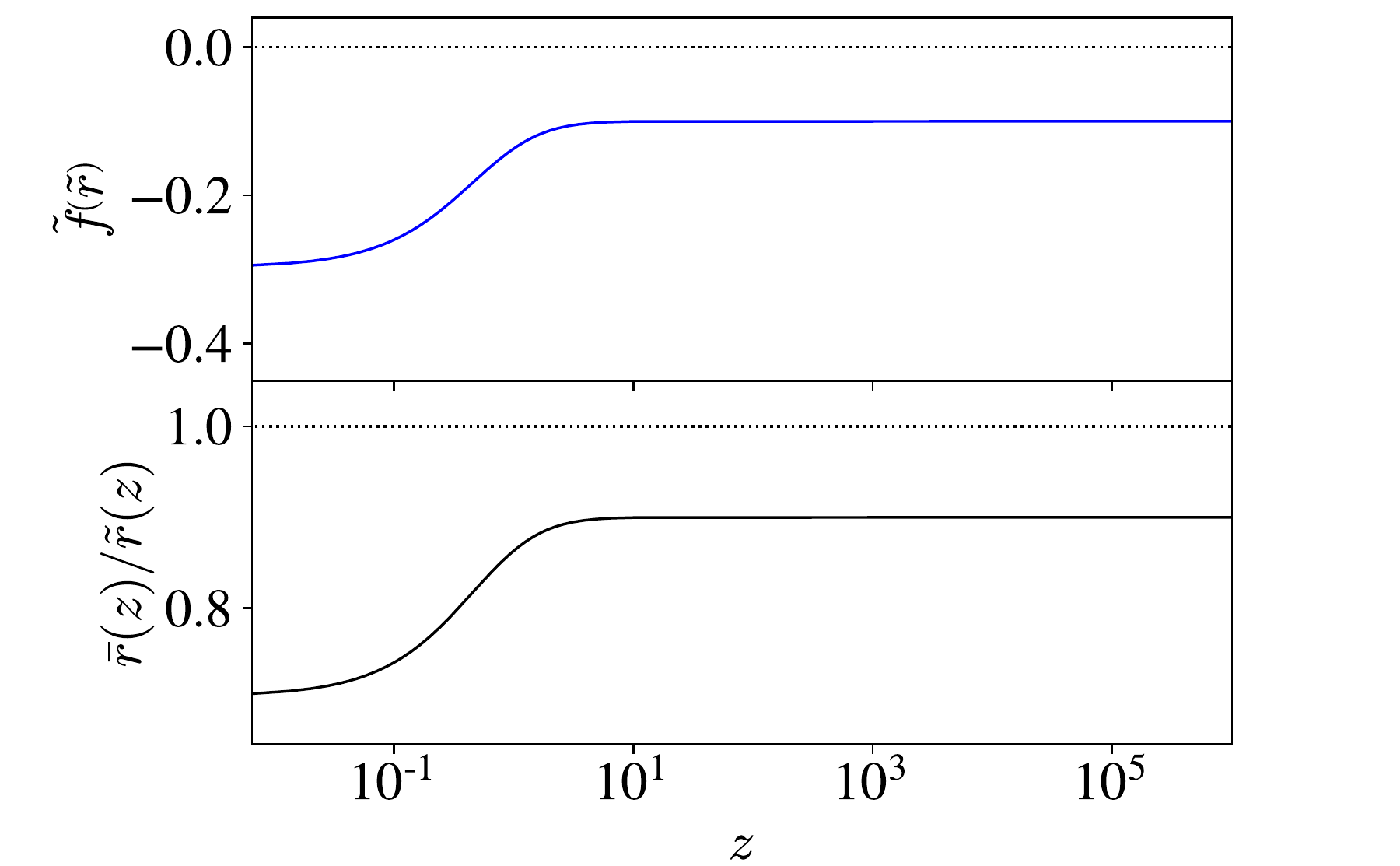} 
\caption{Background quantities for the $w$CDM model in both approaches (dynamical DE and interacting scenario). The cosmological parameters are fixed as in Eq.~\eqref{parameters}. \textbf{Top left panel:} Time evolution of the energy density parameter $\Omega_i(z)$ for radiation (dash-dotted lines), total matter (dashed lines) and DE (solid lines). \textbf{Bottom left panel:} Ratio between the energy density parameters. \textbf{Top right panel:} Interacting function associated to the $w$CDM model. The dotted line corresponds to the non-interacting case ($\Lambda$CDM model). \textbf{Bottom right panel:} Ratio between $\bar{r}\left(a\right)$ and $\tilde{r}\left(a\right)$. The dotted line has no physical meaning here and is only shown as a visual guide to indicate the axis.
}
\label{bgwcdm}
\end{figure}
Fig.~\ref{bgwcdm} illustrates the main background physical quantities with which to assess differences between the dynamical DE and interacting approach in the case of a $w$CDM model. In the top left panel we show the time evolution of the energy density parameter for all matter components in both scenarios.
One observes that the interaction only strongly affects the dynamics of the dark sector at late times. This is not surprising, since the $w$CDM model only deviates from $\Lambda$CDM at low redshifts, but it is particularly interesting because both approaches behave in the same manner at last scattering and at the time of matter-radiation equality, and hence no significant changes are expected in the CMB power spectrum and in the position of the maximum of the matter power spectrum (cf., e.g.,~\cite{Lombriser:2011tj}). The only differences we expect are due to CMB lensing, and, at low $\ell$, to the integrated Sachs-Wolfe effect (ISW). Both are integrated effects, and in the first case the difference is related to the fact that the matter distribution is different between the two approaches whereas the second concerns the different DE evolutions. Of course these effects are in some sense related, since a different DE evolution also causes different matter distribution.
For comparison, the bottom panel on the left-hand side shows the ratio between the energy density parameter of the matter species in each scenario. In the top panel on the right-hand side we illustrate the interaction function $f(r)$ obtained from Eq.~\eqref{frwcdm} for the $w$CDM model. Lastly, the bottom panel on the right-hand side depicts the ratio between $\bar{r}\left(a\right)$ and $\tilde{r}\left(a\right)$ obtained from Eq.~\eqref{rwcdm}.

\begin{figure}
\centering
\includegraphics[width=0.49\columnwidth]{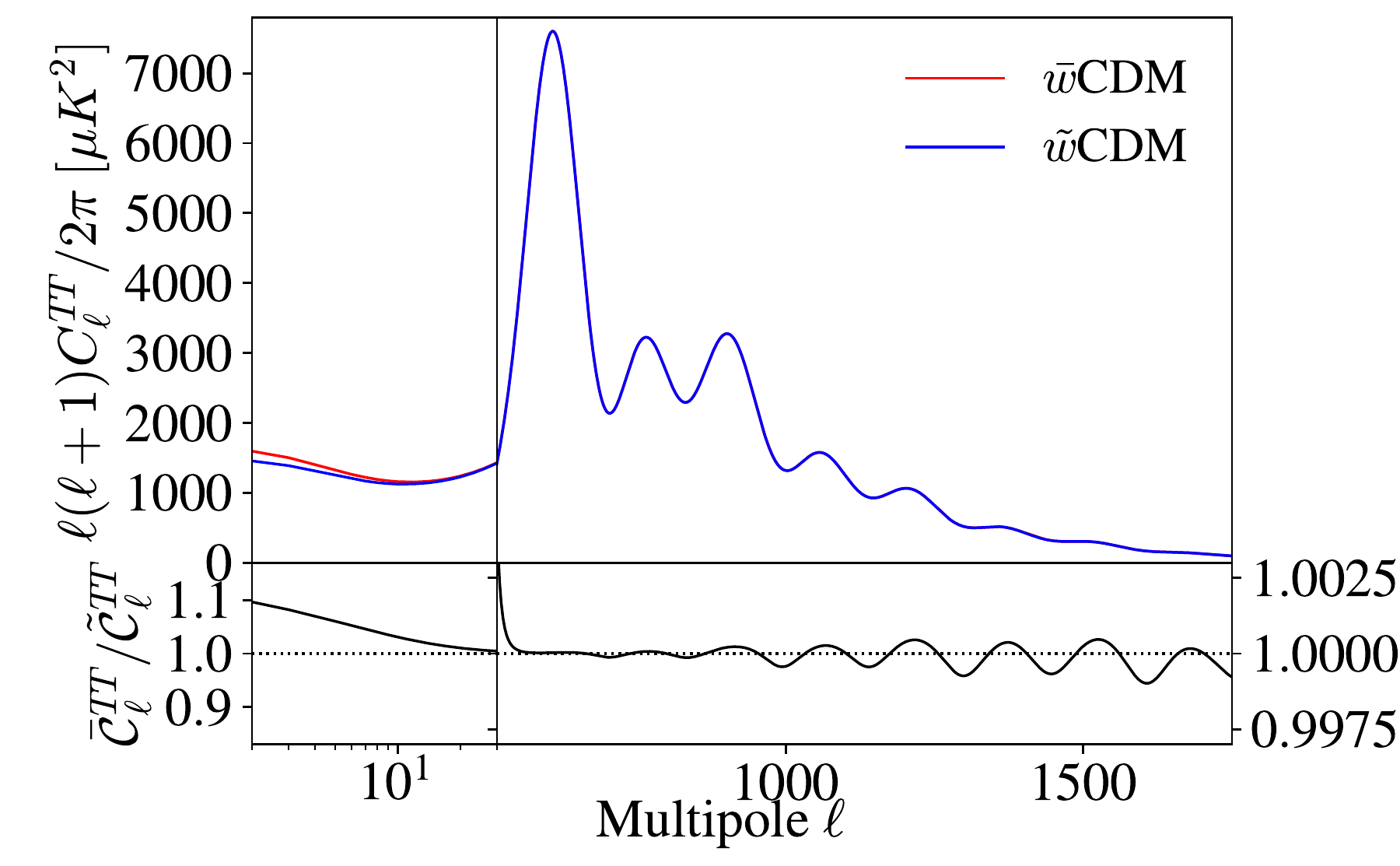} 
\includegraphics[width=0.49\columnwidth]{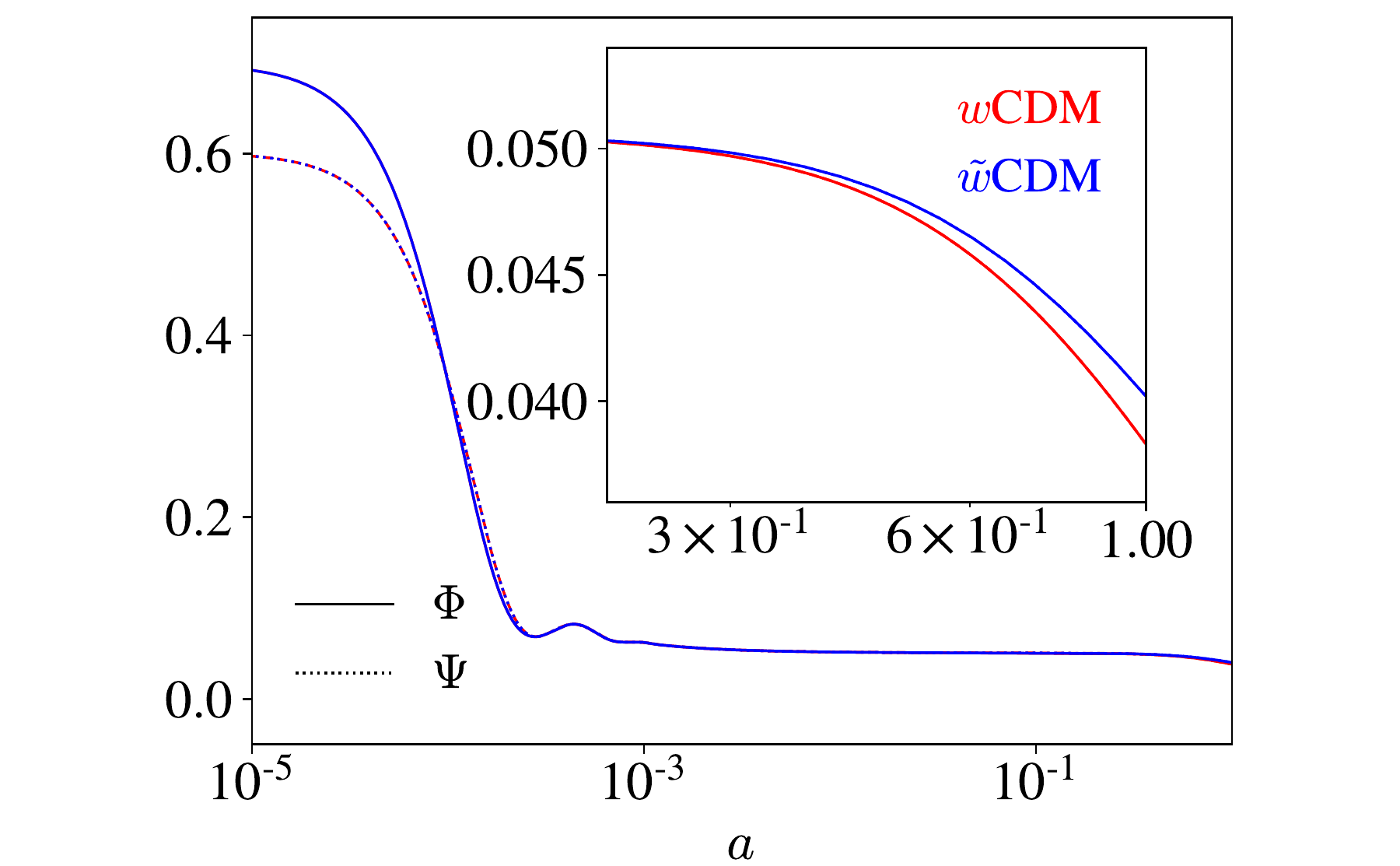}
\includegraphics[width=0.49\columnwidth]{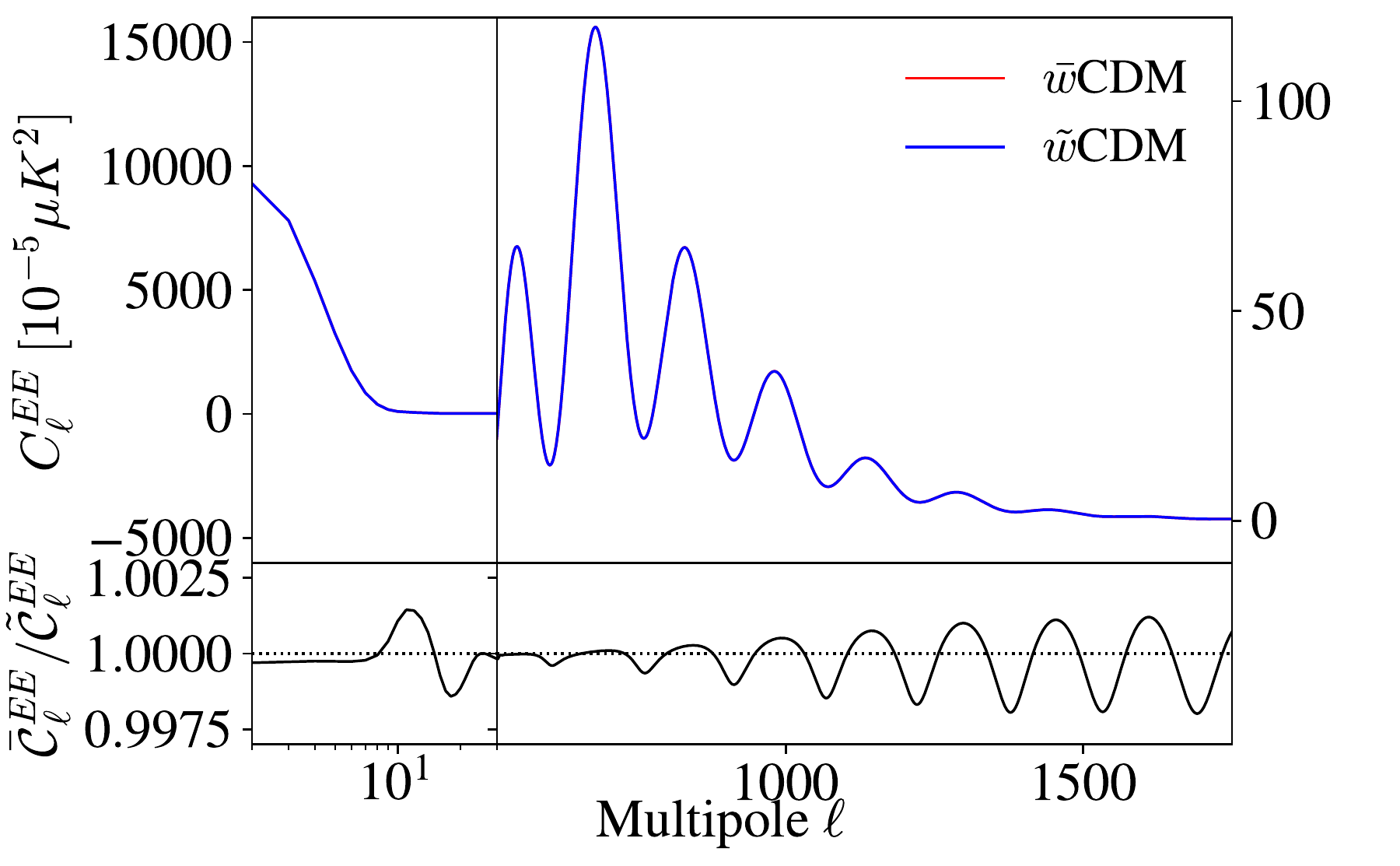} 
\includegraphics[width=0.49\columnwidth]{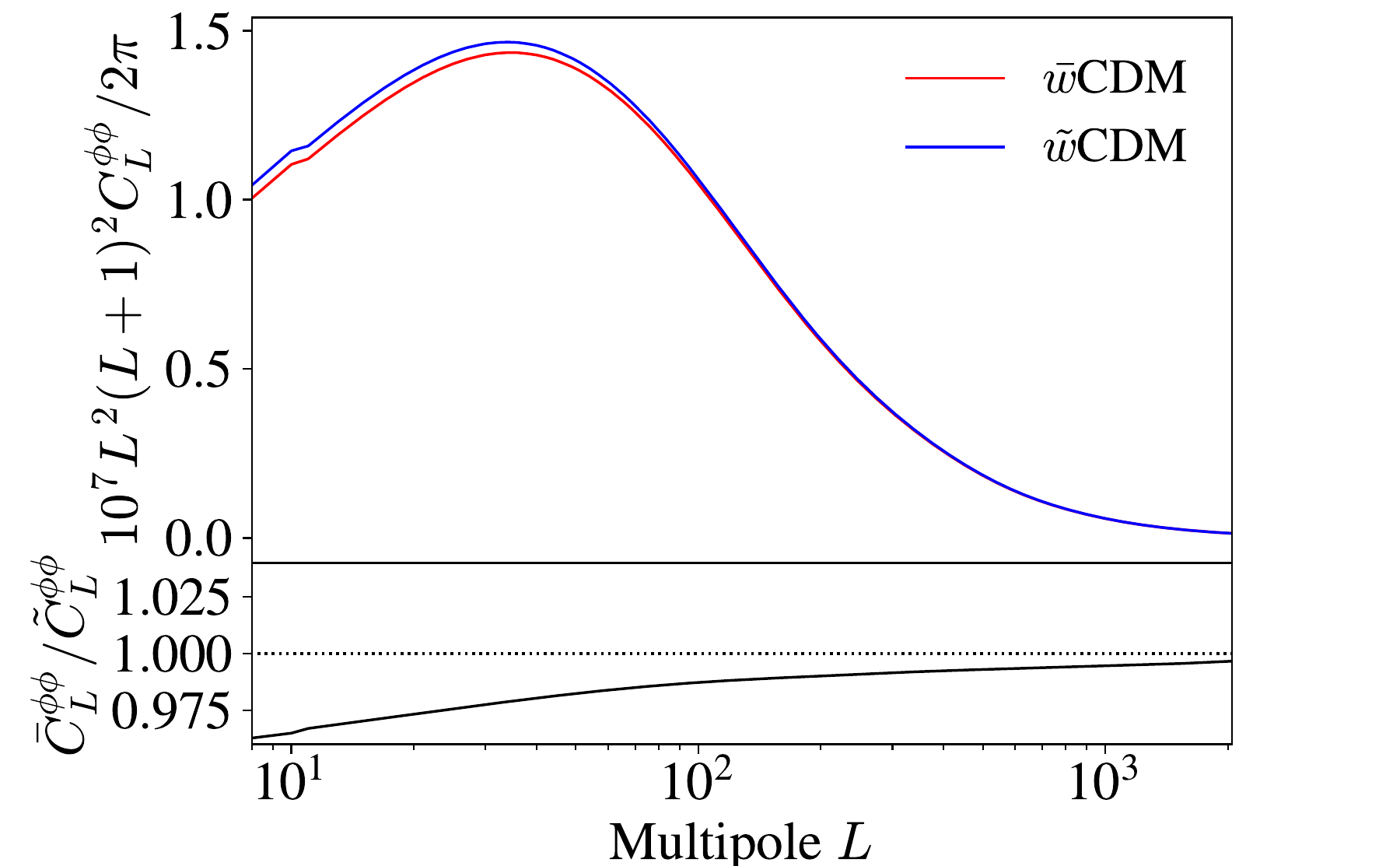}
\caption{Perturbative CMB quantities for the $w$CDM model in both approaches (dynamical DE and interacting scenario). The cosmological parameters are fixed as in Eq.~\eqref{parameters}. \textbf{Top left panel:} Lensed CMB temperature anisotropy power spectrum. \textbf{Bottom left panel:} Lensed CMB polarization (EE) power spectrum. \textbf{Top right panel:} Gauge-invariant gravitational potentials $\Psi$ and $\Phi$ at $k=0.1\,h/$Mpc. \textbf{Bottom right panel:} CMB lensing-potential power spectrum.}
\label{cmbwcdm}
\includegraphics[width=0.49\columnwidth, trim={0.8cm 0 1.7cm 0}, clip]{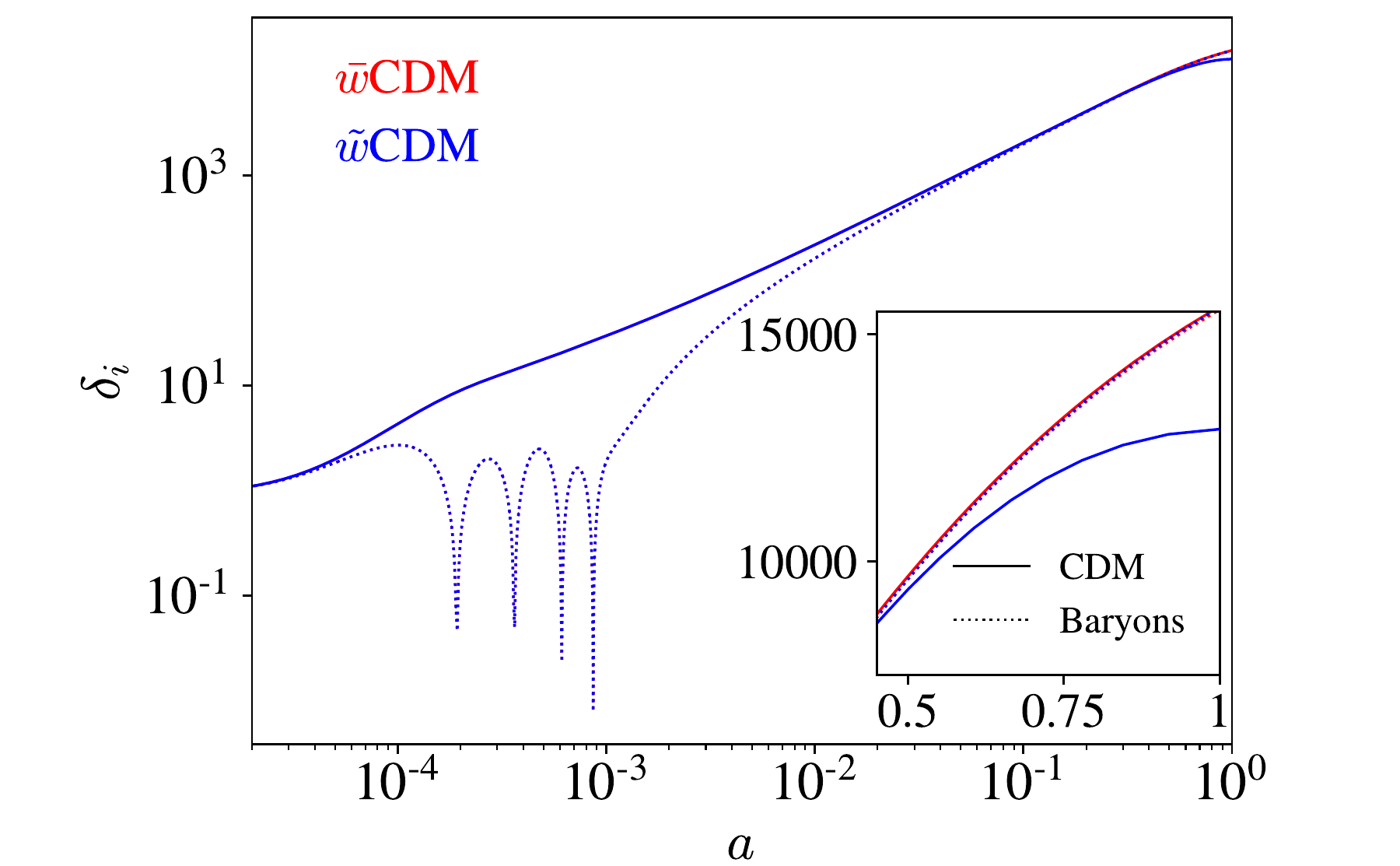} 
\includegraphics[width=0.49\columnwidth, trim={0.8cm 0 1.7cm 0}, clip]{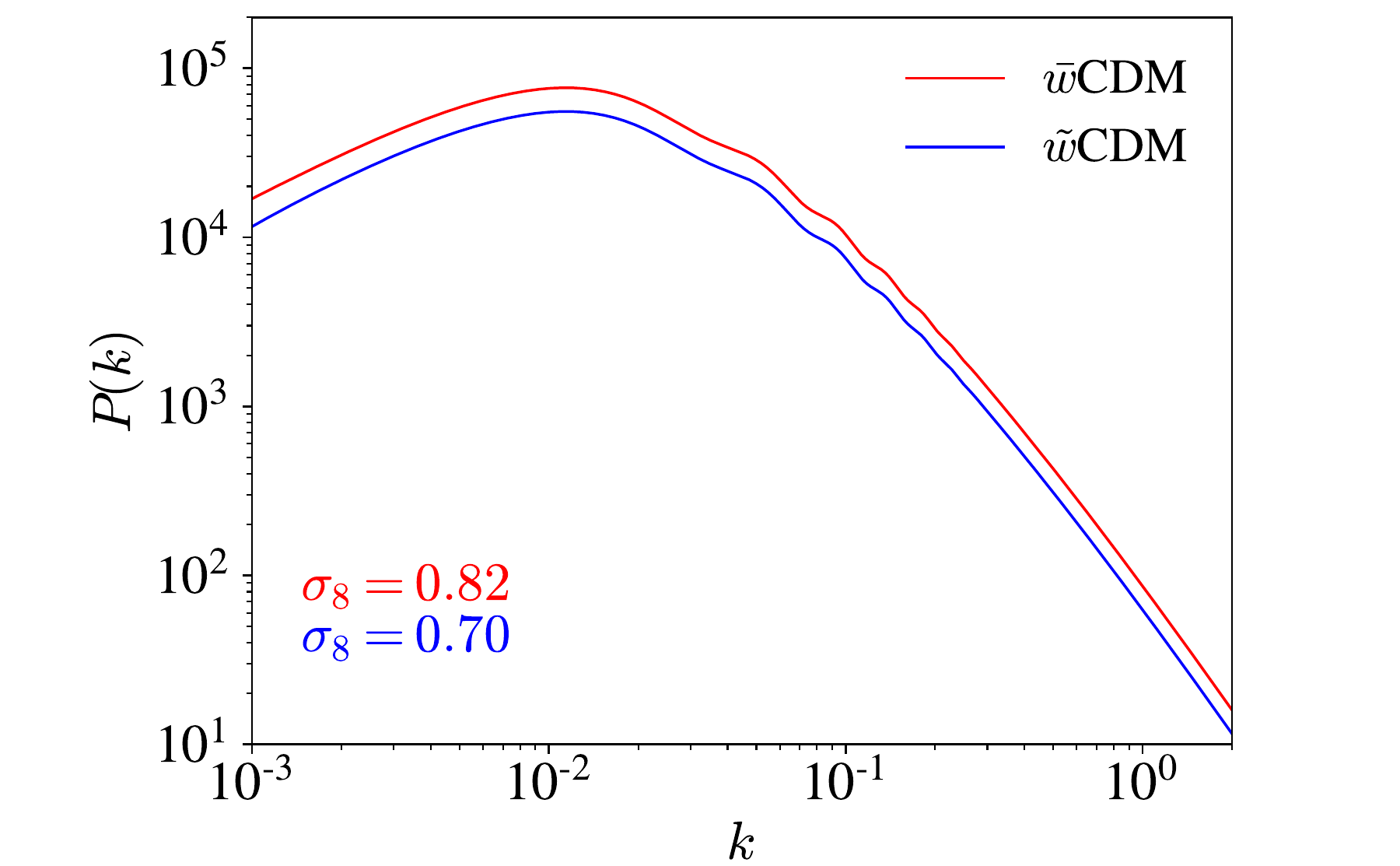} 
\caption{Perturbative matter quantities for the $w$CDM model in both approaches (dynamical DE and interacting scenario). The cosmological parameters are fixed as in Eq.~\eqref{parameters}. \textbf{Left panel:} Density contrast for baryons and CDM components at $k=0.1\,h/$Mpc. \textbf{Right panel:} Total matter power spectrum at $z=0$.}
\label{matterwcdm}
\end{figure}

In Fig.~\ref{cmbwcdm} we show the most important CMB spectra and the gravitational potentials. In the top panel on the left-hand side, we provide the lensed CMB temperature anisotropy power spectrum. As expected, we can see the differences between the dynamical DE and interacting approaches due to CMB lensing and ISW. The CMB lensing effect affects the spectrum for high values of $\ell$. At first glance, it appears that this effect is subdominant, however due to the high precision in the CMB data at high multipoles, this difference may be relevant for parameter constraints~\cite{Smidt:2009qa}. On the other hand, the ISW is responsible for the difference at low values of $\ell$. In agreement with this, the top right-hand panel shows the gauge-invariant gravitational potentials $\Psi$ and $\Phi$ at $k=0.1\,h/$Mpc, where differences between the two approaches only appear at late times. Since we consider a vanishing late-time anisotropic stress, the potentials coincide in the matter-dominated era.
The small departure at late times between the potentials of the two scenarios gives rise to the low $\ell$ difference visible in the CMB spectrum, which is due to the ISW effect. In the bottom left-hand panel we show the lensed CMB polarization (EE) power spectrum where, similarly to the temperature power spectrum, only small differences between the two approaches appear. Finally, the lensing-potential CMB power spectrum is shown on the bottom right-hand side\footnote{In this work we follow the usual convention of using $L$ instead of $\ell$ for lensing multipoles.}. As expected, because CMB lensing is an integrated effect that must be considered from the last-scattering surface until today, the difference in matter evolution along the expansion history of the Universe provides different lensing effects in either case.

Besides the differences in the CMB power spectra, one might expect also differences between the dynamical DE and interacting approaches in the structure formation, i.e., differences in the CDM density contrast, and consequently differences in the total matter power spectrum. Note that the change in the CDM density contrast is a natural consequence of the presence of the interaction/source function in its perturbative energy conservation. However, the changes in the total matter power spectrum only occur because the linear degeneracy is broken, otherwise, it is straightforward to see from Eq.~\eqref{eq00} that the total density contrast is identical in the degenerate case. Essentially, this difference appears due to the choice of the DE pressure perturbation, which allows DE clustering to compensate these changes, or not.

Fig.~\ref{matterwcdm} shows the matter perturbation quantities in both approaches. The left-hand panel illustrates the time evolution of the baryon and CDM density contrast in both approaches at $k=0.1\,h/$Mpc. From this plot, one observes that they coincide at early times whereas at late times the CDM interacting density contrast undergoes a suppression, which can be seen in the inset plot. In the right-hand panel of Fig.~\ref{matterwcdm}, we show the total matter power spectrum at $z=0$. Indeed the position of the peak of the matter power spectrum does not change, but its amplitude is affected, which can be related to the difference between $\bar{\Omega}_{c0}$ and $\tilde{\Omega}_{c0}$ as the dark energy fluid is not able to cluster inside the horizon due to the high sound speed.

\subsection{CPL parameterization}
\label{ssec.cpl}

The second case analyzed in this work is the well-established CPL parameterization \cite{Chevallier:2000qy,Linder:2002et}, which is characterized by the DE EoS parameter $\bar{w}_{x}\left(a\right)=w_{0}+w_{a}\left(1-a\right)$. This is a standard parametrization often used in data analysis, e.g.\ in \cite{Aghanim:2018eyx}, and in forecasts, due to its link with the DETF figure of merit \cite{Albrecht:2006um}. For this model, the energy densities of the dark matter and the dark energy are given by,
\begin{equation} \label{rhocplbar}
\bar{\rho}_{c}=\dfrac{3H_{0}^{2}}{8\pi G}\bar{\Omega}_{c0}a^{-3}\qquad\mbox{and}\qquad\bar{\rho}_{x}=\dfrac{3H_{0}^{2}}{8\pi G}\bar{\Omega}_{x0}a^{-3\left(1+w_{0}+w_{a}\right)}\exp\left[3w_{a}\left(a-1\right)\right] \,,
\end{equation}
which leads to the interacting energy densities
\begin{eqnarray}
\tilde{\rho}_{c}&=&\dfrac{3H_{0}^{2}}{8\pi G}\left\lbrace \bar{\Omega}_{c0}+\bar{\Omega}_{x0}\left[1+w_{0}+w_{a}\left(1-a\right)\right]\exp\left[-3w_{a}\left(1-a\right)\right]a^{-3\left(w_{0}+w_{a}\right)}\right\rbrace a^{-3} \,, \label{rhoc2cpl} \\
\tilde{\rho}_{x}&=&-\dfrac{3H_{0}^{2}}{8\pi G}\bar{\Omega}_{x0}\left[w_{0}+w_{a}\left(1-a\right)\right]\exp\left[-3w_{a}\left(1-a\right)\right]a^{-3\left(1+w_{0}+w_{a}\right)} \,. \label{rhox2cpl}
\end{eqnarray}

Using these relations, we obtain the following expressions for the ratio between CDM and DE energy densities in the two approaches,
\begin{equation} \label{rcpl}
\bar{r}\left(a\right)=\bar{r}_{0}\exp\left[3w_{a}\left(1-a\right)\right]a^{3\left(w_{0}+w_{a}\right)}\quad\mbox{and}\quad\tilde{r}\left(a\right)=-\dfrac{1+w_{0}+w_{a}\left(1-a\right)+\bar{r}_{0}\exp\left[3w_{a}\left(1-a\right)\right]a^{3\left(w_{0}+w_{a}\right)}}{w_{0}+w_{a}\left(1-a\right)} \,.
\end{equation}

The interacting term associated to the CPL parameterization can then be obtained from combining Eq.~\eqref{ftilde} with the second relation of Eq.~\eqref{rcpl},
\begin{equation} \label{frcpl}
\tilde{f}\left(\tilde{r}\right)=\dfrac{\left\lbrace3w_{0}\left(1+w_{0}\right)+w_{a}\left[3+6w_{0}-2a\left(1+3w_{0}\right)\right]+3w_{a}^{2}\left(1-a\right)^{2}\right\rbrace\left\lbrace\bar{r}_{0}a^{3\left(w_{0}+w_{a}\right)}+\exp\left[-3w_{a}\left(1-a\right)\right]\right\rbrace}{3\left[w_{0}+w_{a}\left(1-a\right)\right]\left\lbrace\bar{r}_{0}a^{3\left(w_{0}+w_{a}\right)}+\left[1+w_{0}+w_{a}\left(1-a\right)\right]\exp\left[-3w_{a}\left(1-a\right)\right]\right\rbrace}\,.
\end{equation}

\begin{figure} 
\centering
\includegraphics[width=0.49\columnwidth, trim={0.7cm 0 2.1cm 0}, clip]{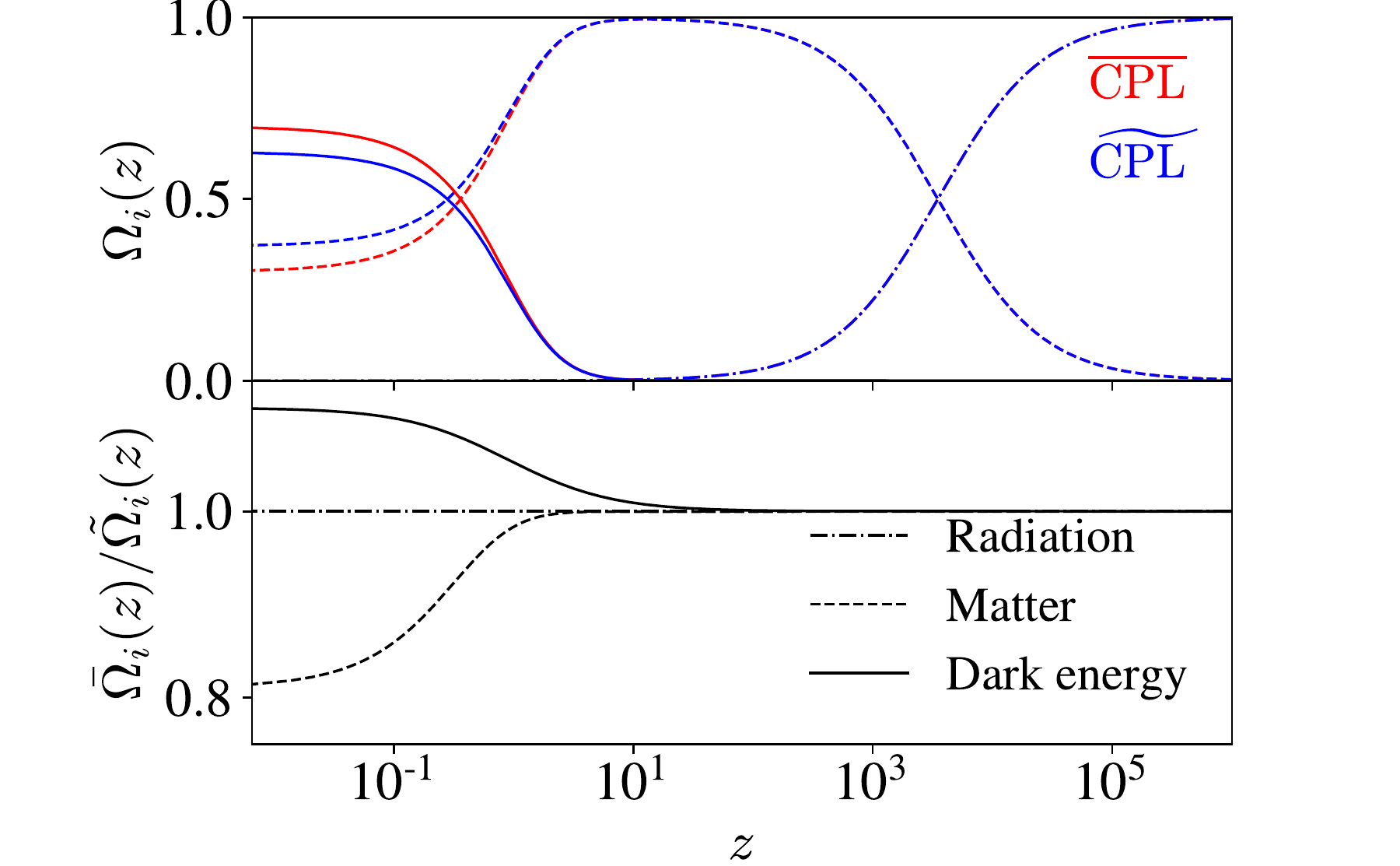} 
\includegraphics[width=0.49\columnwidth, trim={0.7cm 0 2.1cm 0}, clip]{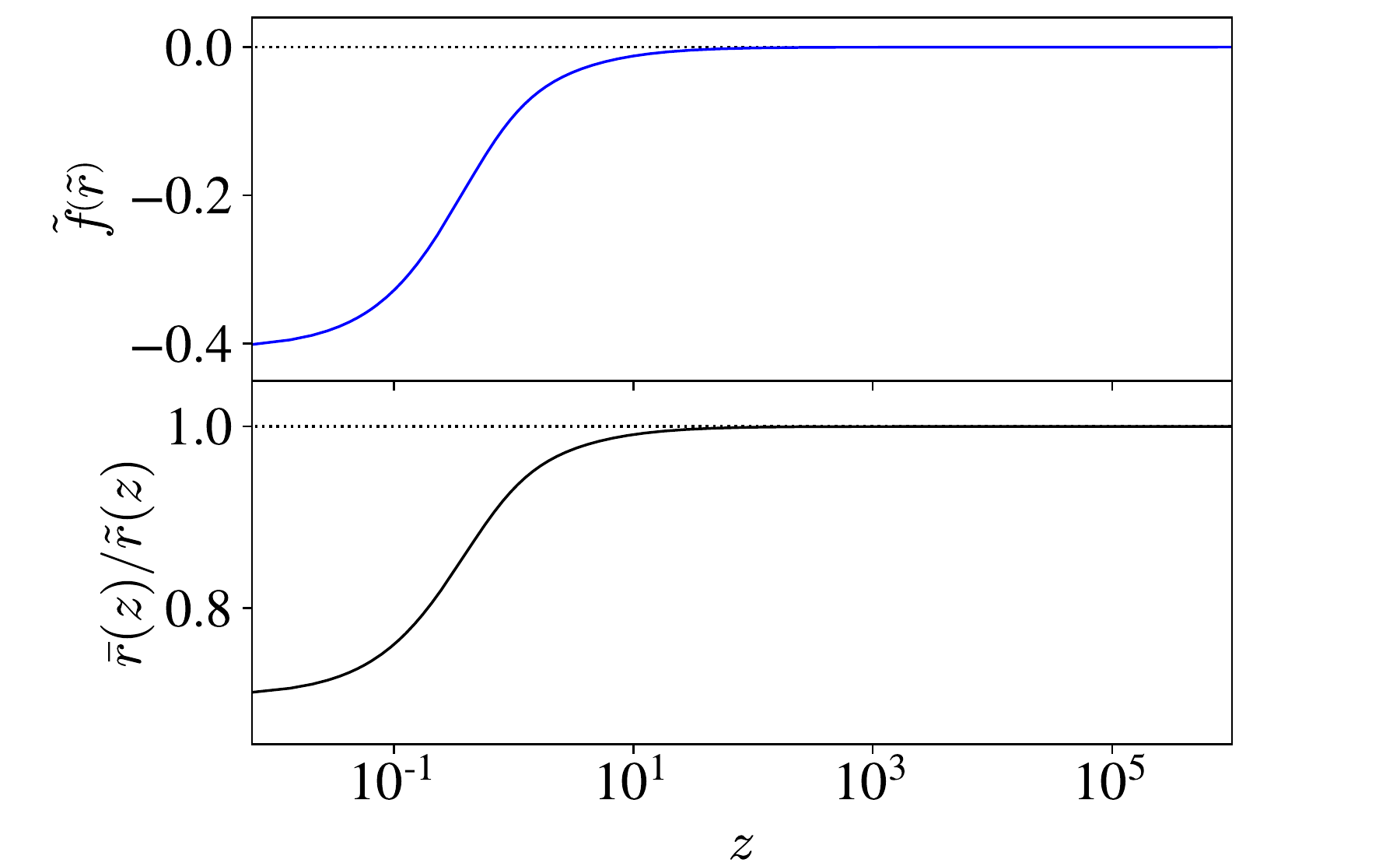} 
\caption{Background quantities for the CPL parameterization in both approaches (dynamical DE and interacting scenario). The cosmological parameters were fixed as in Eq.~\eqref{parameters}. \textbf{Top left panel:} Time evolution of the energy density parameter for radiation (dash-dotted lines), total matter (dashed lines) and DE (solid lines). \textbf{Bottom left panel:} Ratio between the density parameters. \textbf{Top right panel:} Interacting function associated to the CPL parameterization. The dotted line corresponds to the non-interacting case ($\Lambda$CDM model). \textbf{Bottom right panel:} Ratio between $\bar{r}\left(a\right)$ and $\tilde{r}\left(a\right)$. The dotted line has no physical meaning here and is only shown as a visual guide to indicate the axis
}
\label{bgcpl}
\end{figure}

In Fig.~\ref{bgcpl}, we provide the analogous version of Fig.~\ref{bgwcdm} for the CPL parameterization. As in the previous case, one can see that the interaction only strongly affects the background dynamics at late times, keeping unchanged the background dynamics at the last scattering surface and at the time of matter-radiation equality. 

Due to the fact that in this case the interaction is described by two free parameters, the corresponding interacting model is more complex. In this case for example, depending on the choice of the parameters $w_{0}$ and $w_{a}$, the sign of the function $\tilde{f}\left(\tilde{r}\right)$ can change with time, which implies that the direction of the energy transfer changes, as can be seen from Eq.~\eqref{fR}. Models with this particular feature are called ``sign-changeable'' models, and recently, some particular cases have been proposed in order to resolve the $H_{0}$ tension~\cite{Arevalo:2019axj,Pan:2019jqh}.
In particular, if the condition $w_{0}+w_{a}=-1$ is satisfied, the interaction is negligible at early times, becoming dynamically relevant only at late times, as the dark energy approaches a cosmological constant for $a\rightarrow 0$.
This condition is satisfied for our choice of model parameters ($w_{0}=-0.9$ and $w_{a}=-0.1$).
Correspondingly, we observe that $\tilde{f}\left(\tilde{r}\right)$ only becomes non-vanishing for $z\lesssim10^{1}$.

\begin{figure}
\centering
\includegraphics[width=0.49\columnwidth]{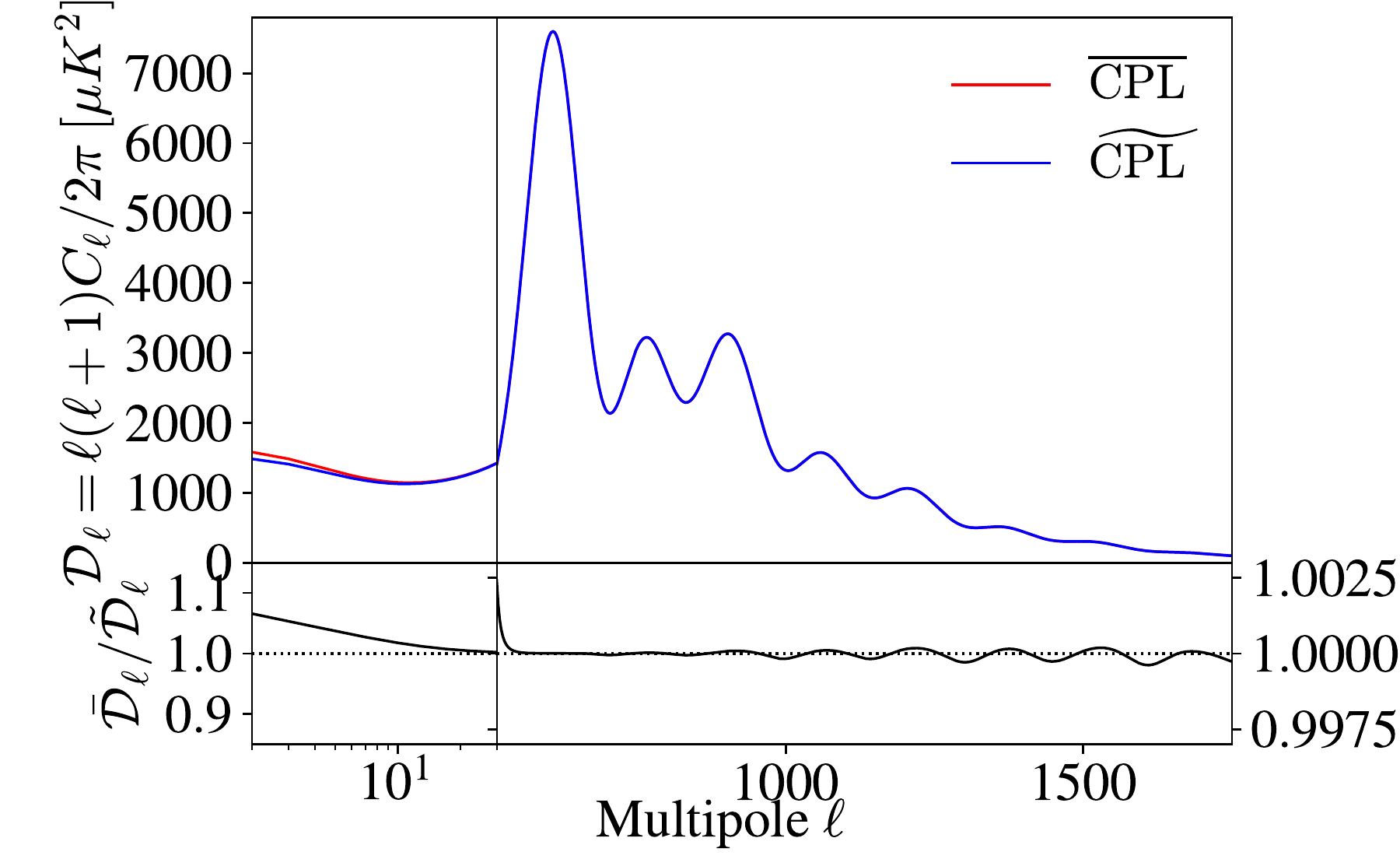} 
\includegraphics[width=0.49\columnwidth]{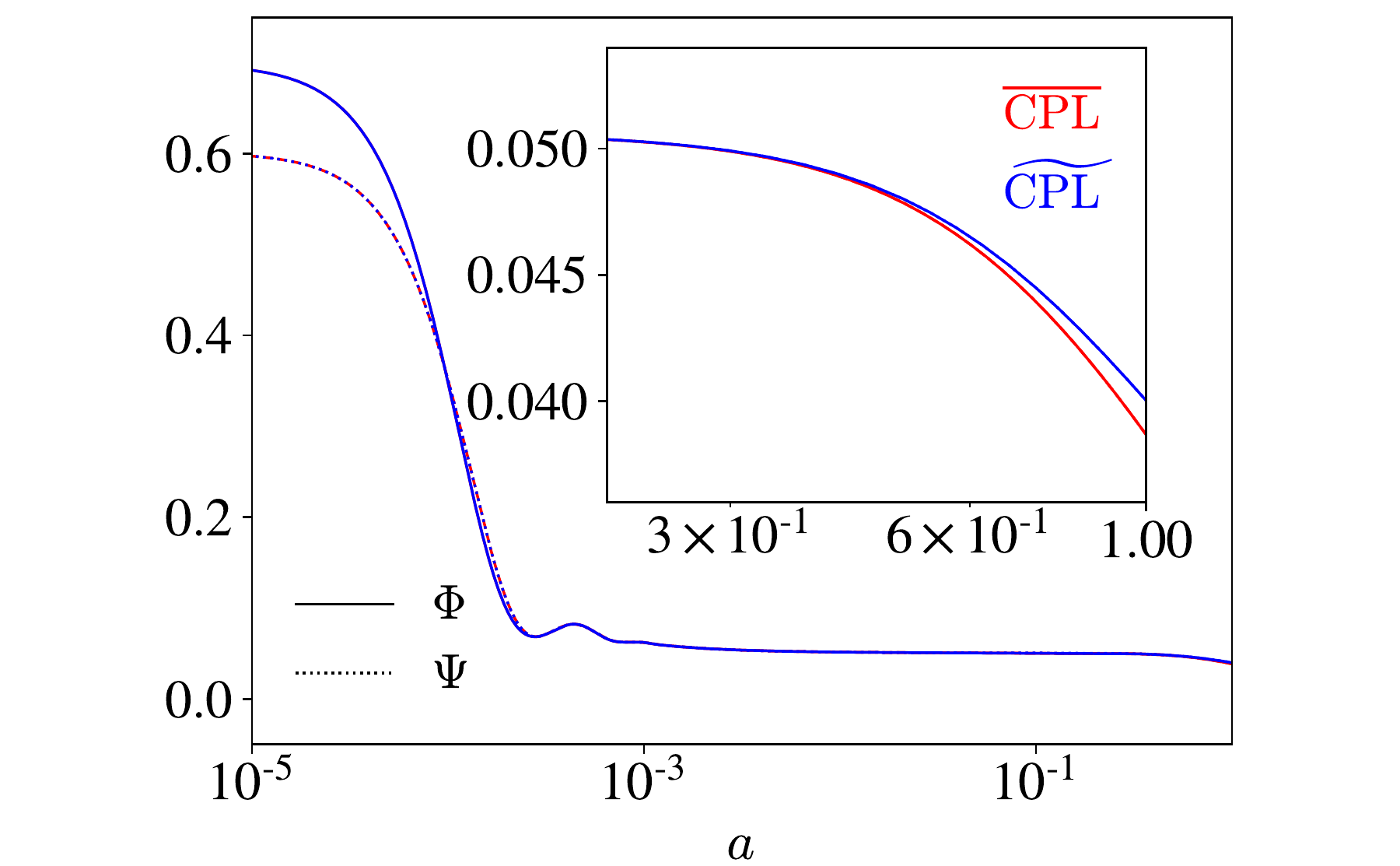} 
\includegraphics[width=0.49\columnwidth]{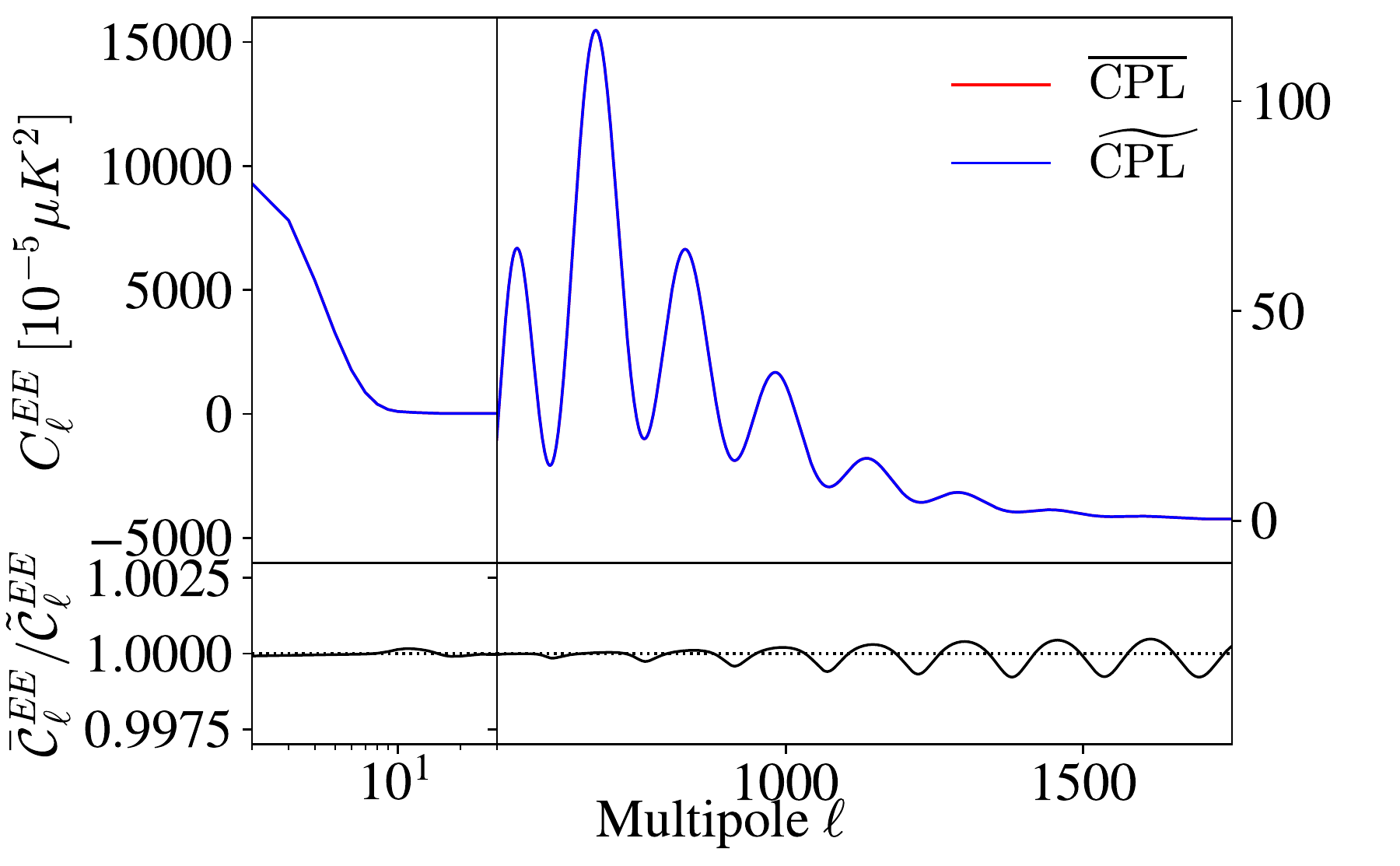} 
\includegraphics[width=0.49\columnwidth]{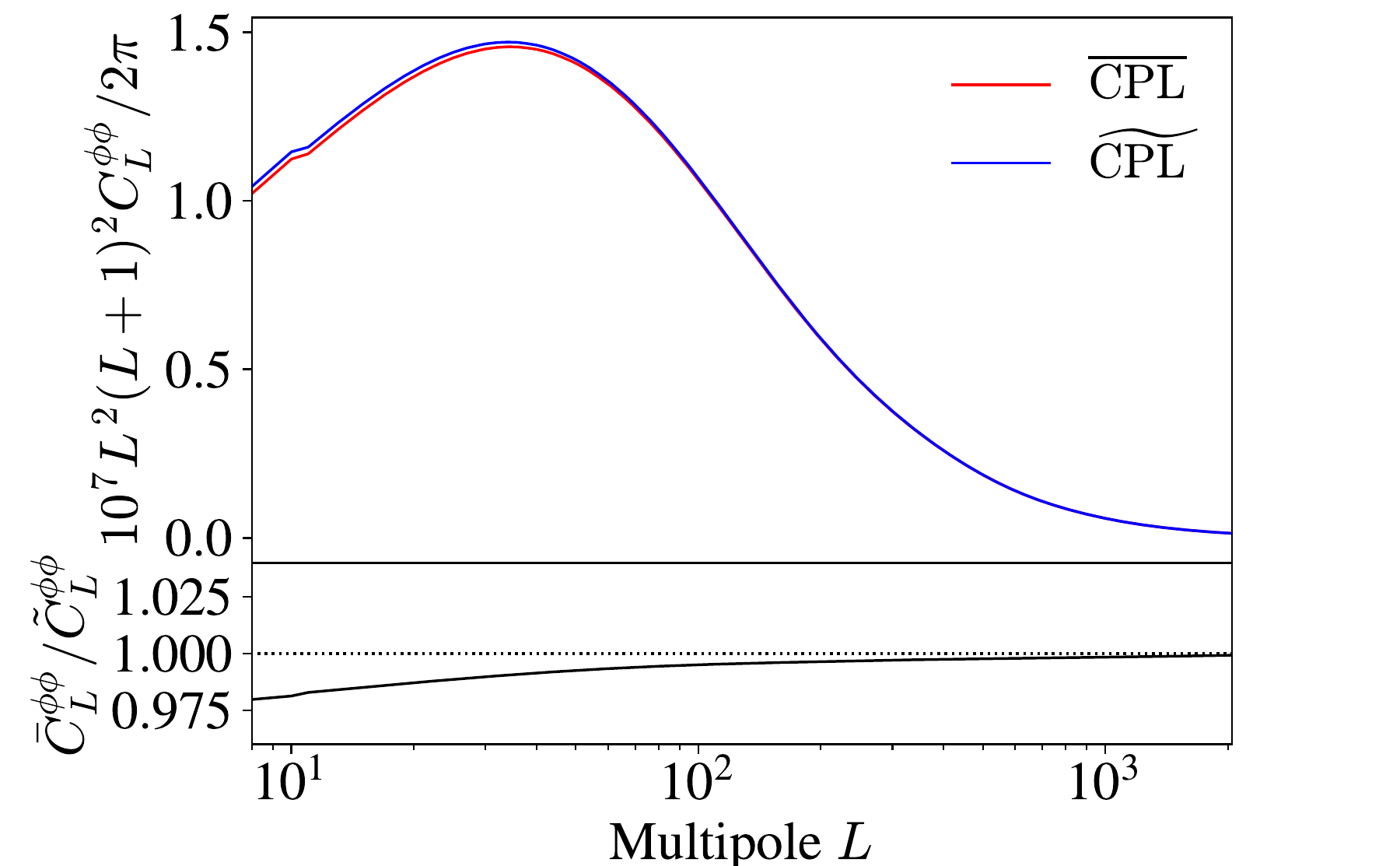}
\caption{Perturbative CMB quantities for the CPL parameterization in both approaches (dynamical DE and interacting scenario). The cosmological parameters were fixed as in Eq.~\eqref{parameters}. \textbf{Top left panel:} Lensed CMB temperature anisotropy power spectrum. \textbf{Bottom left panel:} Lensed CMB polarization (EE) power spectrum. \textbf{Top right panel:} Gauge-invariant gravitational potentials $\Psi$ and $\Phi$ at $k=0.1\,h/$Mpc. \textbf{Bottom right panel:} CMB lensing-potential power spectrum.}
\label{cmbcpl}
\includegraphics[width=0.49\columnwidth, trim={0.8cm 0 1.7cm 0}, clip]{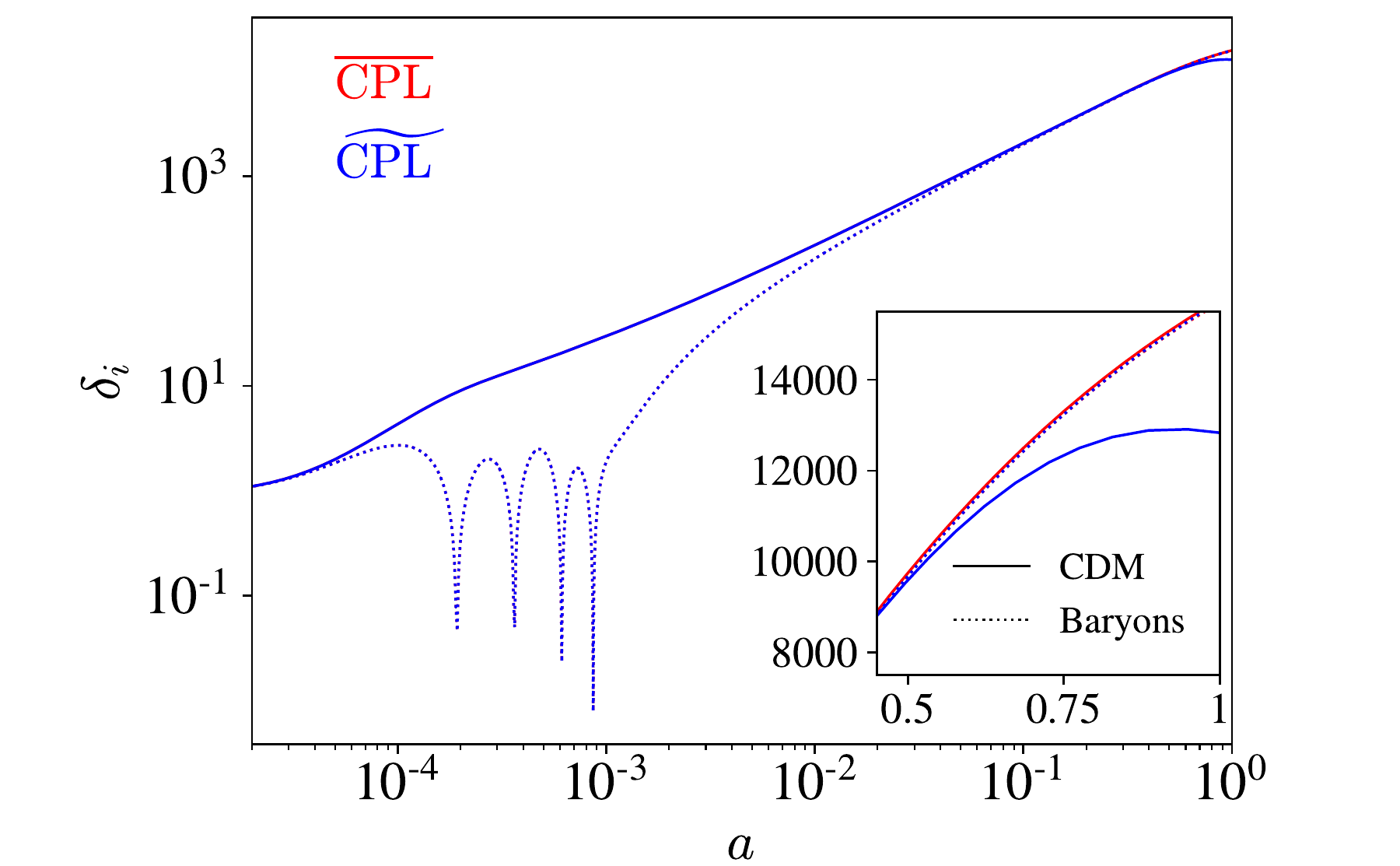} 
\includegraphics[width=0.49\columnwidth, trim={0.8cm 0 1.7cm 0}, clip]{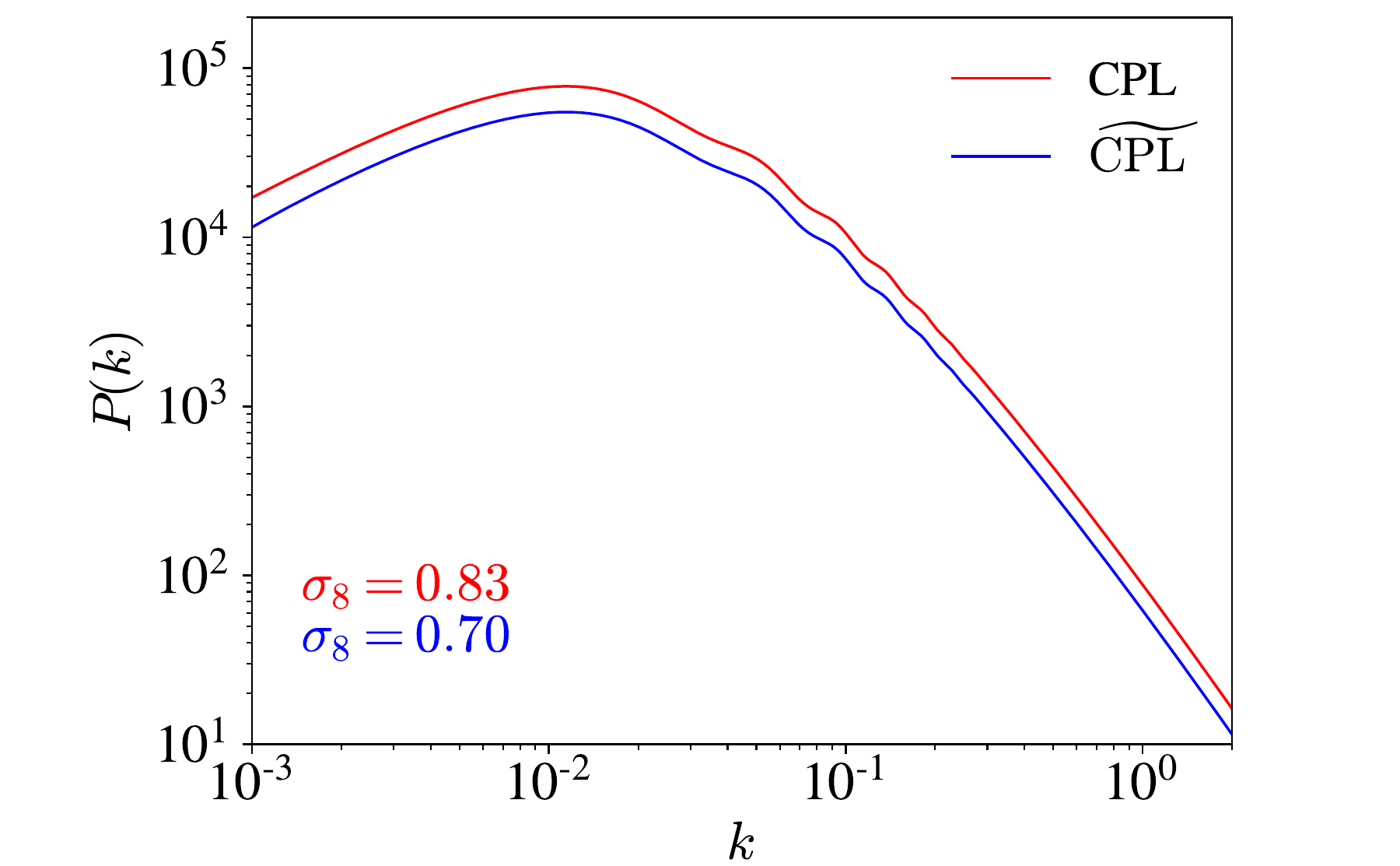} 
\caption{Perturbative matter quantities for the CPL parameterization in both approaches (dynamical DE and interacting scenario). The cosmological parameters were fixed as in Eq.~\eqref{parameters}. \textbf{Left panel:} Density contrast for baryons and CDM components at $k=0.1\,h/$Mpc. \textbf{Right panel:} Total matter power spectrum at $z=0$.}
\label{mattercpl}
\end{figure}

Fig.~\ref{cmbcpl} is the analog to Fig.~\ref{cmbwcdm} for the CPL parameterization. Again, for our parameter choices given in Eq.~\eqref{parameters}, CPL gives similar results to $w$CDM for all CMB power spectra and for the gravitational potentials, but with a slightly smaller difference between the two approaches. Likewise, Fig.~\ref{mattercpl} is the analog to Fig.~\ref{matterwcdm} with the same qualitative behavior, including a late-time suppression exhibited in the interacting CDM clustering.
%

\subsection{BA parameterization}
\label{ssec.ba}

For our last dynamical DE parameterization, we consider the model proposed in Ref.~\cite{Barboza:2008rh}, where the DE EoS is given by
\begin{equation} \label{wba}
\bar{w}_{x}\left(a\right)=w_{0}+w_{a}\dfrac{\left(1-a\right)}{1+2a\left(a-1\right)}\,.
\end{equation}
The parameterization is particularly interesting because in contrast to CPL it is a limited function of redshift throughout the entire evolution of the Universe since $\bar{w}_{x}\left(a\right)$ does not diverge in the limit $a\rightarrow\infty$. In the dynamical DE parameterization approach, the energy densities of the dark components become
\begin{equation} \label{rhobabar}
\bar{\rho}_{c}=\dfrac{3H_{0}^{2}}{8\pi G}\bar{\Omega}_{c0}a^{-3}\qquad\mbox{and}\qquad\bar{\rho}_{x}=\dfrac{3H_{0}^{2}}{8\pi G}\bar{\Omega}_{x0}a^{-3\left(1+w_{0}\right)}\left[1+\left(\dfrac{1-a}{a}\right)^{2}\right]^{3w_{a}/2} \,,
\end{equation}
which leads to the following solutions for the interacting background energy densities, 
\begin{eqnarray} 
\tilde{\rho}_{c}&=&\dfrac{3H_{0}^{2}}{8\pi G}\left\lbrace\bar{\Omega}_{c0}+\bar{\Omega}_{x0}a^{-3w_{0}}\left[1+w_{0}+w_{a}\dfrac{\left(1-a\right)}{1+2a\left(a-1\right)}\right]\left[1+\left(\dfrac{1-a}{a}\right)^{2}\right]^{3w_{a}/2}\right\rbrace a^{-3}\,, \label{rhoctildeba} \\
\tilde{\rho}_{x}&=&-\dfrac{3H_{0}^{2}}{8\pi G}\bar{\Omega}_{x0}a^{-3\left(1+w_{0}\right)}\left[w_{0}+w_{a}\dfrac{\left(1-a\right)}{1+2a\left(a-1\right)}\right]\left[1+\left(\dfrac{1-a}{a}\right)^{2}\right]^{3w_{a}/2} \,. \label{rhoxtildeba}
\end{eqnarray}

From these relations, the ratio between the CDM and DE energy densities in the dynamical DE and interacting scenarios are given by
\begin{equation} \label{rba}
\bar{r}\left(a\right)=\bar{r}_{0}a^{3w_{0}}\left[1+\left(\dfrac{1-a}{a}\right)^{2}\right]^{-3w_{a}/2}\quad\mbox{and}\quad\tilde{r}\left(a\right)=-\dfrac{1+w_{0}+w_{a}\frac{\left(1-a\right)}{1+2a\left(a-1\right)}+\bar{r}_{0}a^{3w_{0}}\left[1+\left(\frac{1-a}{a}\right)^{2}\right]^{-3w_{a}/2}}{w_{0}+w_{a}\frac{\left(1-a\right)}{1+2a\left(a-1\right)}} \,.
\end{equation}

Lastly, using the second term in Eq.~\eqref{rba} in Eq.~\eqref{ftilde}, one obtains the interacting term associated to the BA parameterization,
\begin{eqnarray} \label{frba}
\tilde{f}\left(\tilde{r}\right)&=&-1+\Bigg\lbrace w_{a}\left[1+2a\left(a-2\right)\right]\Bigg(2-\dfrac{2}{a}+\dfrac{1}{a^{2}}\Bigg)^{3w_{a}/2}a^{1-3w_{0}}+\bar{r}_{0}\Bigg[-3\bigg(w_{0}+2w_{0}a\left(a-1\right)\bigg)^{2}-6w_{0}w_{a} \nonumber \\ 
&&+w_{a}a\bigg(1+18w_{0}+2a\left(a-2\right)\left(1+6w_{0}\right)\bigg)-3w_{a}^{2}\left(a-1\right)^{2}\Bigg]\Bigg\rbrace\Bigg\lbrace3\left[w_{0}+2w_{0}a\left(a-1\right)+w_{a}\left(1-a\right)\right] \nonumber \\
&&\Bigg[\bigg(1+2a\left(a-1\right)\bigg)\bar{r}_{0}+\bigg(1+w_{0}+2a\left(a-1\right)\left(1+w_{0}\right)+w_{a}\left(1-a\right)\bigg)\Bigg(2-\dfrac{2}{a}+\dfrac{1}{a^{2}}\Bigg)^{3w_{a}/2}a^{-3w_{0}}\Bigg]\Bigg\rbrace^{-1} \,.
\end{eqnarray}

\begin{figure} 
\centering
\includegraphics[width=0.49\columnwidth, trim={0.7cm 0 2.1cm 0}, clip]{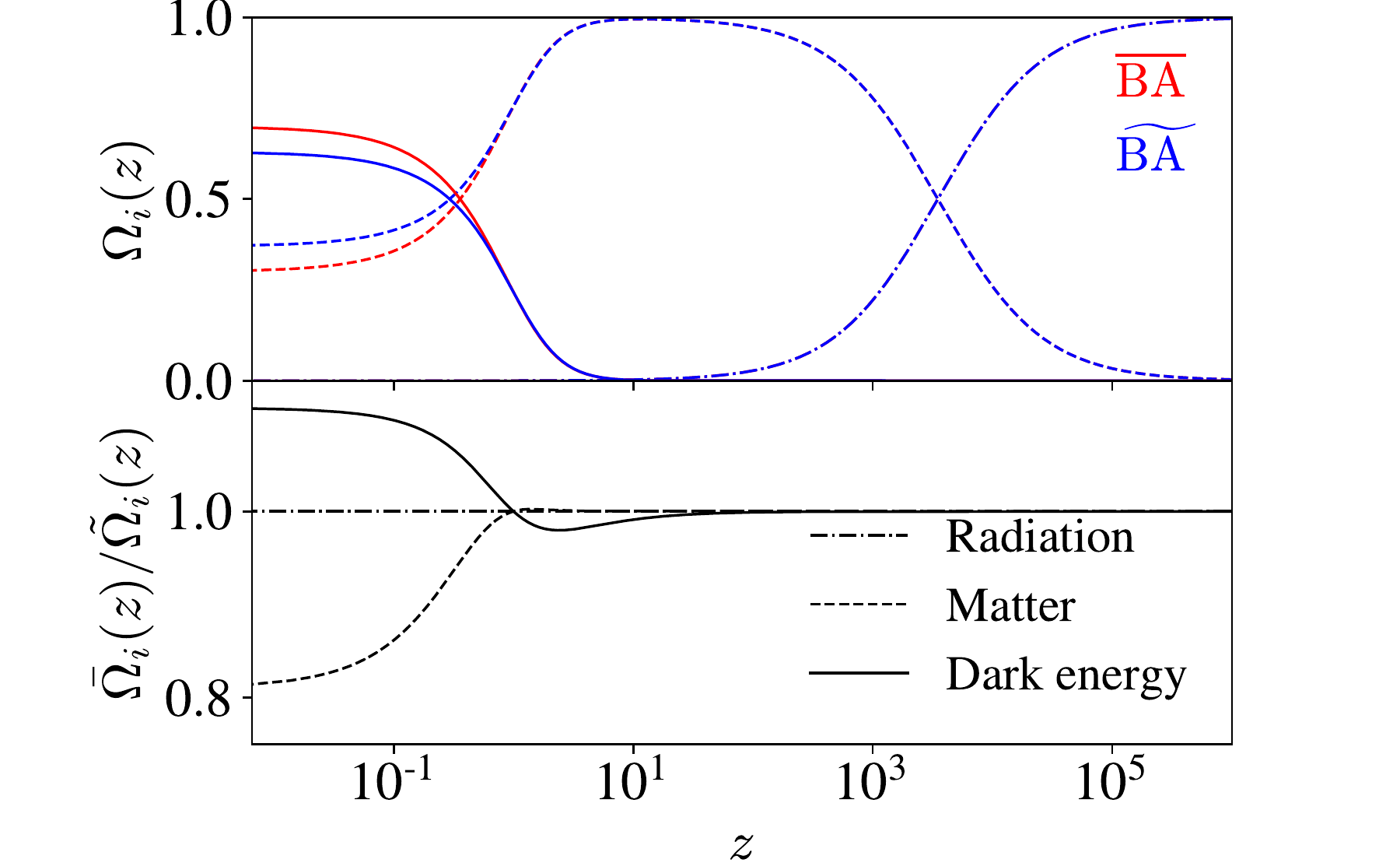} 
\includegraphics[width=0.49\columnwidth, trim={0.7cm 0 2.1cm 0}, clip]{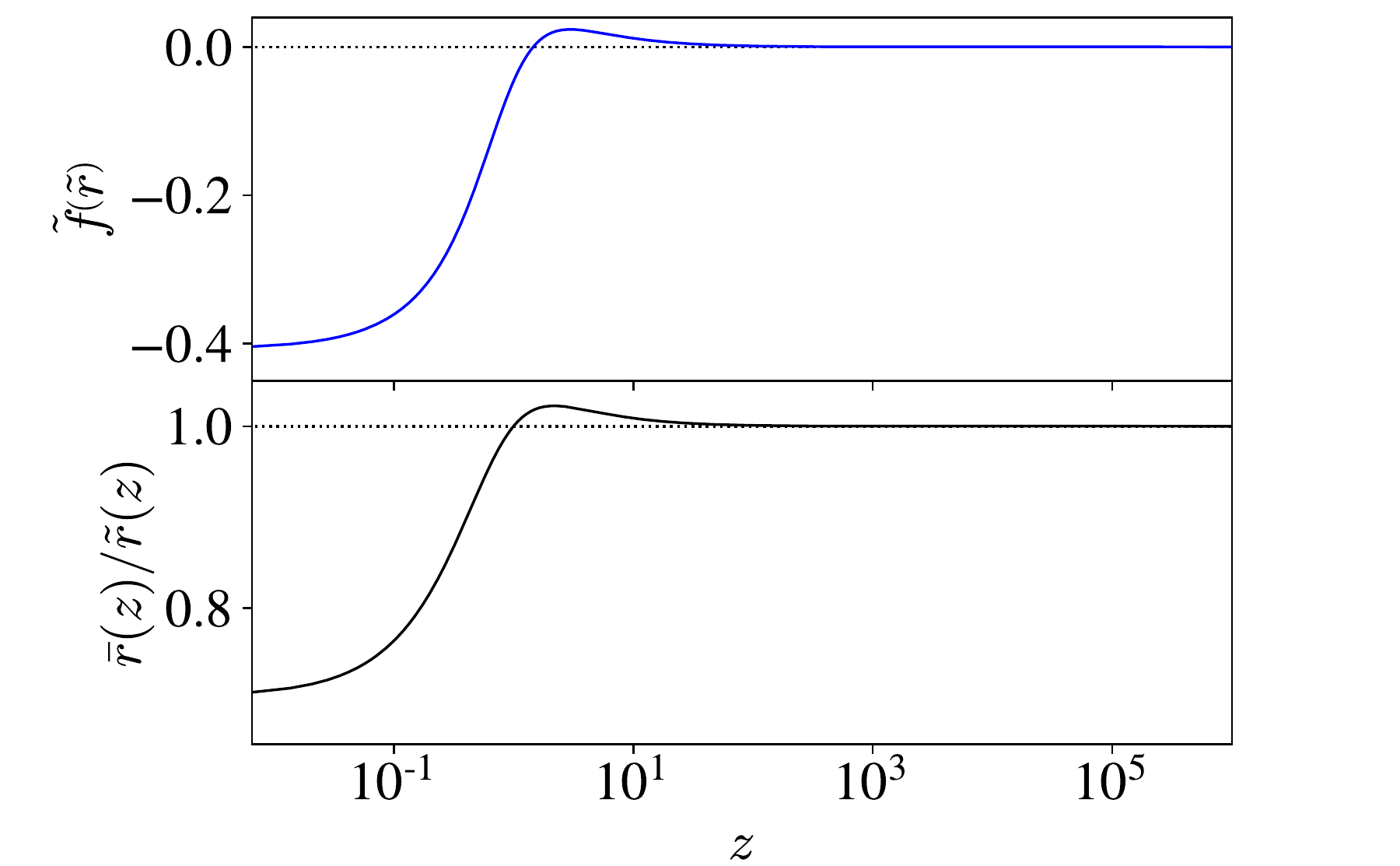} 
\caption{Background quantities for the BA parameterization in both approaches (dynamical DE and interacting scenario). The cosmological parameters were fixed as in Eq.~\eqref{parameters}. \textbf{Top left panel:} Time evolution of the energy density parameter for radiation (dash-dotted lines), total matter (dashed lines) and DE (solid lines). \textbf{Bottom left panel:} Ratio between the density parameters. \textbf{Top right panel:} Interacting function associated to the BA parameterization. The dotted line corresponds to the non-interacting case ($\Lambda$CDM model). \textbf{Bottom right panel:} Ratio between $\bar{r}\left(a\right)$ and $\tilde{r}\left(a\right)$. The dotted line has no physical meaning here and is only shown as a visual guide to indicate the axis.
}
\label{bgba}
\end{figure}

Analogously to the analysis performed in Secs.~\ref{ssec.wcdm} and~\ref{ssec.cpl} for the $w$CDM and CPL parameterizations, in Fig.~\ref{bgba} we show the dynamically relevant background physical quantities to assess the difference between the dynamical DE and interacting approaches for the BA  parameterization. One can see from the top right-hand panel that in this case, adopting $w_{0}=-0.9$ and $w_{a}=-0.1$, the model produces a sign-changeable interacting scenario. Physically, this means that the direction of the background energy transfer changes in time. Once again, the condition $w_{0}+w_{a}=-1$ leads to a vanishing interaction at early times.

\begin{figure}
\centering
\includegraphics[width=0.45\columnwidth]{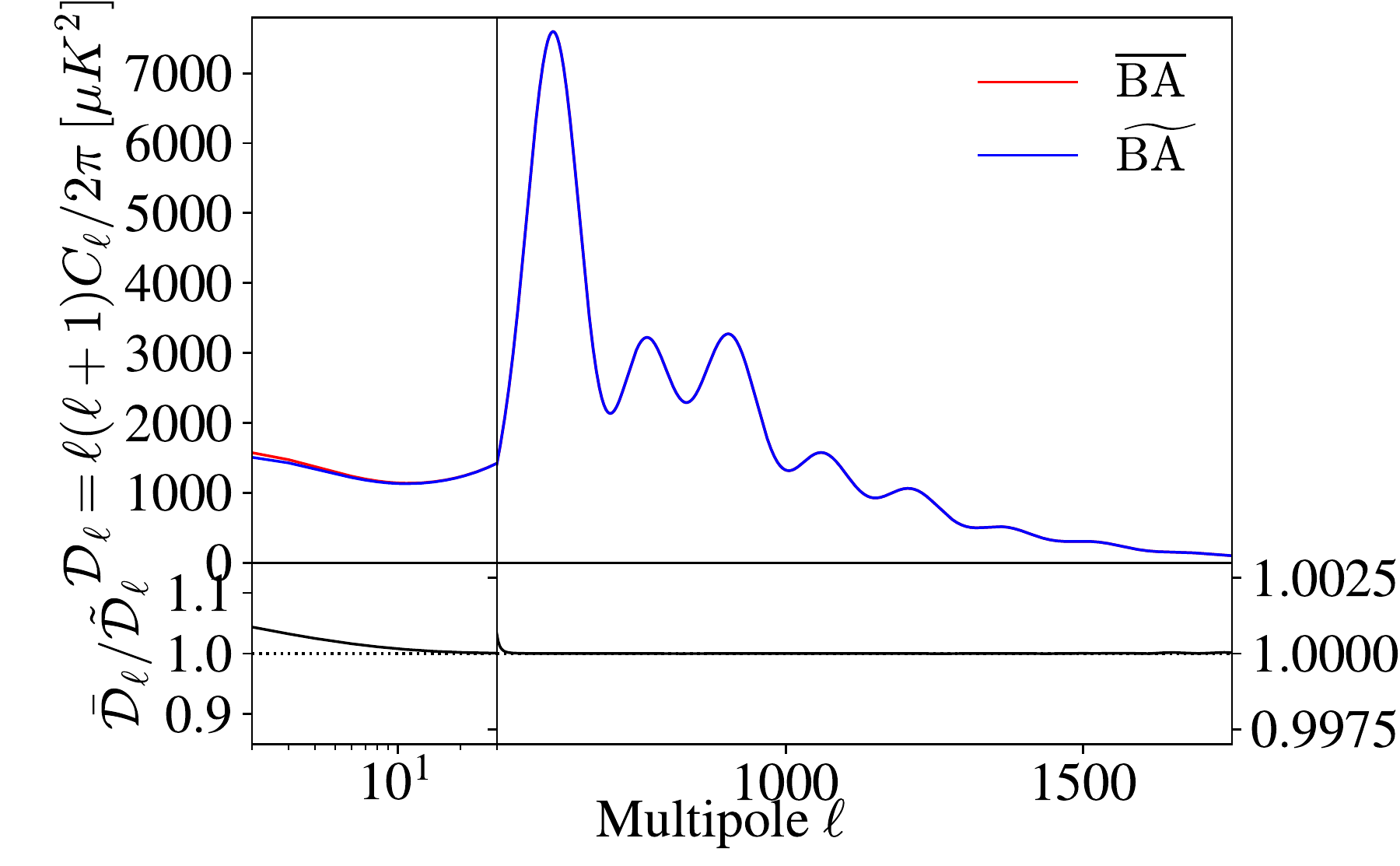} 
\includegraphics[width=0.45\columnwidth]{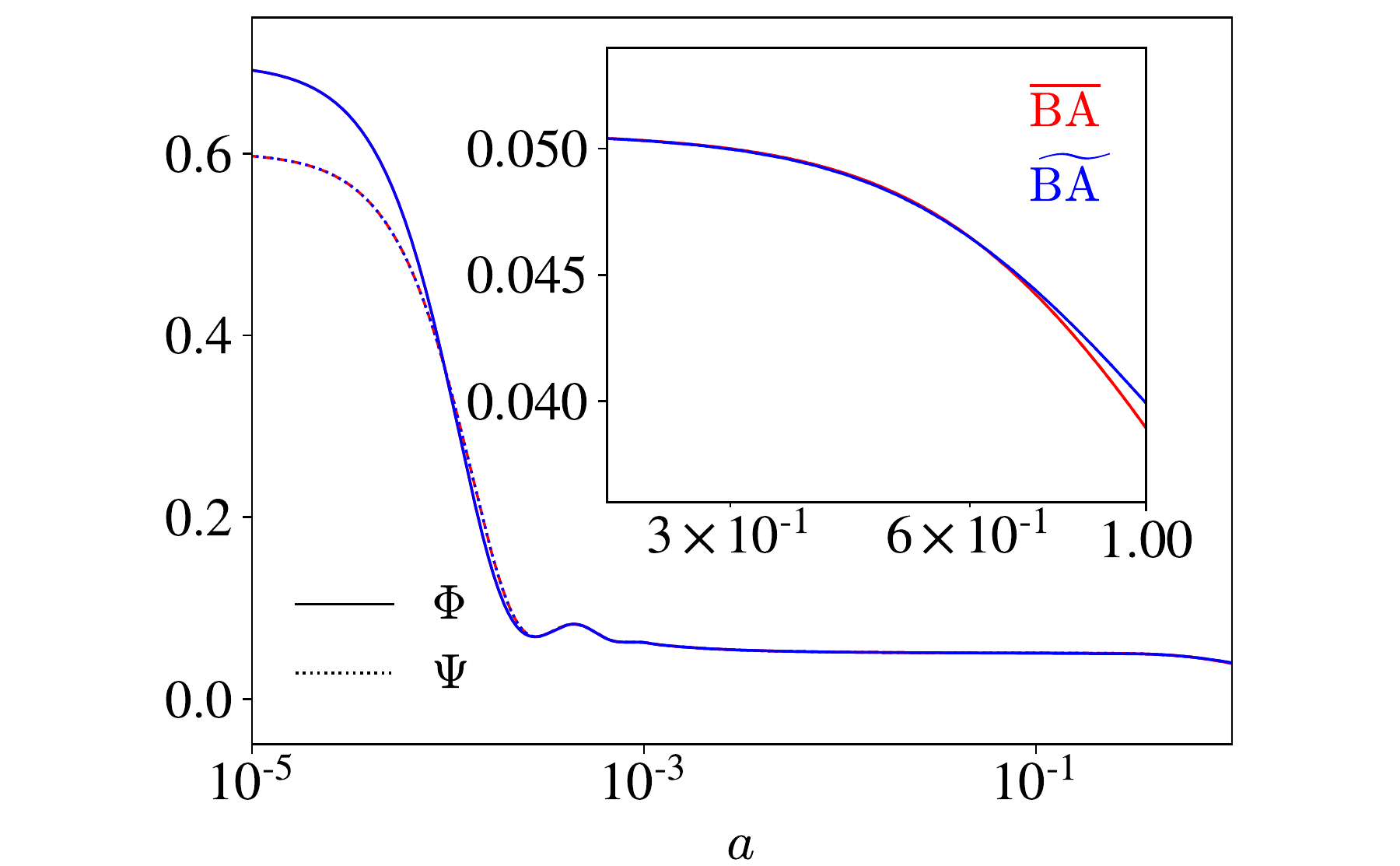} 
\includegraphics[width=0.45\columnwidth]{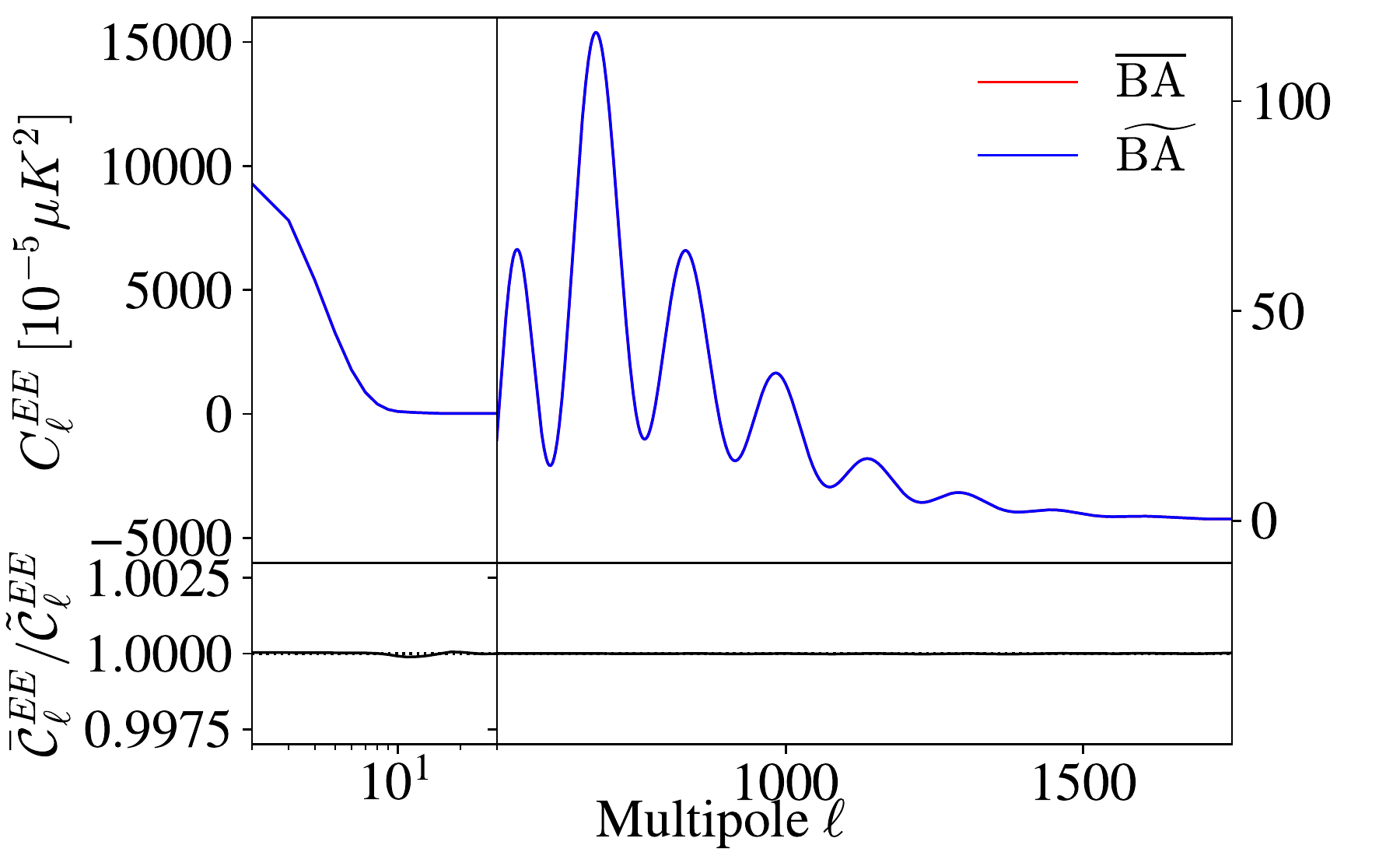} 
\includegraphics[width=0.45\columnwidth]{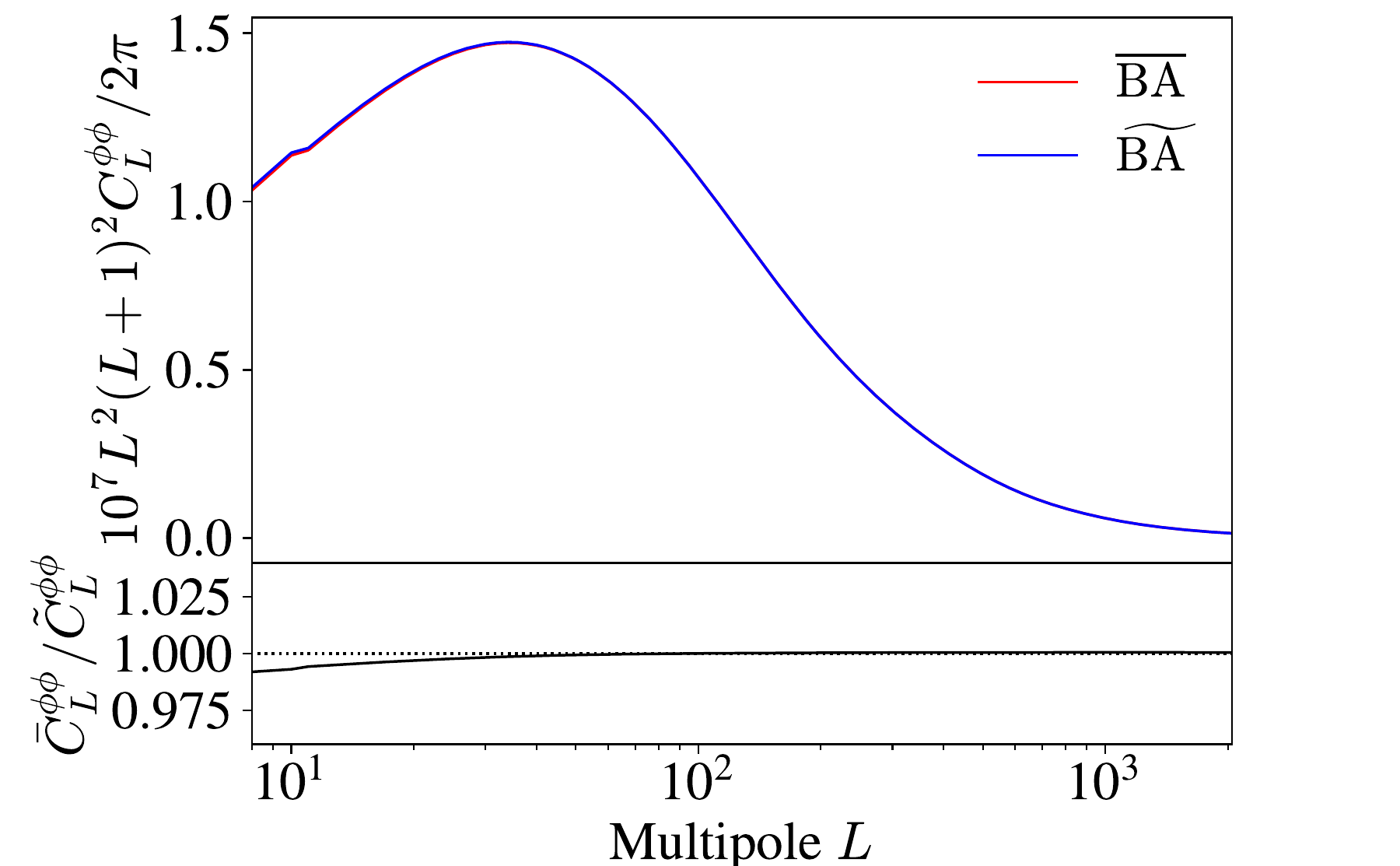}
\caption{Perturbative CMB quantities for the BA  parameterization in both approaches (dynamical DE and interacting scenario).  The cosmological parameters were fixed as in Eq.~\eqref{parameters}. \textbf{Top left panel:} Lensed CMB temperature anisotropy power spectrum. \textbf{Bottom left panel:} Lensed CMB polarization (EE) power spectrum. \textbf{Top right panel:} Gauge-invariant gravitational potentials $\Psi$ and $\Phi$ at $k=0.1\,h/$Mpc. \textbf{Bottom right panel:} CMB lensing-potential power spectrum.}
\label{cmbba}
\includegraphics[width=0.49\columnwidth, trim={0.8cm 0 1.7cm 0}, clip]{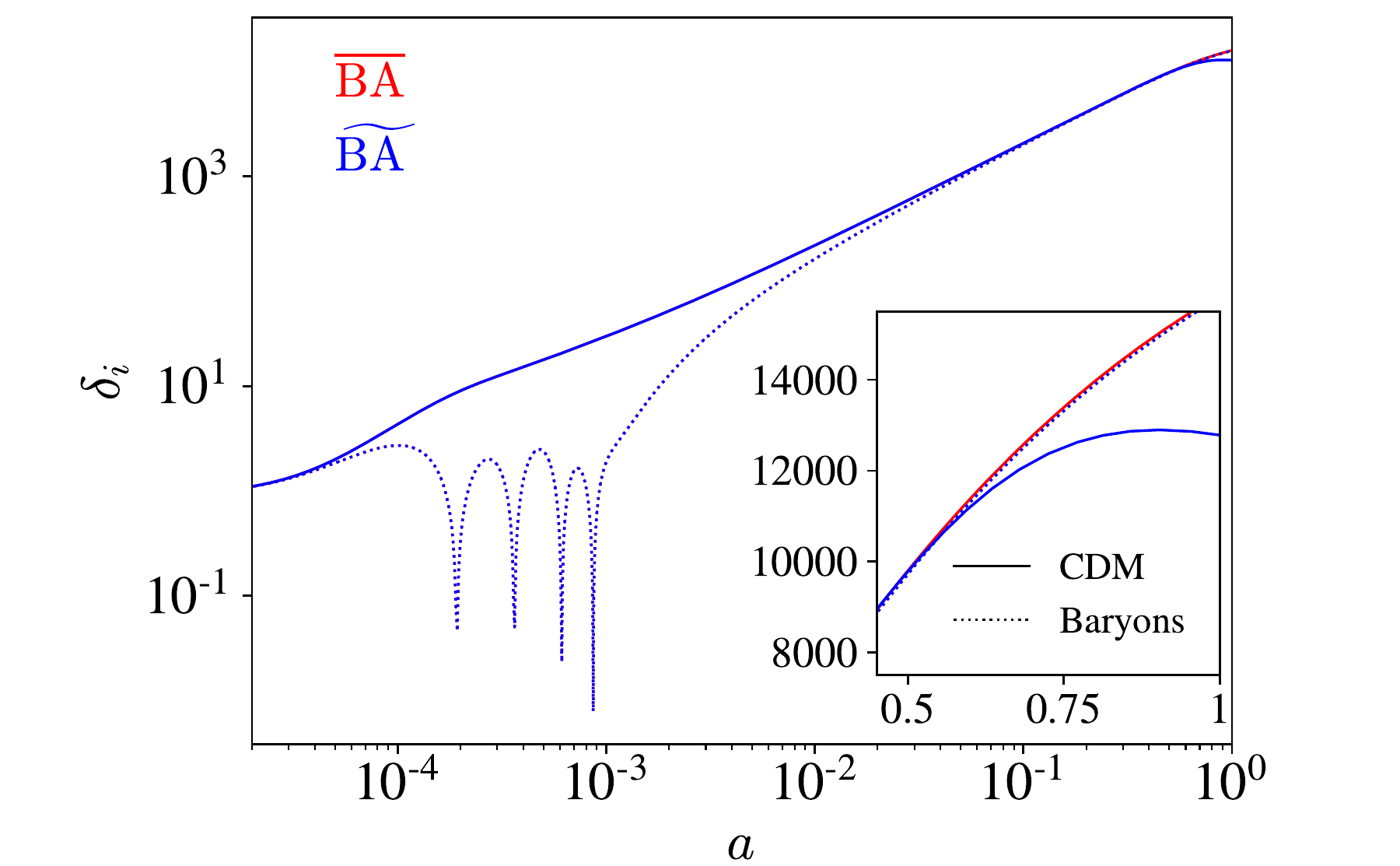} 
\includegraphics[width=0.49\columnwidth, trim={0.8cm 0 1.7cm 0}, clip]{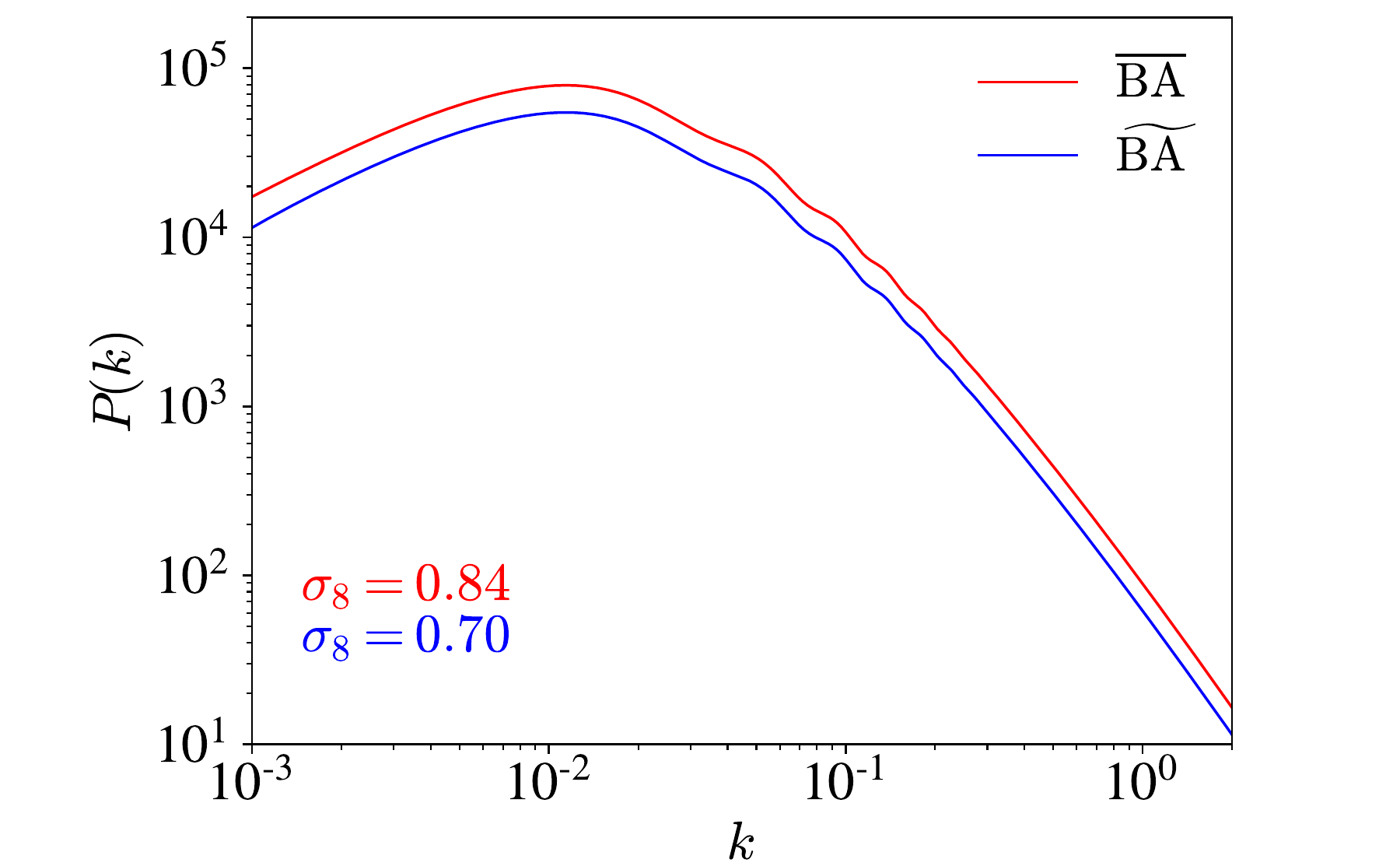} 
\caption{Perturbative matter quantities for the BA  parameterization in both approaches (dynamical DE and interacting scenario).  The cosmological parameters were fixed as in Eq.~\eqref{parameters}. \textbf{Left panel:} Density contrast for baryons and CDM components at $k=0.1\,h/$Mpc. \textbf{Right panel:} Total matter power spectrum at $z=0$.}
\label{matterba}
\end{figure}

Analogously to Figs.~\ref{cmbwcdm} and \ref{cmbcpl}, the same CMB quantities are shown in Fig.~\ref{cmbba}. In this particular case, for our choice of cosmological parameters given in Eq.~\eqref{parameters}, the difference between the two approaches in the ISW is of comparable magnitude to the other cases whereas CMB lensing effects are considerably weaker in comparison to $w$CDM and CPL. The difference due to the ISW effect can be seen in the low $\ell$ region of the top left-hand panel of Fig.~\ref{cmbba}, which is in agreement with the panel on the top right-hand side. On the other hand, the tiny difference due to lensing effects can be seen in the bottom right-hand panel. As a consequence of this small effect, the difference in the temperature CMB power spectrum (for high values of $\ell$) and for the polarization (EE) CMB power spectrum are also tiny. Lastly, analogously to Figs.~\ref{matterwcdm} and \ref{mattercpl}, Fig.~\ref{matterba} shows the matter perturbation quantities for the BA parameterization with comparable results.

%

\section{Statistical analysis}
\label{sec.stat}

With the dark degeneracy at the level of the cosmological background and linear perturbations elaborated in Sec.~\ref{sec.equiv} and the introduction of three example parametrizations of the dark sector that break the degeneracy at the linear level in Sec.~\ref{sec.specific}, we shall now turn to performing a Bayesian statistical parameter estimation analysis of the three parametrizations in each of the two scenarios, dynamical DE and the interacting counterpart.
For this purpose, we employ data from type Ia Supernovae (SNe Ia), Baryonic Acoustic Oscillations (BAO), Cosmic Chronometers (CC) and the full CMB data from Planck, which encompasses information from temperature and polarization maps and the lensing reconstruction, Planck(TT,TE,EE+lowE+lensing).

For this analysis we use a modified version of the {\sc class} code and its embedding in {\sc MontePython}~\cite{Audren:2012wb,Brinckmann:2018cvx} to conduct the Markov Chain Monte Carlo (MCMC) sampling of the parameter space.
The statistical analysis of the chains is performed with the integrated {\sc GetDist} code, which we also employ to produce figures representing the resulting likelihoods in the parameter space.
For all our chains, we require that the Gelman-Rubin convergence parameter satisfies the condition $\hat{R}-1<0.01$~\cite{gelman1992inference}. 

We divide our statistical analysis into two parts: in the first stage we consider only background data from SNe Ia, BAO and CC, whereas for the second stage we follow the analysis of the latest Planck paper~\cite{Aghanim:2018eyx}, adopting the background parameterization ($w_{0}$,$w_{a}$) and employing the Planck data (TT,TE,EE+lowE+lensing) combined with the SNe Ia and BAO data. In the first analysis, since the background data cannot distinguish baryons and CDM even in the interacting approach, the baryonic component is fixed according to the latest Planck results~\cite{Aghanim:2018eyx}. Thus, the set of free cosmological parameters we consider is \{$\bar{\Omega}_{c0}$/$\tilde{\Omega}_{c0}$,$H_{0}$,$w_{0}$,$w_{a}$\}\footnote{For the $w$CDM model $w_{a}$ is zero.}. Since the background degeneracy is maintained and all the background data we have used are based on measure distances, in principle the statistical analysis should give identical results for both approaches, i.e., they should lead to the same best-fit and posterior distributions. However for finite MCMC chains there will be tiny differences in the parameter constraints. In agreement with this expectation, the background corner plots in Appendix~\ref{ap.corner}, Figs.~\ref{bgcorner_wcdm},~\ref{bgcorner_cpl} and~\ref{bgcorner_ba} , show that the results of the two approaches are indeed compatible. Note that the same result for both approaches means that they should have exactly the same best-fit/posterior distribution for $H_{0}$, $w_{0}$ and $w_{a}$ but not for $\bar{\Omega}_{c0}$/$\tilde{\Omega}_{c0}$. Since $\bar{w}_{x}\left(a=1\right)=w_{0}$ for all cases analyzed here, Eq.~\eqref{rhoc2} leads to the condition $\tilde{\Omega}_{c0}=\bar{\Omega}_{c0}+\bar{\Omega}_{x0}\left(1+w_{0}\right)$, where, neglecting radiation, $\bar{\Omega}_{x0}=1-\bar{\Omega}_{c0}-\bar{\Omega}_{b0}$. This relation must be verified for the best-fit values and the posterior distributions must satisfy the error propagation.

In the second analysis the set of free cosmological parameters we consider is \{$\omega_{b}$, $\omega_{c}$, $H_{0}$, $\tau_{reio}$, $\ln\left(10^{10}A_{s}\right)$, $n_{s}$, $w_{0}$, $w_{a}$\}, where $\omega_{b}\equiv\Omega_{b0}h^{2}$, $\omega_{c}\equiv\Omega_{c0}h^{2}$, $\tau_{reio}$ is the reionization optical depth, $A_{s}$ is the initial super-horizon amplitude of curvature perturbations at $k_{\rm pivot}=0.05\ {\rm Mpc}^{-1}$ and $n_{s}$ is the primordial spectral index. In order to assess LSS tensions, the matter fluctuation amplitude $\sigma_{8}$ is also computed as a derived parameter. In the perturbative case, since the degeneracy is broken, no identical results/algebraic relations are expected \textit{a priori}. In both analyses we have used wide flat priors.

An important difference between the dynamical and interacting approaches is that $w_{0}$ and $w_{a}$ imply different phenomenology, for instance, in the interacting approach they affect the CDM evolution. This implies that parameter uncertainties and correlations between different parameters can be different. In particular, we expect a stronger correlation between $w_{0}$ and $w_{a}$ and the CDM density parameter in the interacting approach, but as a consequence of this, we also expect that uncertainties in $w_{0}$ and $w_{a}$ propagate to the CDM component, yielding larger errors on $\tilde{\Omega}_{c0}$ and $\tilde{\sigma}_{8}$.

\subsection{Data}
\label{ssec.data}

Before presenting the results of the parameter estimation analysis in Sec.~\ref{ssec.results}, we shall give a brief discription of the datasets we employ. More precisely, the datasets employed are the following:
\begin{itemize}
\item \textbf{SNe Ia:} For the type Ia SNe data, we use the 1048 distance moduli
from the Pantheon sample~\cite{Jones:2017udy,Scolnic:2017caz} with its covariance matrix  (taking into account the statistical and systematic errors)\footnote{The Pantheon data as well as its covariance matrix can be downloaded from \href{https://github.com/dscolnic/Pantheon}{github.com/dscolnic/Pantheon}.}.
The Pantheon catalog contains data points of peak magnitudes in the rest frame of the B band $m_{B}$, which is related to the distance modulus
as
$\mu=m_{B}+M$.
The theoretical prediction for the distance modulus is given by
\begin{equation} \label{mu}
\mu\left(z\right)=5\log\left[\frac{d_{L}\left(z\right)}{{\rm 1Mpc}}\right]+25\,,
\end{equation}
where $d_{L}$ is the luminosity distance
\begin{equation} \label{dl}
d_{L}\left(z\right)=\left(1+z\right)\int_{0}^{z}\frac{d\tilde{z}}{H\left(\tilde{z}\right)}\,.
\end{equation}

\item \textbf{Cosmic Chronometers:} We use 30 data points obtained from differential ages of passively evolving old galaxies whose redshifts are known. Combining these quantities, it is possible to obtain model-independent measurements of the Hubble  expansion rate at different redshifts~\cite{Zhang:2012mp,Simon:2004tf,Moresco:2012jh,Stern:2009ep,Moresco:2015cya,Moresco:2016mzx}.
The collection of data points is summarized in Ref.~\cite{Moresco:2016mzx}.

\item \textbf{BAO:} Following the latest Planck results~\cite{Aghanim:2018eyx}, we use BAO data from the 6dF Galaxy Survey~\cite{Beutler:2011hx} and BOSS-DR12~\cite{Alam:2016hwk}. We avoid using further available dataset, e.g., the BAO data from WiggleZ~\cite{Kazin:2014qga},
since there is some overlap in the region of sky with the data we already use with unknown correlations.
For the BAO data, the important physical quantities are the sound horizon of the primordial photon-baryon plasma at the drag epoch, the angular diameter distance and the dilation scale~\cite{Eisenstein:2005su}.
These are respectively given by 
\begin{equation} \label{baoquantities}
r_{s}=\int_{z_{drag}}^{\infty}\frac{c_{s}\left(z\right)}{H\left(z\right)}dz\quad,\quad d_{A}=\frac{1}{1+z}\int_{0}^{z}\frac{d\tilde{z}}{H\left(\tilde{z}\right)}\quad\mbox{and}\quad D_{V}=\left[\left(1+z\right)^{2}d_{A}\frac{z}{H\left(z\right)}\right]^{1/3} \,,
\end{equation}
where
$c_{s}$ denotes the sound speed in the primordial photon-baryon plasma.
Since the baryon and radiation components are unaffected in the models we consiser, $c_{s}$ can be obtained from the usual expression valid for the $\Lambda$CDM model,
\begin{equation} \label{cs}
c_{s}=\frac{1}{\sqrt{3\left[1+\frac{4\Omega_{b0}}{4\Omega_{r0}}\left(1+z\right)^{-1}\right]}}\,.
\end{equation}

\item \textbf{Planck 2018(TT, TE, EE+lowE+lensing):} Finally, we employ the most recent full CMB data from the temperature maps, polarization maps and lensing reconstruction through the Commander and Plik codes\footnote{Likelihood codes and the data can be downloaded from \href{http://pla.esac.esa.int/pla/}{pla.esac.esa.int/pla}.}~\cite{Aghanim:2019ame}. In general, Planck data cannot strongly constrain the $w_{0}$ and $w_{a}$ parameters of the dynamical DE scenario (which is the reason we decide to combine the data with SNe Ia and BAO data). In the interacting approach, however, we expect stronger constraints from the CMB since there these parameters affect the CDM evolution.
\end{itemize}

\subsection{Results}
\label{ssec.results}

Having discussed the models tested in Sec.~\ref{sec.specific} and the observational data employed in Sec.~\ref{ssec.data}, we now present the results of our parameter estimation analysis. 
For simplicity, when results are presented for both the dynamical DE scenarios and their interacting counterparts in combined tables and figures, we allow ourselves to avoid the bar/tilde notation, but it is implicit that each result corresponds to the respective scenario. For example in Tab.~\ref{tabbg}, the column $\Omega_{c0}$ corresponds to $\bar{\Omega}_{c0}$ when it refers to a dynamical DE model and to $\tilde{\Omega}_{c0}$ for the interacting dark sector models.
We will first present the parameter constraints obtained from the background data in Sec.~\ref{ssec.backgroundconstraints} and then in Sec.~\ref{ssec.CMBconstraints} present the results inferred from the combination with CMB data.

\subsubsection{Background analysis} 
\label{ssec.backgroundconstraints}

\begin{figure}
\centering
\includegraphics[scale=0.35]{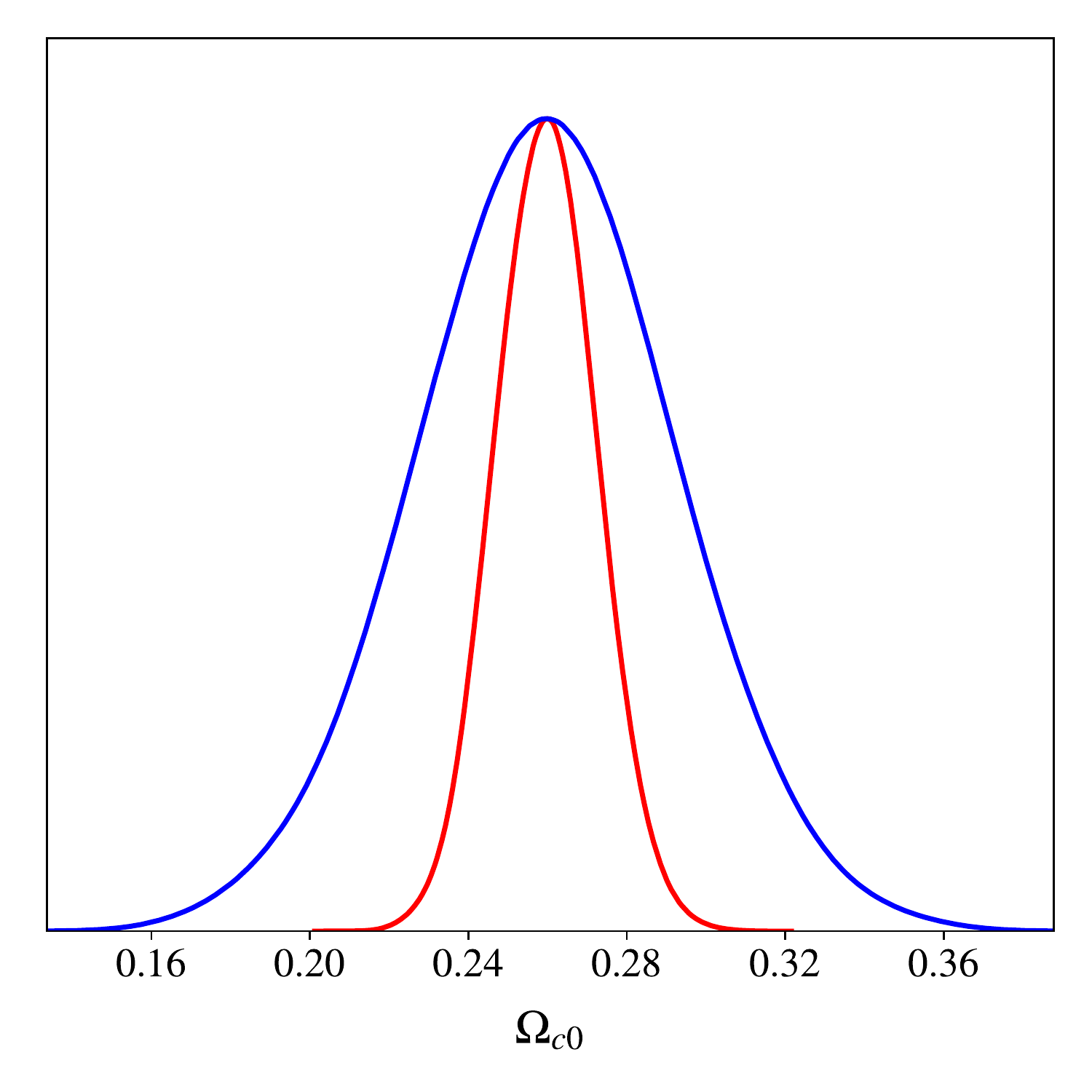} 
\includegraphics[scale=0.35]{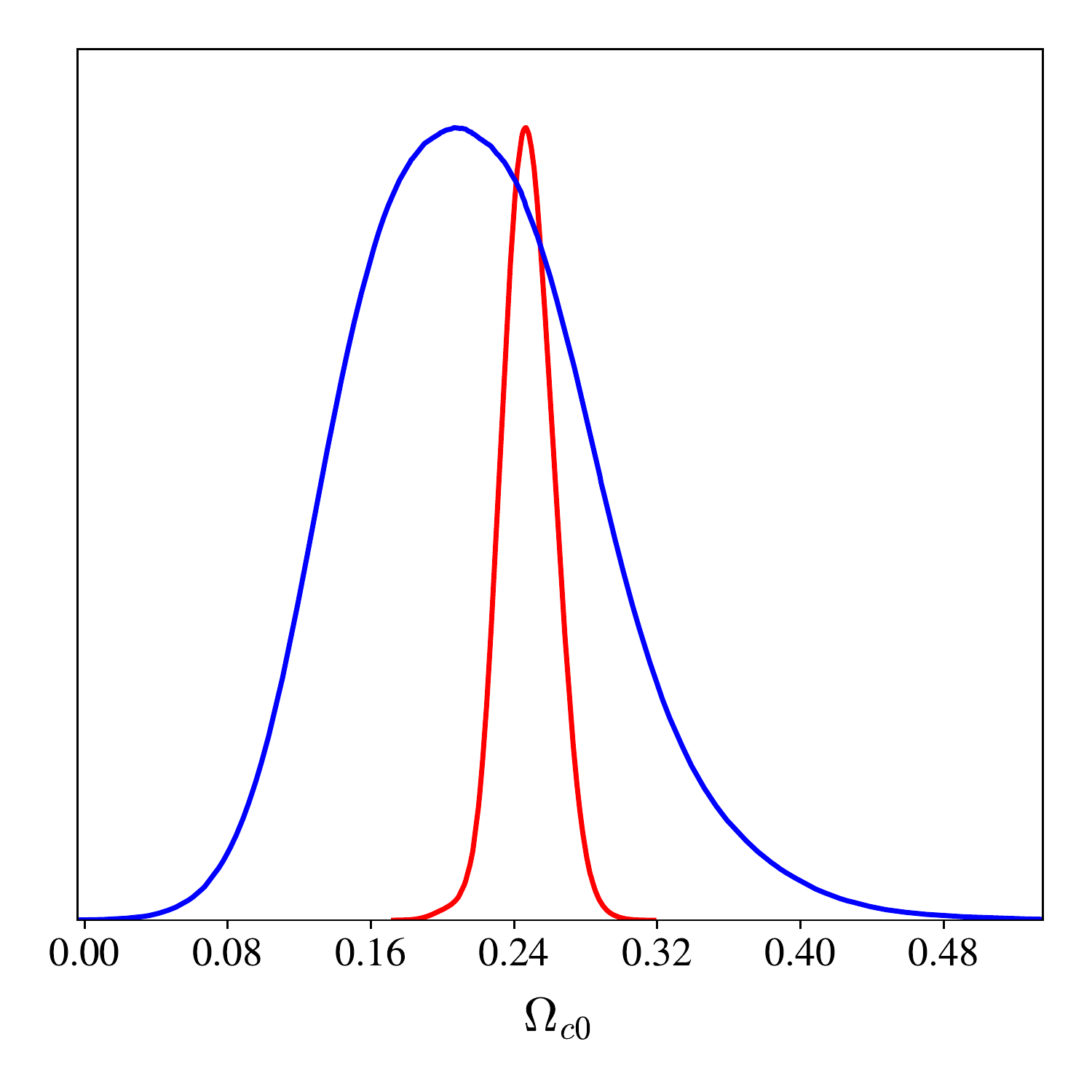} 
\includegraphics[scale=0.35]{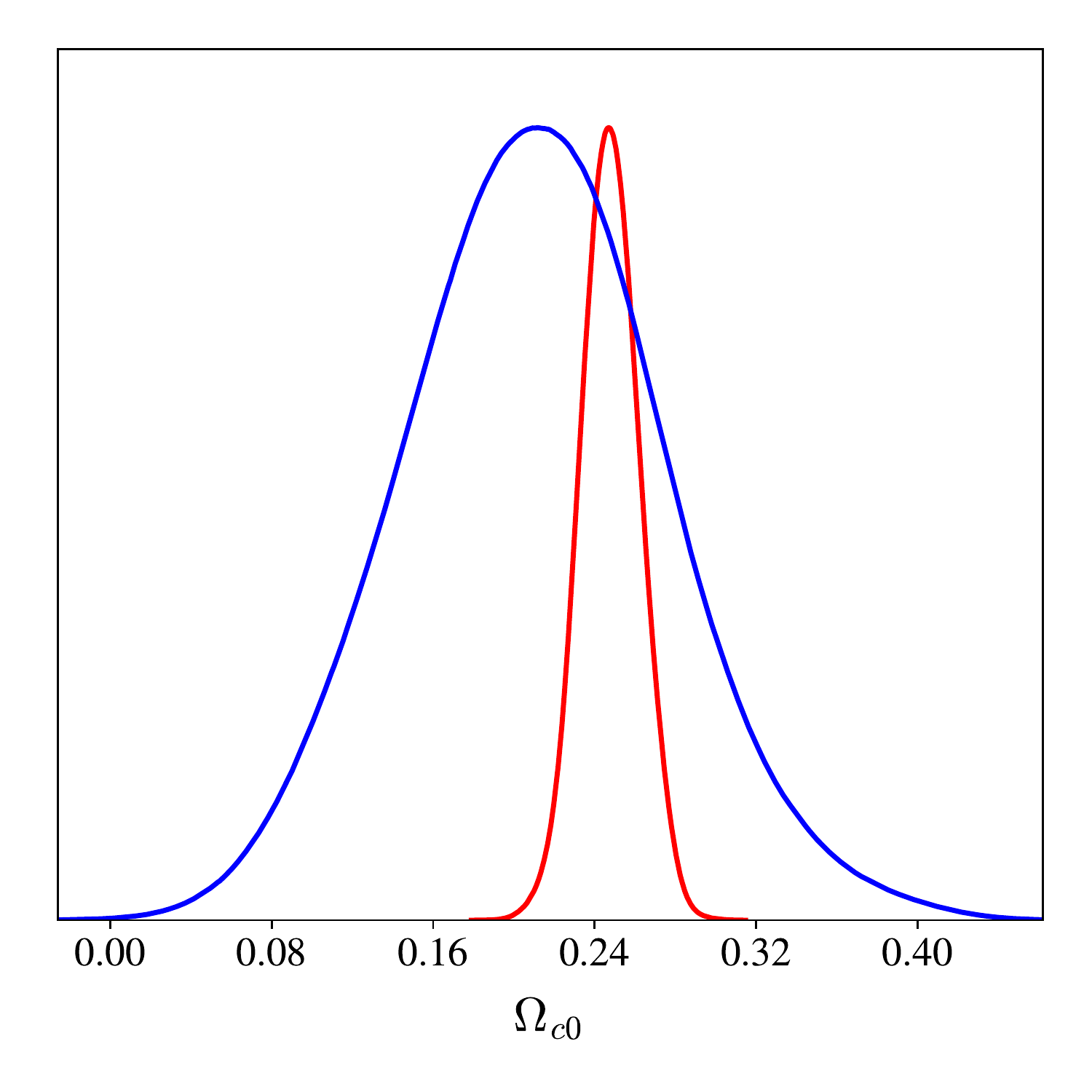} 
\caption{Probability distribution functions for the CDM energy density parameter today. Red lines indicate results for the dynamical DE scenarios whereas blue lines show the interacting counterparts. In all cases, constraints are weaker for the interacting scenarios because of the influence of $w_{0}$ and $w_{a}$ on the CDM dynamics, which implies that uncertainties in $w_{0}$ and $w_{a}$ propagate to $\tilde{\Omega}_{c0}$. For all models the relation $\tilde{\Omega}_{c0}=\bar{\Omega}_{c0}+\bar{\Omega}_{x0}\left(1+w_{0}\right)$ is satisfied. \textbf{Left panel:}~$w$CDM parameterization. \textbf{Middle panel:}~CPL. \textbf{Right panel:}~BA .}
\label{omegacbg}
\end{figure}

As discussed in Sec.~\ref{ssec.data}, our statistical analysis of the background employs the SNe~Ia, BAO, and CC data.
The resulting parameter constraints on all our models is presented in Tab.~\ref{tabbg}. The probability density functions (PDFs) for $\Omega_{c0}$ obtained from the background analysis are shown in Fig.~\ref{omegacbg}, and the complete corner plots can be found in App.~\ref{ap.corner}. Note that due to the degeneracy of the dark sector at the background level, for a given model only the results for $\Omega_{c0}$ change satisfying Eq.~\eqref{rhoc2}. From a physical perspective, Eq.~\eqref{rhoc2} reflects the fact that matter creation/decaying caused by the interaction changes the amount of CDM today $\tilde{\Omega}_{c0}$. Such an effect is very common in interacting models, and have been widely discussed in literature \cite{Yang:2019uzo,Paliathanasis:2019hbi,Kumar:2016zpg}.
\begin{table}
\centering
\begin{tabular}{l|c|c|c|c|c}
\hline\hline
Model                   & $\Omega_{c0}$ 			& $H_{0}$              & $w_{0}$					& $w_{a}$					& $\chi^{2}_{min}$              \\ \hline
$\Lambda$CDM 	        & $0.303_{-0.012}^{+0.013}$ & $69.3_{-1.4}^{+1.3}$ & *                          & *	                        & $1046.9$    \\
$\bar{w}$CDM 	        & $0.303_{-0.013}^{+0.013}$ & $68.8_{-1.7}^{+1.7}$ & $-0.999_{-0.044}^{+0.045}$	& \multirow{2}{*}{*}        & \multirow{2}{*}{$1046.7$} \\
$\tilde{w}$CDM 	        & $0.303_{-0.029}^{+0.034}$ & $69.0_{-1.7}^{+1.8}$ & $-0.998_{-0.042}^{+0.046}$ &                           &                           \\ 
$\overline{{\rm CPL}}$	& $0.298_{-0.016}^{+0.014}$ & $68.8_{-1.6}^{+1.8}$ & $-1.043_{-0.100}^{+0.072}$ & $0.39_{-0.30}^{+0.69}$    & \multirow{2}{*}{$1045.5$} \\
$\widetilde{{\rm CPL}}$	& $0.261_{-0.086}^{+0.053}$ & $68.8_{-1.7}^{+1.7}$ & $-1.055_{-0.110}^{+0.069}$ & $0.38_{-0.25}^{+0.74}$    &                           \\
$\overline{{\rm BA}}$ 	& $0.298_{-0.015}^{+0.014}$ & $68.6_{-1.8}^{+1.8}$ & $-1.057_{-0.088}^{+0.079}$	& $0.32_{-0.27}^{+0.36}$    & \multirow{2}{*}{$1045.5$} \\
$\widetilde{{\rm BA}}$ 	& $0.256_{-0.071}^{+0.063}$ & $68.7_{-1.8}^{+1.8}$ & $-1.058_{-0.088}^{+0.079}$	& $0.32_{-0.27}^{+0.38}$    &                           \\ \hline\hline
\end{tabular}
\caption{Results of the background statistical analysis. As expected, since by construction the models in the dynamical and interacting approaches yield equivalent distance measures, constraints on the background parameters, except for $\Omega_{c0}$, agree between the two approaches.
The differences in $\Omega_{c0}$ are consistent with $\tilde{\Omega}_{c0}=\bar{\Omega}_{c0}+\bar{\Omega}_{x0}\left(1+w_{0}\right)$. The corner plots associated with this analysis are shown in App.~\ref{ap.corner}.}
\label{tabbg}
\end{table}

One can also see that for the interacting dark sector approach these are considerably larger. As expected, in the interacting approach we have a weakening of the parameter constraints for the CDM density contrast. Physically, this is related to the fact that $w_{0}$ and $w_{a}$ are no longer restricted to changing the DE component, but they directly affect both the CDM and DE evolutions. Therefore, the uncertainties related to $w_{0}$ and $w_{a}$ also propagate to $\tilde{\Omega}_{c0}$.

\subsubsection{CMB analysis} 
\label{ssec.CMBconstraints}

\begin{figure}
\centering
\includegraphics[width=0.3\textwidth]{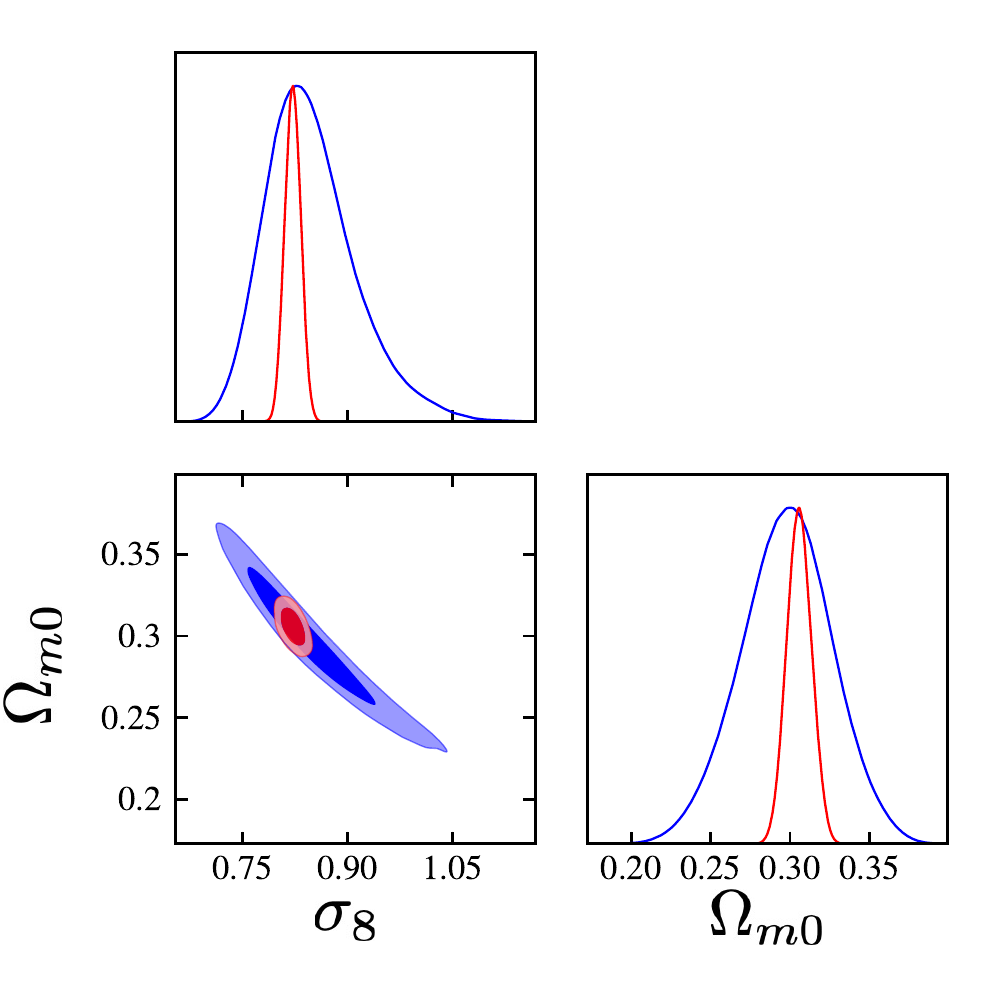} 
\includegraphics[width=0.3\textwidth]{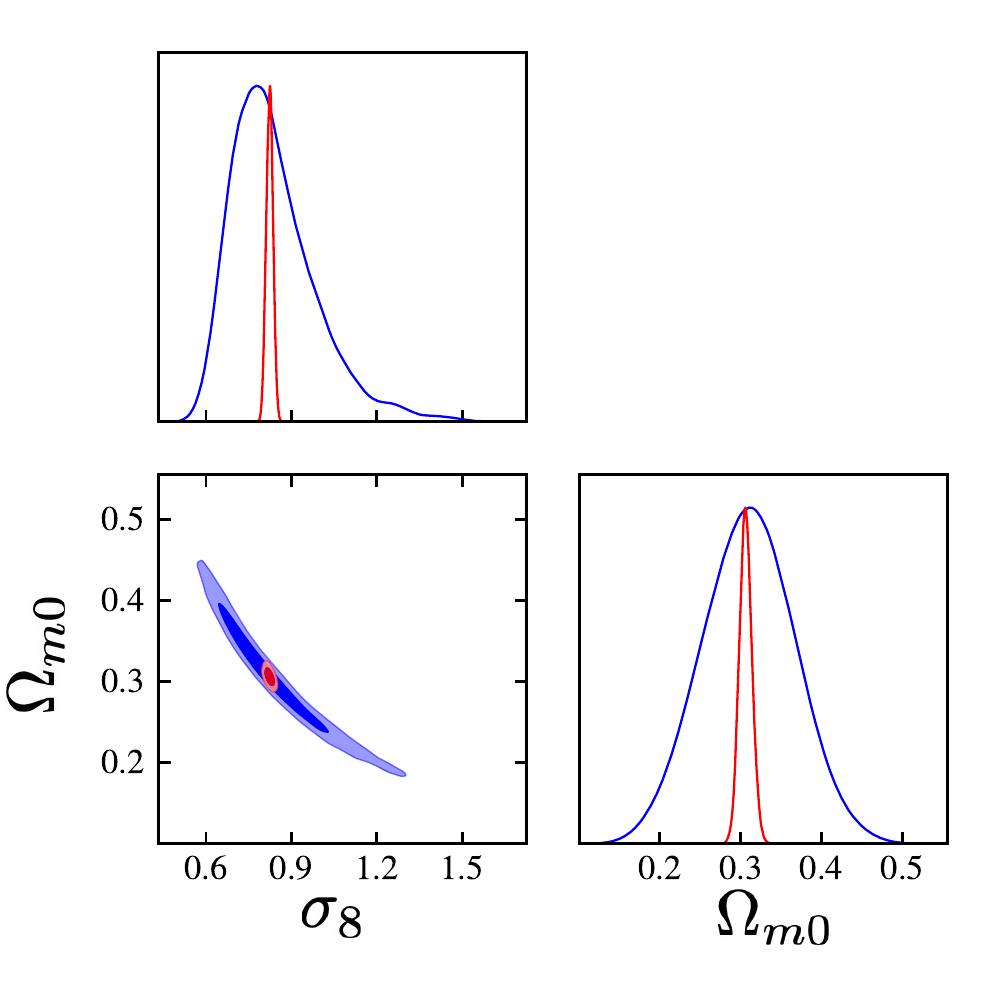} 
\includegraphics[width=0.3\textwidth]{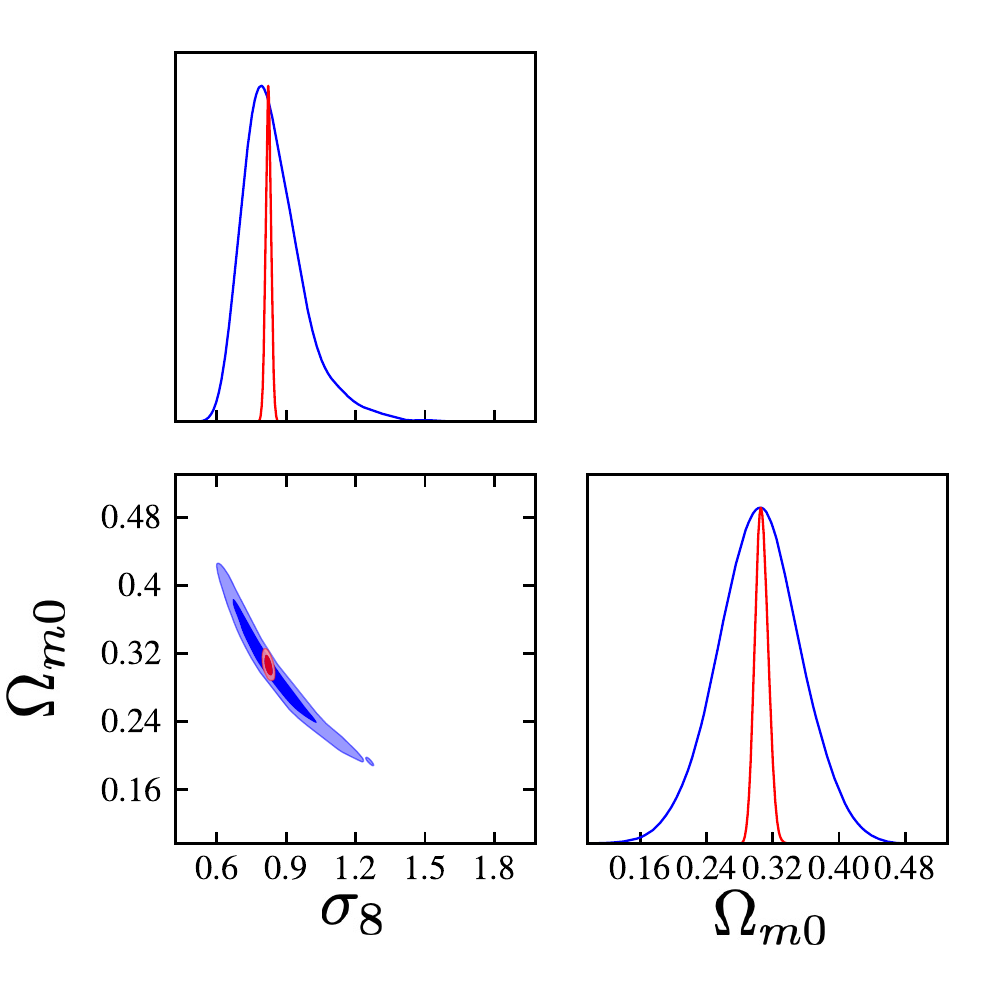} 
\caption{Contour plots for $\Omega_{m0}-\sigma_{8}$. Results of the dynamical DE approach are presented in red whereas those of the interacting counterpart are shown in blue. As in the background case, the fact that $w_{0}$ and $w_{a}$ also affect the CDM evolution in the interacting approach results in a considerable widening of the parameter uncertainties for $\Omega_{m0}$ and $\sigma_{8}$. \textbf{Left panel:}~$w$CDM parametrization. \textbf{Middle panel:}~CPL. \textbf{Right panel:}~BA .}
\label{omsig8}
\end{figure}

\begin{figure}
\centering
\includegraphics[width=0.48\textwidth]{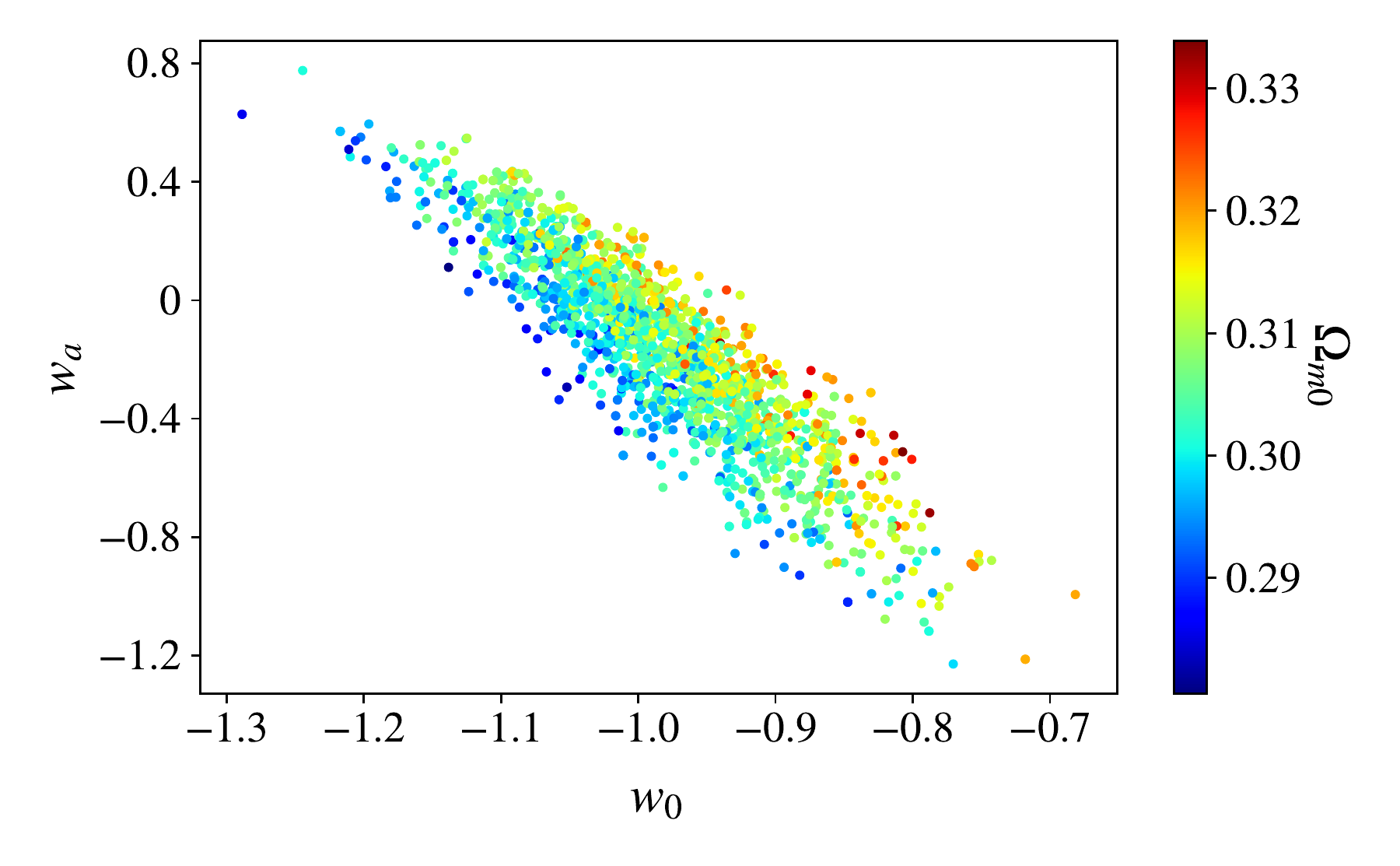} 
\includegraphics[width=0.48\textwidth]{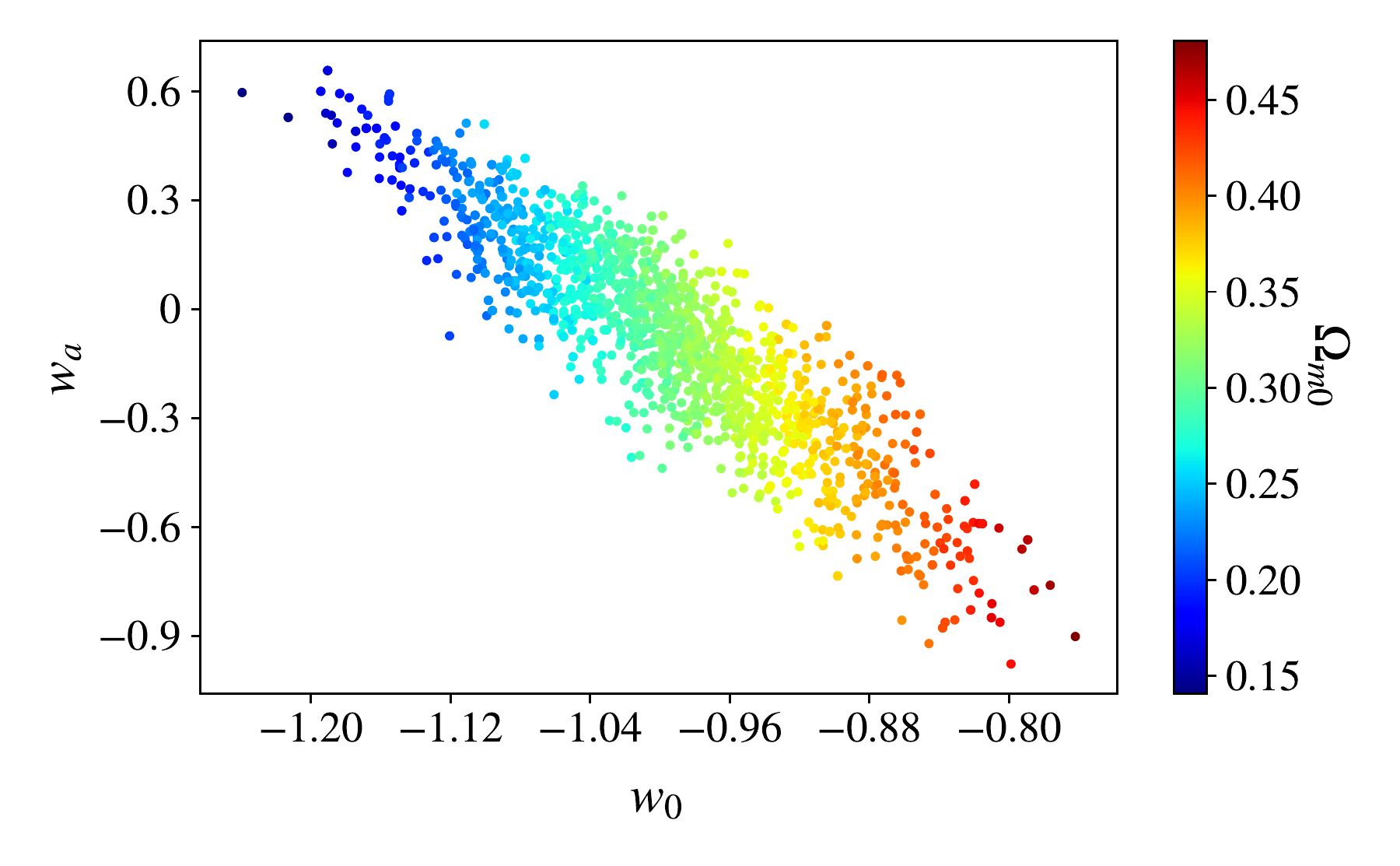} 
\includegraphics[width=0.48\textwidth]{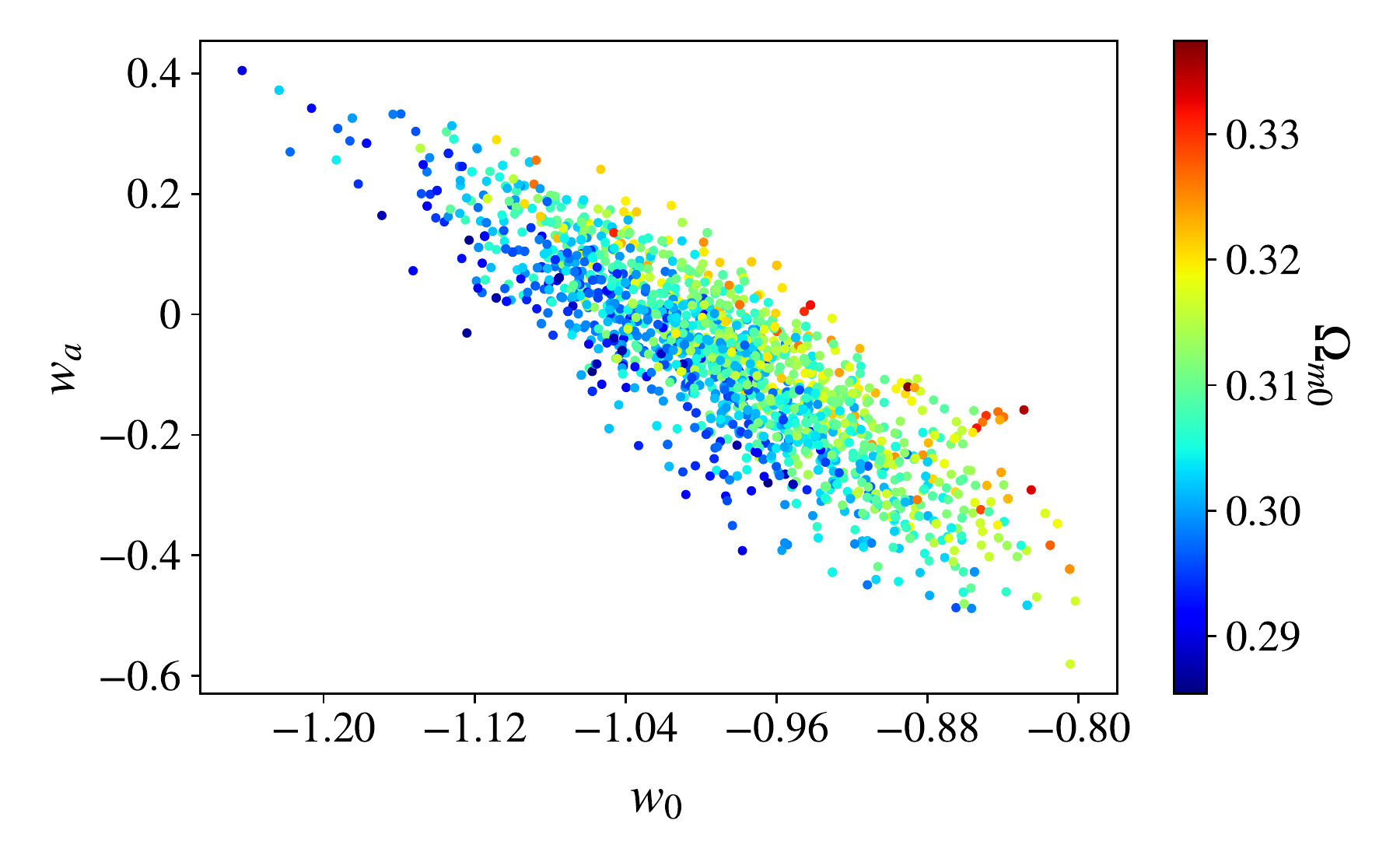} 
\includegraphics[width=0.48\textwidth]{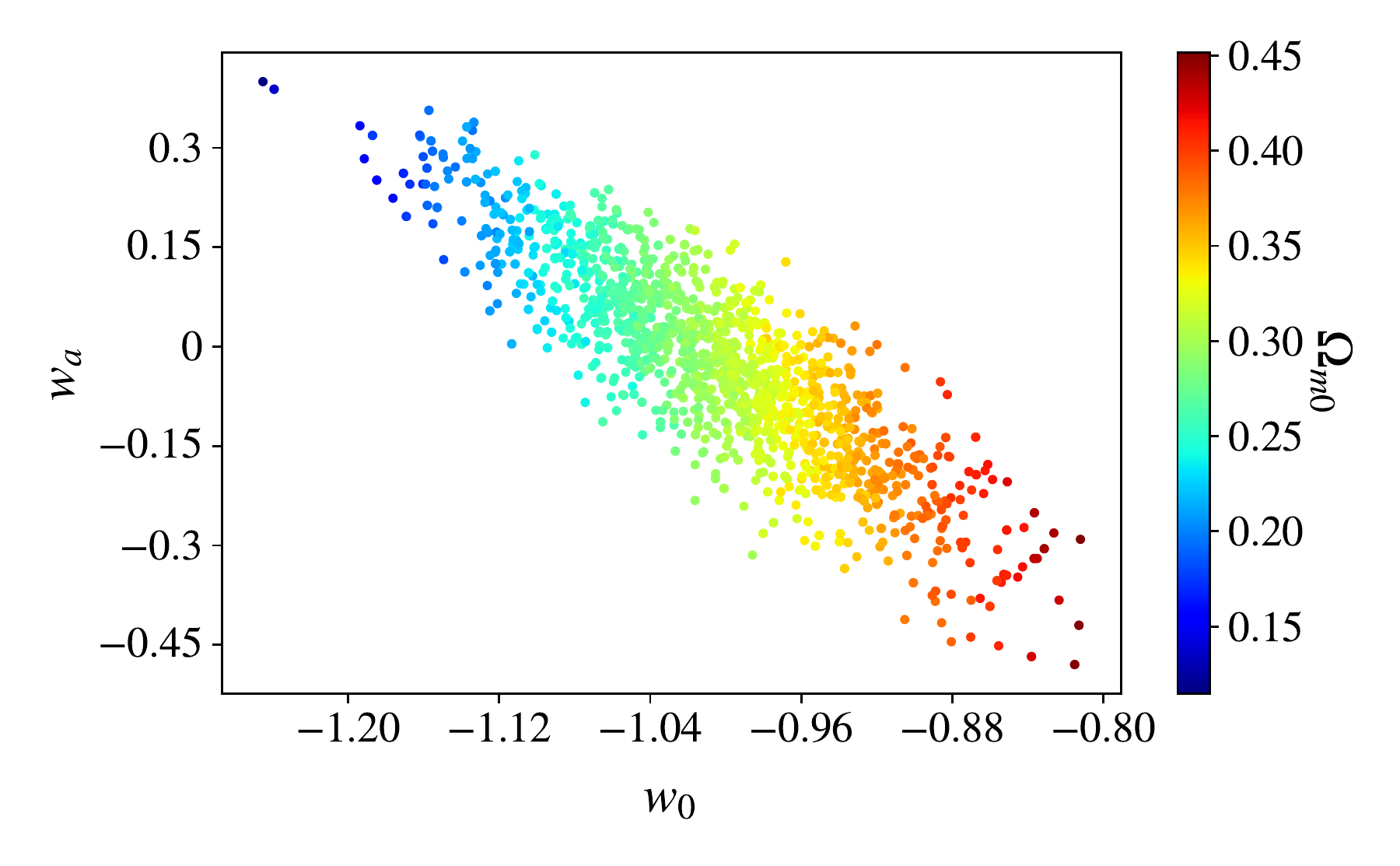}
\caption{Contour plot $w_{0}$-$w_{a}$, colour-coded by $\Omega_{m0}$. The difference in the parameter correlations between the dynamical and interacting approaches becomes explicit. In the dynamical DE picture (right-hand panels) one can see a weak (diffuse) correlation between the parameters, whereas in interacting scenario (left-hand panels) one can see a strong correlation between them. This correlation makes it easier to distinguish the interacting models from $\Lambda$CDM as compared to the dynamical DE models. \textbf{Top left-hand panel: }$\overline{{\rm CPL}}$. \textbf{Top right-hand panel: }$\widetilde{{\rm CPL}}$. \textbf{Bottom left-hand panel: }$\overline{{\rm BA}}$. \textbf{Bottom right-hand panel: }$\widetilde{{\rm BA}}$.}
\label{w0wa}
\end{figure}

Next, we present our results from the analysis including the Planck CMB data (TT, TE, EE+lowE+lensing), which is employed in combination with SNe Ia+BAO.
A summary of the parameter constraints inferred is given in Tab.~\ref{tabplanck}.
\begin{table}
\centering
\footnotesize
\begin{tabular}{l|c|c|c|c|c|c|c}
\hline\hline
Parameter						& $\Lambda$CDM                      & $\bar{w}$CDM					& $\tilde{w}$CDM				& $\overline{{\rm CPL}}$			                        & $\widetilde{{\rm CPL}}$		& $\overline{{\rm BA}}$			& $\widetilde{{\rm BA}}$        \\ \hline
$10^{-2}\omega_{b}$			    & $2.243_{-0.013}^{+0.014}$         & $2.238_{-0.015}^{+0.015}$		& $2.239_{-0.015}^{+0.014}$ 	& $2.236_{-0.014}^{+0.013}$ 	    & $2.235_{-0.015}^{+0.013}$		& $2.225_{-0.015}^{+0.015}$	& $2.228_{-0.016}^{+0.015}$     \\
$\omega_{c}$					& $0.1193_{-0.00091}^{+0.00094}$    & $0.1198_{-0.0011}^{+0.0011}$	& $0.1126_{-0.0090}^{+0.0120}$	& $0.1201_{-0.0012}^{+0.0011}$	& $0.1208_{-0.0240}^{+0.0250}$    & $0.1194_{-0.0014}^{+0.0013}$	& $0.1175_{-0.0210}^{+0.0215}$  \\
$H_{0}$							& $67.71_{-0.42}^{+0.42}$      	    & $68.39_{-0.81}^{+0.78}$      	& $68.17_{-0.90}^{+0.83}$   	& $68.32_{-0.85}^{+0.78}$   	& $68.17_{-0.83}^{+0.86}$ 		& $68.15_{-0.83}^{+0.84}$		& $68.08_{-0.80}^{+0.79}$       \\
$\tau_{reio}$				    & $0.0571_{-0.0076}^{+0.0071}$ 	    & $0.0556_{-0.0080}^{+0.0068}$  & $0.0566_{-0.0085}^{+0.0066}$ 	& $0.0541_{-0.0074}^{+0.0070}$  & $0.0553_{-0.0074}^{+0.0070}$	& $0.064_{-0.014}^{+0.015}$		& $0.068_{-0.013}^{+0.013}$     \\
$\ln\left(10^{10}A_{s}\right)$  & $3.049_{-0.015}^{+0.014}$		    & $3.047_{-0.016}^{+0.013}$		& $3.048_{-0.016}^{+0.014}$ 	& $3.044_{-0.015}^{+0.013}$ 	& $3.046_{-0.014}^{+0.012}$ 	    & $3.061_{-0.027}^{+0.027}$	    & $3.069_{-0.024}^{+0.023}$     \\
$n_{s}$							& $0.9663_{-0.0038}^{+0.0039}$	    & $0.9652_{-0.0040}^{+0.0042}$	& $0.9654_{-0.0039}^{+0.0043}$	& $0.9645_{-0.0041}^{+0.0039}$	& $0.9646_{-0.0045}^{+0.0043}$	& $0.9648_{-0.0047}^{+0.0047}$  & $0.9658_{-0.0046}^{+0.0045}$  \\ 
$w_{0}$							& *	                                & $-1.031_{-0.029}^{+0.034}$	& $-1.021_{-0.032}^{+0.035}$	& $-0.963_{-0.076}^{+0.081}$    & $-0.997_{-0.076}^{+0.077}$  	& $-0.987_{-0.071}^{+0.071}$	& $-1.004_{-0.065}^{+0.066}$    \\
$w_{a}$							& *								    & *								& *								& $-0.27_{-0.27}^{+0.30}$   	& $-0.12_{-0.26}^{+0.34}$ 	    & $-0.08_{-0.15}^{+0.16}$	    & $-0.03_{-0.14}^{+0.15}$       \\ \hline
$\sigma_{8}$					& $0.8106_{-0.0063}^{+0.0060}$ 	    & $0.8197_{-0.0120}^{+0.0110}$ 	& $0.8571_{-0.079}^{+0.051}$ 	& $0.822_{-0.012}^{+0.011}$ 	& $0.832_{-0.170}^{+0.095}$		& $0.823_{-0.012}^{+0.012}$		& $0.848_{-0.153}^{+0.084}$     \\ \hline
$\chi^{2}_{min}$				& $3810$     					    & $3808$     					& $3808$						& $3806$						& $3806$  						& $3807$						& $3807$                        \\
\hline\hline
\end{tabular}
\caption{Results of the statistical analysis employing Planck CMB data. Even though the linear degeneracy is broken by setting the DE sound speed to unity, the parameter estimates and constraints between the dynamical and interacting approaches only change slightly in all models. More precisely, the difference is more pronounced for $w$CDM model, where changes are within $0.7\sigma$, whereas for CPL and BA changes are within $0.3\sigma$.
This is due to the main difference between the two approaches arising from the lensing and ISW effects, which are small in magnitude.
The uncertainties in the cosmological parameters associated to the CDM component are also considerably widened for the interacting approach in all scenarios. This also applies to the current CDM abundance, indicated here by $\omega_{c}$ and the matter fluctuation amplitude $\sigma_{8}$.}
\label{tabplanck}
\end{table}
As already discussed in connection to Figs.~\ref{cmbwcdm}, \ref{cmbcpl} and \ref{cmbba}, the two main effects of the interactions on the CMB are on the lensing and the ISW effect. Since the differences in CMB lensing are small and the large cosmic variance cannot strongly constrain the CMB power spectrum at low $\ell$, no relevant difference appears for the best-fits. However, since $w_{0}$ and $w_{a}$ govern the interacting CDM component, two new features appear in the CMB results when compared to the dynamical DE scenario. The first feature is the significant widening of the uncertainties on $\tilde{\Omega}_{c0}$ and $\tilde{\sigma}_{8}$, which as a consequence is likely to soften the tension between CMB and LSS data.
Fig.~\ref{omsig8} shows the $\Omega_{m0}-\sigma_{8}$ corner plot for the three different parametrizations in both the dynamical DE and interacting scenarios.
The second feature is the correlation between $\tilde{\Omega}_{c0}$ and the parameters $w_{0}$ and $w_{a}$.
This correlation makes it easier to distinguish the interacting models from $\Lambda$CDM than for the dynamical DE models with future data, e.g., from the next-generation LSS surveys. This particularly holds for the CPL and BA parameterizations, which have two free parameters. Fig.~\ref{w0wa} shows the correlation between the parameters $\Omega_{m0}$, $w_{0}$ and $w_{a}$ for the dynamical DE and interacting scenarios. For all parametrizations, this correlation can also be observed in the full CMB triangle plots presented in App.~\ref{ap.corner}.

\section{Conclusions}
\label{sec.conclusions}

In this first paper of a series on the dark degeneracy, we have explored the degeneracy between dynamical DE parameterizations and vacuum interacting DE models ($\tilde{w}_{x}=-1$). We have first derived a general mapping between the two. At the background level, the dynamical DE approach is characterized by the DE EoS parameter $w_{x}\left(a\right)$, whereas the dark sector interaction is characterized by the ratio $r$ between the CDM and DE energy densities.
At linear level of scalar perturbations we have specified the DE pressure perturbation (or equivalently the DE sound speed) and anisotropic stress that causes a degeneracy. For tensor perturbations, the dark degeneracy is tied to the condition that the anisotropic stress is identical between the two approaches.

For illustration of the degeneracy, we adopted three particular interacting models based on three well-established DE parameterizations: $w$CDM, CPL and BA.
We then broke the dark degeneracy at the linear level by setting the comoving DE sound speed to unity ($c_{s}^{2}=1$) in both approaches.
Other parameter values of these models were chosen such that the dark sector interaction only becomes dynamically relevant at late times. As a consequence, the early time physics, e.g., the last scattering and the matter-radiation equality occur at the same epoch in both approaches. Because of this ``equivalence'' at early times, therefore only integrated effects (lensing and ISW) should distinguish between the dynamical and interacting approaches in the CMB power spectra. Similarly, the positions of the maximum of the matter power spectrum would not differ.
We have then computed the CMB temperature, polarization (EE) and lensing-potential power spectra, finding only small differences between the lensing and ISW effects of the two approaches.
Futhermore, we separately analyzed the respective behavior of the gauge-invariant metric potentials, the matter density contrast for CDM and baryons and the total matter power spectrum.
For all models, the CDM density contrast undergoes a suppression at late times in the interacting scenario. This also reflects on the current total matter power spectrum, where the interacting approach shows a considerable decrease in amplitude.

Lastly, we have performed a full Bayesian statistical parameter estimation analysis of all three models in each of the two scenarios, dynamical DE and the interacting case. We divided our analysis into two parts: in the first analysis we only employed background data from type Ia Supernovae (SNe Ia), Baryonic Acoustic Oscillations (BAO) and Cosmic Chronometers (CC), and in the second analysis,
we used the Planck CMB data (TT,TE,EE+lowE+lensing) in combination with the SNe Ia and BAO data.
The two analyses led to qualitatively similar results: the dynamical and interacting descriptions yield almost the same best-fit values and exhibit similar posterior distributions for almost all parameters, except for $\Omega_{c0}$ and $\sigma_{8}$, for which the constraints are considerably weakened in the interacting vacuum DE picture. For the background analysis this agreement in the constraints is expected by construction since both approaches give the same Hubble rate. However, for the CMB analysis, the results imply that the degeneracy we broke by setting $c_{s}^{2}=1$ for both approaches is not sufficient for distinguishing the two descriptions.

The difference in the constraints on $\Omega_{c0}$ and $\sigma_{8}$ arise from the fact that $w_{0}$ and $w_{a}$ do not only affect the DE dynamics in the interacting description but also change the CDM dynamics. For that reason, the uncertainties associated with $w_{0}$ and $w_{a}$ also propagate to $\Omega_{c0}$ and $\sigma_{8}$.
The weakening of these parameter constraints can be used to ease tensions between the KiDS and DES measurements of the matter perturbations with respect to their counterparts from Planck.
In particular, one might argue that the error bars may be underestimated in non-interacting approaches. The ambiguity on the strength of the constraints on $\Omega_{m0}$ and $\sigma_8$ shows that one has to be careful when assessing the significance of a tension. Indeed, as we have shown here, models with very similar phenomenology may yield very different constraints.

Another important issue encountered in our statistical results is that since in the interacting perspective $w_{0}$ and $w_{a}$ affect not only the DE evolution but also the CDM evolution, their correlation with other cosmological parameters depends on the particular approach taken. The main differences appear in the correlation of $w_{0}$ and $w_{a}$ with $\Omega_{c0}$ and $\sigma_{8}$. This can na\"ively be understood from the alignment in the contour of the corner plots presented in Appendix~\ref{ap.corner}. For example, for the dynamical DE approach, all contour plots $w_{0}-\Omega_{c0}$ have vertical alignment, whereas for the interacting approach, the contour plots exhibit a diagonal alignment. This stronger correlation makes it easier to distinguish the interacting approach from $\Lambda$CDM because the deviations from the standard value of $\Omega_{m0}$ lead to different values of $w_{0}$ and $w_{a}$. In this context, the differences between the two approaches we found in the matter perturbations suggest that LSS data, e.g., Redshift-Space Distortions (RSD) and CMB cross-correlations with foreground galaxies through the ISW could provide interesting probes for testing these models and further assess the extent of the dark degeneracy. We leave this task for future work.

\section*{Acknowledgements}
It is a pleasure to thank Saulo Carneiro for useful comments and discussions.
RvM acknowledges support from the Federal Commission for Scholarships for Foreign Students on the Swiss Government Excellence Scholarship (ESKAS No.~2018.0443) for the academic year 2018-2019.
LL was supported by a Swiss National Science Foundation Professorship grant (No.~170547). MK acknowledges funding from the Swiss National Science Foundation. JA acknowledges support from FAPERJ (grant no.  E-26/203.024/2017) and CNPq (grant  no.   310790/2014-0  and 400471/2014-0). The computations were performed on the Baobab cluster of the University of Geneva and on scilab computers of the Observat\'orio Nacional. 

\appendix

\section{General description of the dark degeneracy} 
\label{ap.general}

In Sec.~\ref{sec.equiv} we restricted ourselves to the mapping between dynamical DE and the vacuum interacting DE approaches. However, in the most general case, one can regard scenarios where the DE component is both dynamical and interacting with the CDM component. Such a scenario has recently been proposed in the context of the $H_{0}$ tension~\cite{Yang:2018uae,Pan:2019gop}.

Let us consider two different general descriptions for the dark sector, denoted by the subindex $1$ and $2$. At the background level, each description shall be characterized by a DE EoS parameter $w_{ix}$ and an interaction function $Q_{i}$ (or equivalently, as discussed in Sec.~\ref{ssec.intde}, the ratio between CDM and DE energy densities $r_{i}$) with $i=1,2$. Since the background degeneracy condition follows from the equality in the unified dark EoS parameter (given by the second relation in Eq.~\eqref{EoS}) of the two approaches, if only $w_{1x}$ and $r_{1}$ are known, this single constraint cannot establish a unique degenerate solution for $w_{2x}$ and $r_{2}$.
Instead, for the background degeneracy, three of the time-dependent functions $\{w_{1x}, w_{2x}, r_{1}, r_{2}\}$ must be known. Using the second term of Eq.~\eqref{EoS}, the generalization of the background degeneracy condition ($w_{d1}=w_{d2}$) takes the form
\begin{equation} \label{wrgen}
\dfrac{w_{1x}}{1+r_{1}}=\dfrac{w_{2x}}{1+r_{2}} \,,
\end{equation}
where now the ratio between CDM and DE must satisfy the generalization of Eq.~\eqref{dr},
\begin{equation} \label{frgen}
\dot{r}_{i}+3Hr_{i}\big[f_{i}\left(r_{i}\right)-w_{i}\big]=0 \,.
\end{equation}
In practice, for a given dark sector description ($w_{1x}$ and $r_{1}$), one can always choose a DE EoS parameter $w_{2x}$ and find an interaction $r_{2}$ that satisfies the background degeneracy condition, or in a symmetric way, it is possible to choose an interaction (via $r_{2}$ or via $f_{2}$) and determine the DE EoS parameter $w_{2x}$ that satisfies the background degeneracy condition. Note that, in the last case, if the interaction is chosen via $r_{2}$, $w_{2x}$ can easily be obtained from Eq.~\eqref{wrgen}. In contrast, if the interaction is chosen via $f_{2}$, one must first solve Eq.~\eqref{wrgen} for $w_{2x}$ and then replace it in Eq.~\eqref{frgen} to obtain $r_{2}$.
Subsequently, $w_{2x}$ is again obtained from Eq.~\eqref{wrgen}.

For the background energy density solutions of the dark components, Eqs.~\eqref{rhoc2} and~\eqref{rhox2} can be generalized as
\begin{eqnarray}
\rho_{2c}&=&\rho_{1c}+\rho_{1x}\left(1-\dfrac{w_{1x}}{w_{2x}}\right)\,, \label{rhocgen} \\
\rho_{2x}&=&\dfrac{w_{1x}}{w_{2x}}\rho_{1x}\,. \label{rhoxgen}
\end{eqnarray}

The main difference in this general case lies in the linear-level perturbations because now the DE velocity is dynamically relevant for both approaches. For scalar perturbations, the Einstein equations are still given by Eqs.~\eqref{eq00},~\eqref{eq0i},~\eqref{eqii} and~\eqref{eqij} and energy-momentum conservation is respectively given by Eqs.~\eqref{pertenergy} and~\eqref{pertmomentum}. To obtain identical geometrical quantities at the linear order, the Einstein equations must satisfy the constraints
\begin{eqnarray}
\delta\rho_{1c}+\delta\rho_{1x}&=&\delta\rho_{2c}+\delta\rho_{2x} \,, \label{deltarhodarkgen} \\
\rho_{1c}\theta_{1c}+\rho_{1x}\left(1+w_{1x}\right)\theta_{1x}&=&\rho_{2c}\theta_{2c}+\rho_{2x}\left(1+w_{2x}\right)\theta_{2x} \,, \label{thetadarkgen} \\
\delta p_{1x}&=&\delta p_{2x}\,, \label{deltapdarkgen} \\
\rho_{1x}\left(1+w_{1x}\right)\sigma_{1x}&=&\rho_{2x}\left(1+w_{2x}\right)\sigma_{2x} \,. \label{sigmadarkgen}
\end{eqnarray}
As for the scenario analyzed in Sec.~\ref{sssec.scalar}, the constraints on the DE pressure perturbations and DE anisotropic stress, here given by Eqs.~\eqref{deltapdarkgen} and~\eqref{sigmadarkgen}, are simply algebraic relations.
In this case, for the energy density perturbation and velocity, Eqs.~\eqref{deltarhodarkgen} and~\eqref{thetadarkgen} provide two constraints for four variables ($\delta\rho_{2c}$, $\delta\rho_{2x}$, $\theta_{2c}$ and $\theta_{2x}$)\footnote{Note that the description $1$ is assumed to be known.}.
Hence, to close the system, the energy-momentum conservation must be used. Similarly to the procedure followed in Sec.~\ref{sssec.scalar}, we take the derivatives of Eqs.~\eqref{deltarhodarkgen} and~\eqref{thetadarkgen} and solve them for $\delta\rho_{2c}^{\prime}$ and $\theta_{2c}^{\prime}$ respectively. Afterwards, we replace the results for $\delta\rho_{2c}^{\prime}$/$\theta_{2c}^{\prime}$ in the energy/momentum conservation equations of the CDM component of approach $2$. Using the background degeneracy conditions~\eqref{rhocgen} and~\eqref{rhoxgen} and the relations for the DE pressure perturbation and DE anisotropic stress~\eqref{deltapdarkgen} and~\eqref{sigmadarkgen} combined with the conservation equations for description $1$, it is straightforward to show the DE energy/momentum conservation of approach $2$. A symmetric procedure can be followed in solving Eqs.~\eqref{deltarhodarkgen} and~\eqref{thetadarkgen} for $\delta\rho_{2x}^{\prime}$ and $\theta_{2x}^{\prime}$ as a starting point, which leads to the CDM energy/momentum conservation equations of approach $2$. This means that if the background degeneracy conditions and the conditions on the DE pressure perturbation and DE anisotropic stress are satisfied, the solutions for the energy density perturbation and velocity from the conservation equations automatically satisfy the dark degeneracy at the linear order.

\section{Corner plots} 
\label{ap.corner}

For completeness, we show the corner plots obtained from the statistical analysis presented in Sec.~\ref{sec.stat}. We reiterate that all MCMC processes were performed using {\sc MontePython} and the analysis of the chains as well as the generation of the plots were made using {\sc GetDist}.
All of the analyses satisfy the condition $\hat{R}-1<0.01$ for the Gelman-Rubin convergence parameter.

\subsection{Background corner plots}

\begin{figure}
\centering
\includegraphics[width=\textwidth]{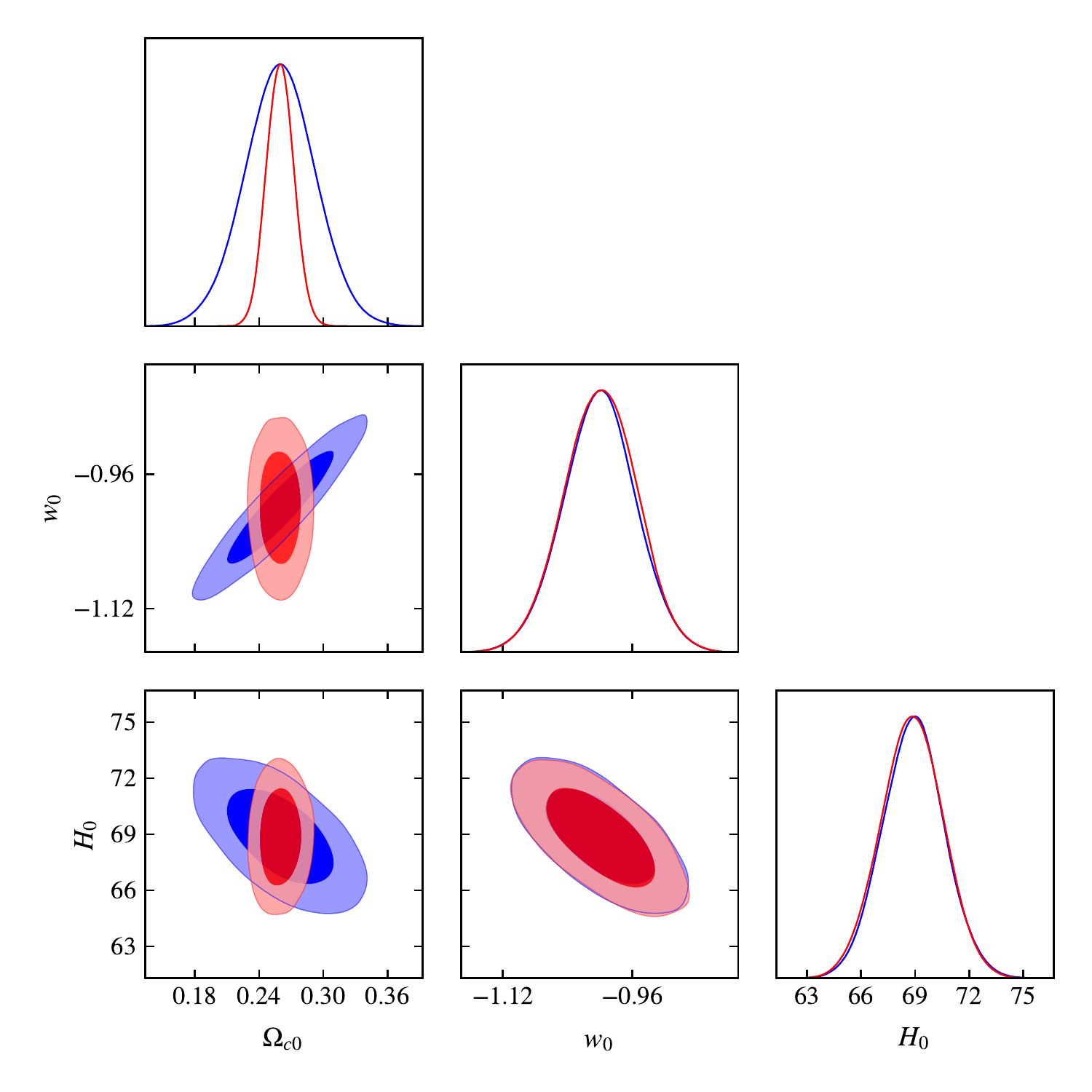} 
\caption{Corner plot for the $w$CDM model using background data from SNe Ia (Pantheon), Cosmic Chronometers and BAO. Results for the dynamical DE approach are presented in red whereas those for the interacting counterpart are shown in blue. Except for the first column, the contours and constraints coincide. In the first column we can see the widening of the uncertainties on $\tilde{\Omega}_{c0}$ and the different correlation between $\Omega_{c0}$ and $w_{0}$ between the dynamical and interacting approaches.}
\label{bgcorner_wcdm}
\end{figure}

\begin{figure}
\centering
\includegraphics[width=\textwidth]{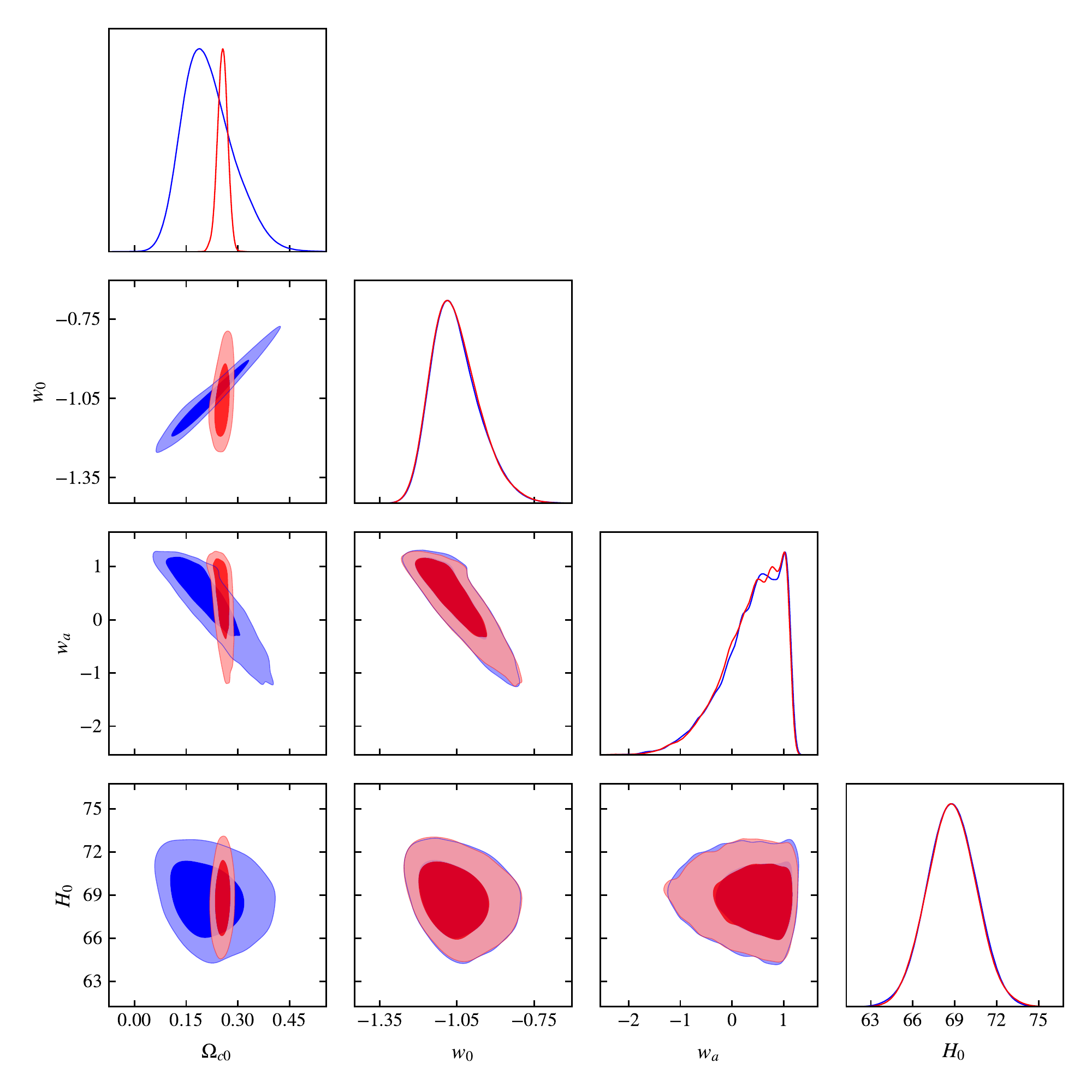} 
\caption{Corner plot for the CPL parameterization using background data from SNe Ia (Pantheon), Cosmic Chronometers and BAO. Results for the dynamical DE approach are presented in red whereas those for the interacting counterpart are shown in blue. Except for the first column, the contours and constraints coincide. In the first column we can see the widening of the uncertainties on $\tilde{\Omega}_{c0}$ and the different correlation between $\Omega_{c0}$ and both $w_{0}$ and $w_{a}$ between the dynamical and interacting approaches.}
\label{bgcorner_cpl}
\end{figure}

\begin{figure}
\centering
\includegraphics[width=\textwidth]{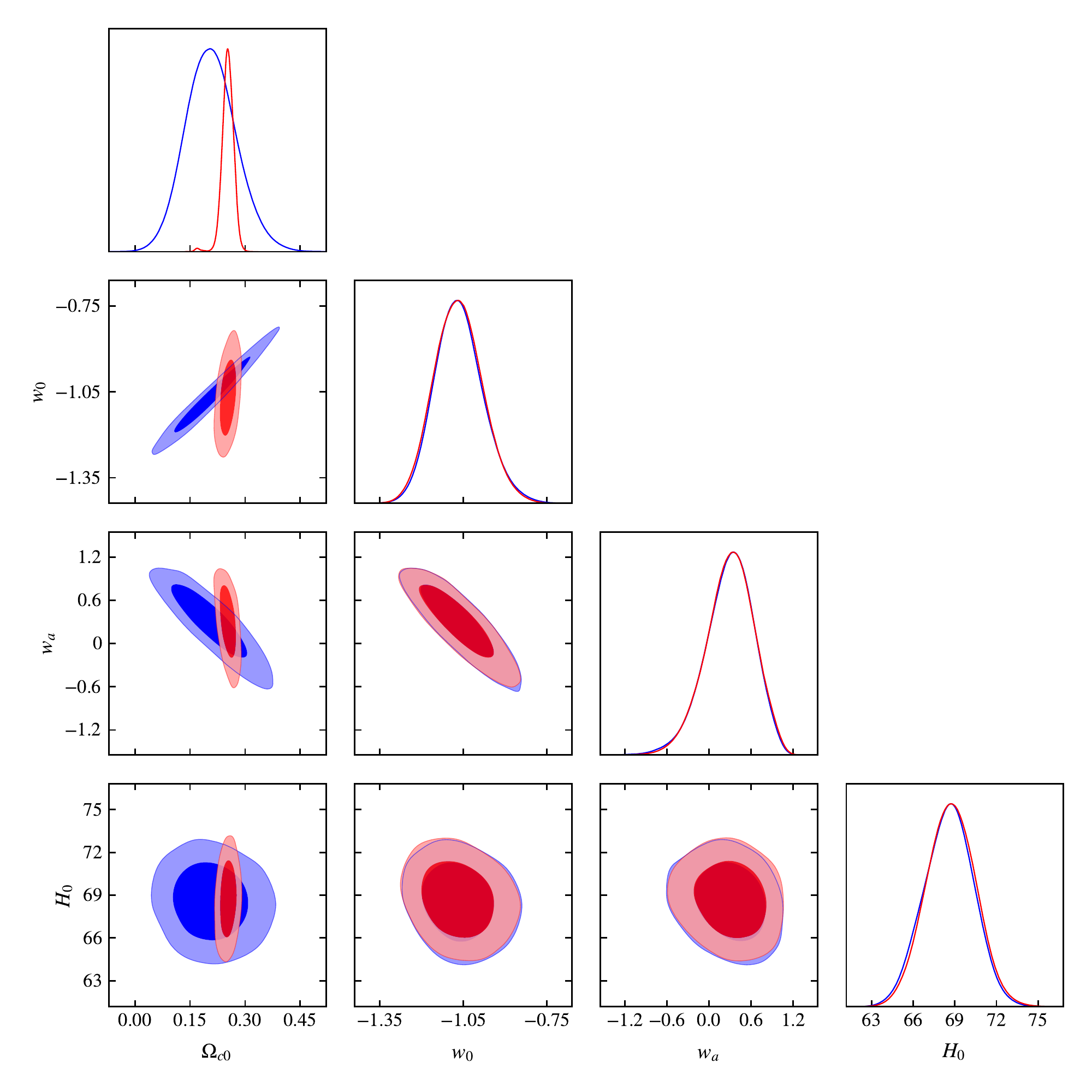} 
\caption{Corner plot for the BA parameterization using background data from SNe Ia (Pantheon), Cosmic Chronometers and BAO. Results for the dynamical DE approach are presented in red whereas those of the interacting counterpart are shown in blue. Except for the first column, the contours and constraints coincide. In the first column we can see the widening of the uncertainties on $\tilde{\Omega}_{c0}$ and the different correlation between $\Omega_{c0}$ and both $w_{0}$ and $w_{a}$ between the dynamical and interacting approaches.}
\label{bgcorner_ba}
\end{figure}

In the background analysis, as already mentioned, the different approaches yield equivalent distance measures by construction. This implies that the same results must be obtained in the background analysis, which means the same best-fits and posterior distributions for all parameters except for $\Omega_{c0}$. For the CDM density parameter the relation $\tilde{\Omega}_{c0}=\bar{\Omega}_{c0}+\bar{\Omega}_{x0}\left(1+w_{0}\right)$ must be satisfied, including error propagation. Figs.~\ref{bgcorner_wcdm},~\ref{bgcorner_cpl} and~\ref{bgcorner_ba} show the background corner plots for the $w$CDM, CPL and BA parameterizations respectively. For all plots, red lines indicate the dynamical DE approach whereas blue lines refer to interacting vacuum DE. For all analyses the results are in good agreement between the two approaches
and tiny differences correspond to fluctuations due to finite MCMC processes.
%

\subsection{Planck corner plots}

\begin{figure}
\centering
\includegraphics[width=\textwidth]{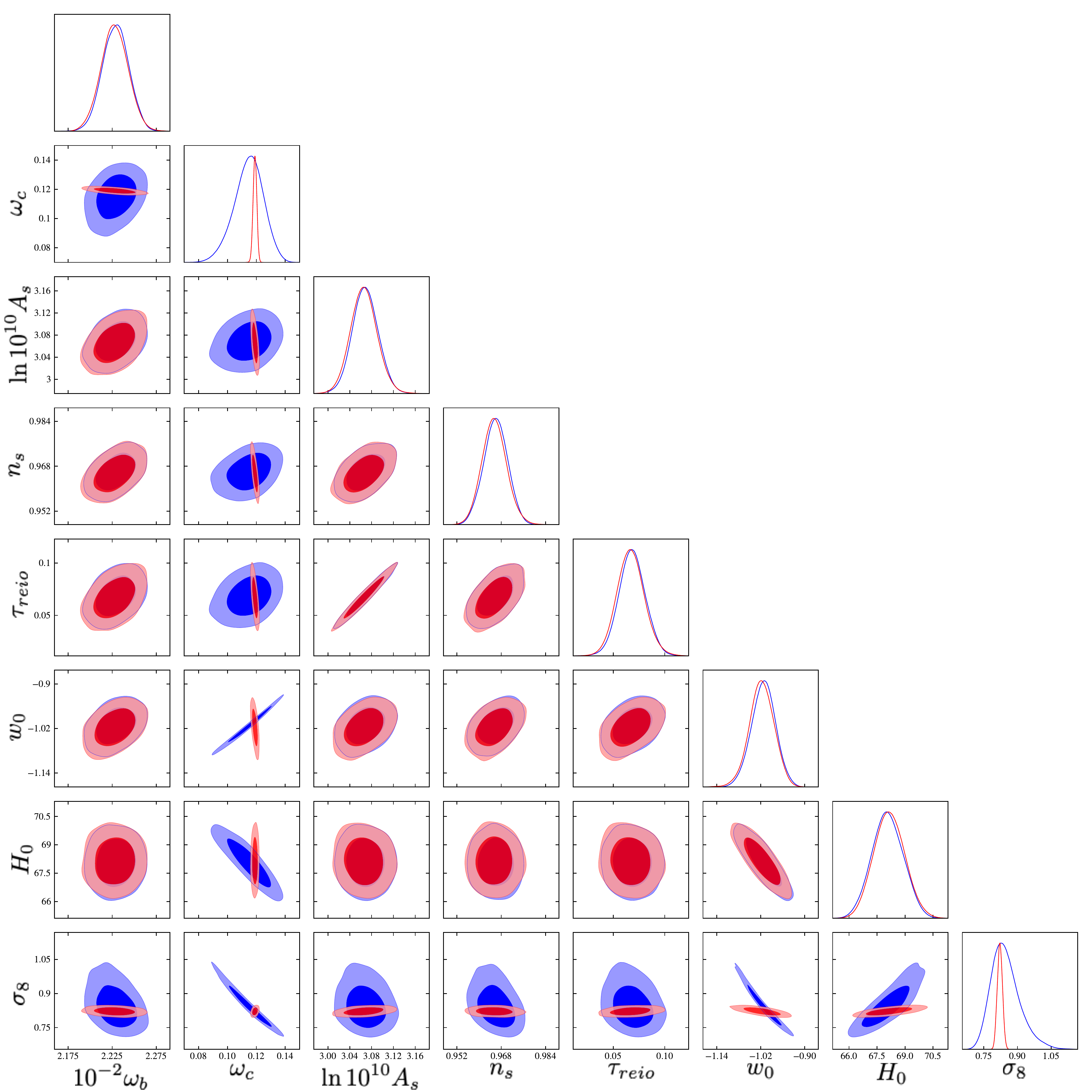} 
\caption{Corner plot for the $w$CDM model using Planck (TT,TE,EE+lowE+lensing), SNe Ia (Pantheon) and BAO data. Results for the dynamical DE approach are presented in red whereas those for the interacting counterpart are shown in blue. Except for the second column, second row and eighth row, the contours and constraints coincide. The second column and the second row are related to $\omega_{c}$, and as for the background analysis we can see the widening of the uncertainties and the different correlations between $\omega_{c}$ and $w_{0}$ between the dynamical and interacting approaches. The eighth row is associated to $\sigma_{8}$, where a widening of the uncertainties and a difference in the correlation with $w_{0}$ is a natural consequence of the differences in $\omega_{c}$.}
\label{planckcorner_wcdm}
\end{figure}

\begin{figure}
\centering
\includegraphics[width=\textwidth]{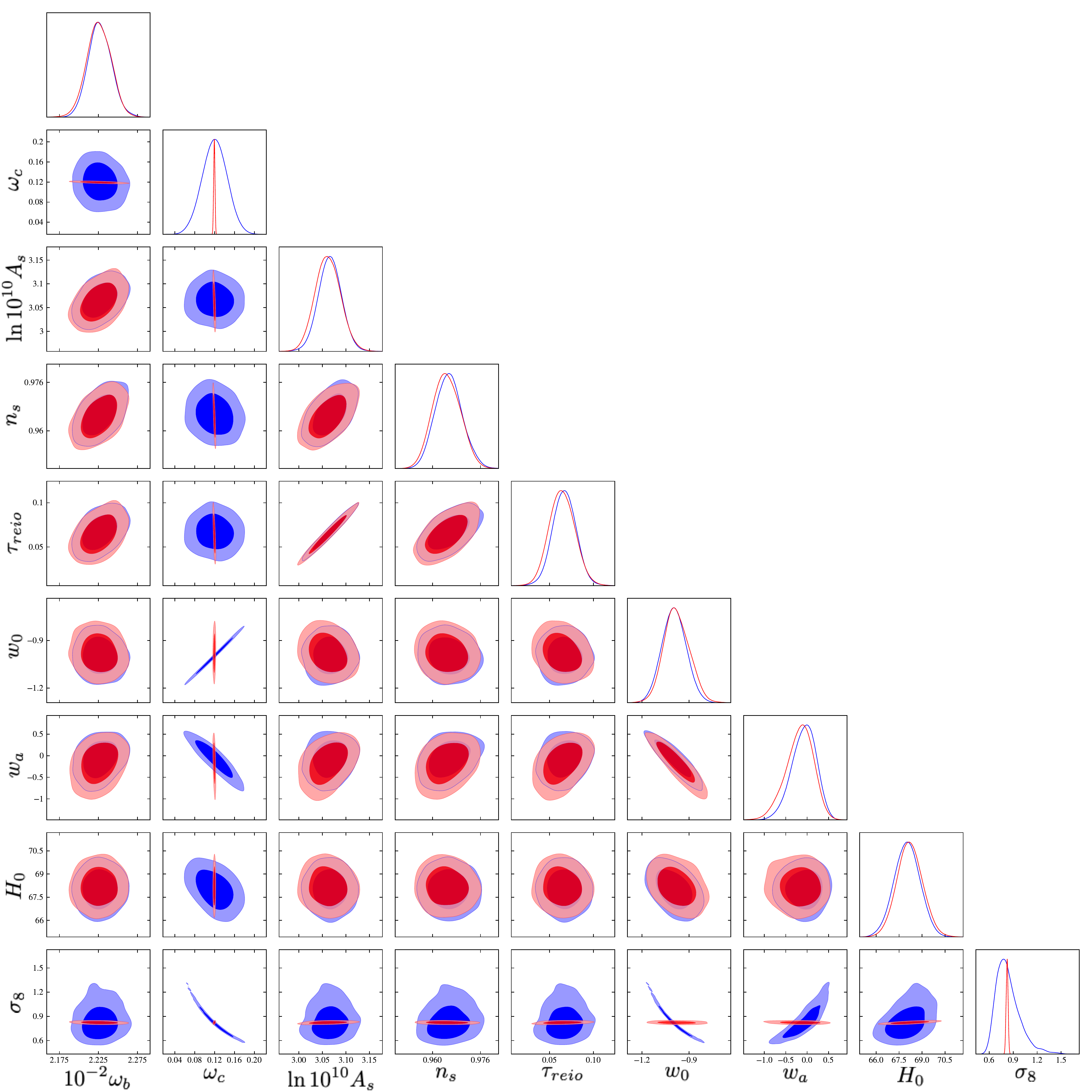} 
\caption{Corner plot for the CPL parameterization using Planck (TT,TE,EE+lowE+lensing), SNe Ia (Pantheon) and BAO data. Results for the dynamical DE approach are presented in red whereas those for the interacting counterpart are shown in blue. Except for the second column, second row and ninth row, the contours and constraints coincide. The second column and the second row are related to $\omega_{c}$, and as for the background analysis, we can see the widening of the uncertainties and the different correlations between $\omega_{c}$ and both $w_{0}$ and $w_{a}$ between the dynamical and interacting approaches. The eighth row is associated to $\sigma_{8}$, where a widening of the uncertainties and a difference in the correlation with both $w_{0}$ and $w_{a}$ is a natural consequence of the differences in $\omega_{c}$.}
\label{planckcorner_cpl}
\end{figure}

\begin{figure}
\centering
\includegraphics[width=\textwidth]{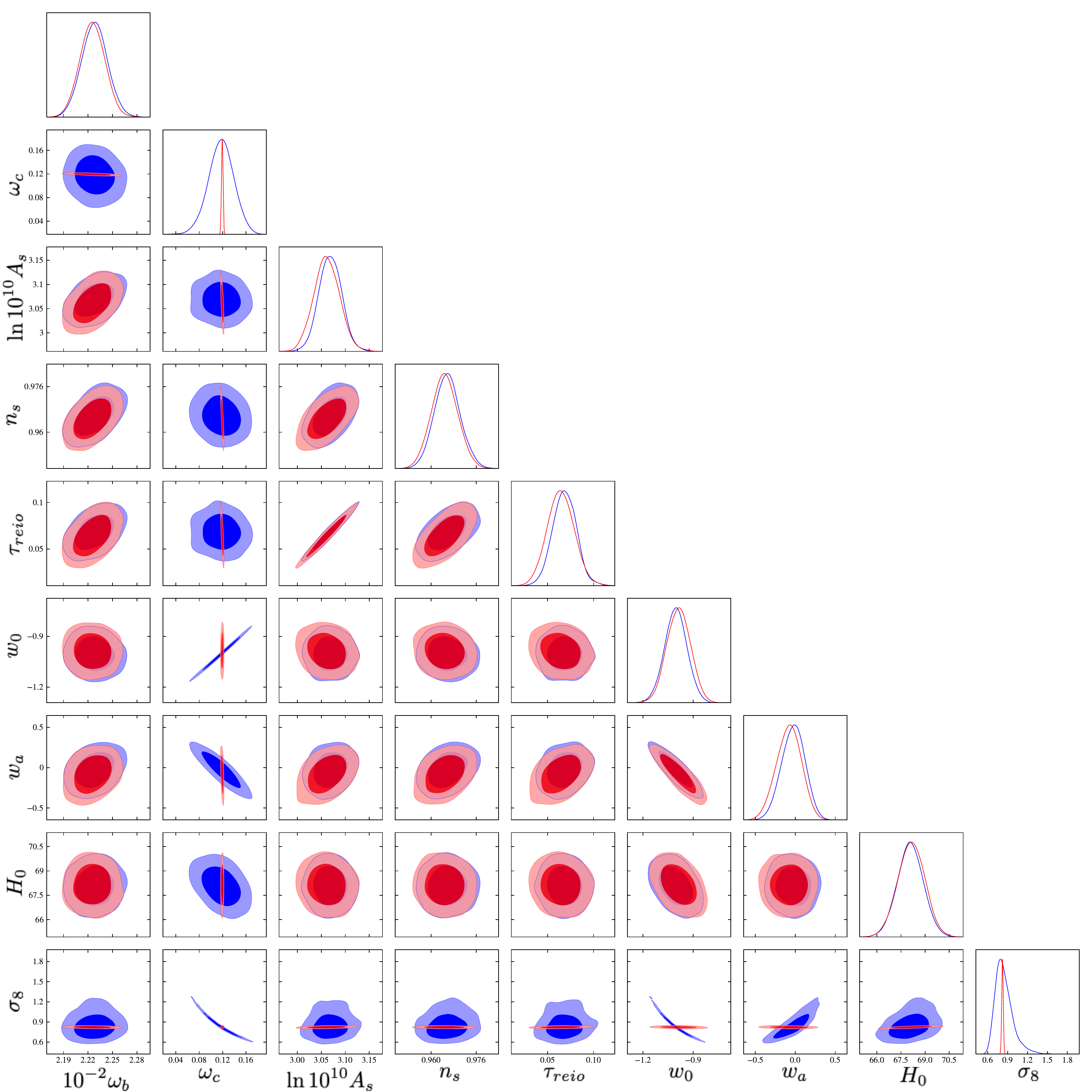} 
\caption{Corner plot for the BA  parameterization using Planck (TT,TE,EE+lowE+lensing), SNe Ia (Pantheon) and BAO data. Results for the dynamical DE approach are presented in red whereas those for the interacting counterpart are shown in blue. Except for the second column, second row and ninth row, the contours and constraints coincide. The second column and the second row are related to $\omega_{c}$, and as for the background analysis we can see the widening of the uncertainties and the different correlations between $\omega_{c}$ and both $w_{0}$ and $w_{a}$ between the dynamical and interacting approaches. The eighth row is associated to $\sigma_{8}$, where a widening of the uncertainties and a difference in the correlation with both $w_{0}$ and $w_{a}$ is a natural consequence of the differences in $\omega_{c}$.}
\label{planckcorner_ba}
\end{figure}

Since the linear degeneracy is broken by the assumption that the DE sound speed is unity  regardless of taking a dynamical or interacting approach, the statistical analysis using CMB data was not \textit{a priori} expected to yield coinciding results. However, given that all models we have analyzed describe the same physics at the last-scattering epoch in either approach, no significant differences translate to the statistical results. This is because the differences in the two integrated effects
on the CMB power spectra are not significant. The first effect is the CMB lensing, which differs between the two approaches because the matter distribution is different. However, as can be seen from the bottom right-hand panels of Figs.~\ref{cmbwcdm},~\ref{cmbcpl} and~\ref{cmbba}, the changes are small. The second is the ISW effect, which is related to the rate at which the lensing potential changes. In this case, although the difference is a bit larger, as can be seen in the region of low $\ell$ in the top left-hand panels of Figs.~\ref{cmbwcdm},~\ref{cmbcpl} and~\ref{cmbba}, the large cosmic variance prevents the CMB data from distinguishing between the dynamical and interacting approaches.
Figs.~\ref{planckcorner_wcdm},~\ref{planckcorner_cpl} and~\ref{planckcorner_ba} show the corner plots for the analysis of the Planck (TT,TE,EE+lowE+lensing) data combined with the SNe Ia and BAO data for the $w$CDM, CPL and BA parameterizations, respectively. 
Red lines again indicate the dynamical DE approach whereas blue lines refer to interacting vacuum DE.


\printcredits

\bibliographystyle{cas-model2-names}

\bibliography{cas-refs}


\end{document}